# WHEN DOES REGULATION BY INSURANCE WORK? THE CASE OF FRONTIER AI

*Cristian Trout*[1]

## ABSTRACT

No one doubts the utility of insurance for its ability to spread risk or streamline claims management; much debated is when and how insurance uptake can improve welfare by *reducing harm*, despite moral hazard. Proponents and dissenters of "regulation by insurance" have now documented a number of cases of insurers succeeding or failing to have such a *net regulatory effect* (in contrast with a *net hazard effect*). Collecting these examples together and drawing on an extensive economics literature, this Article develops a principled framework for evaluating insurance uptake's effect in a given context. The presence of certain distortions – including judgment-proofness, competitive dynamics, and behavioral biases – creates *potential* for a net regulatory effect. How much of that potential gets realized then depends on the type of policyholder, type of risk, type of insurer, and the structure of the insurance market. The analysis suggests regulation by insurance can be particularly effective for catastrophic non-product accidents where market mechanisms provide insufficient discipline and psychological biases are strongest. As a demonstration, the framework is applied to the frontier AI industry, revealing significant potential for a net regulatory effect but also the need for policy intervention to realize that potential. One option is a carefully designed mandate that encourages forming a specialized insurer or mutual, focuses on catastrophic rather than routine risks, and bars pure captives.

---

[1] Artificial Intelligence Underwriting Company. Correspondence to: cristian@aiuc.com. This project has benefitted greatly from insightful comments and suggestions by Peter Wills, Daniel Schwarcz, Kyle Logue, Rajiv Dattani, Iskandar Haykel, Gabriel Weil, Thomas Dorsey; helpful conversations with Ben Garfinkel, Phil Dawson, Aidan Homewood, Halfdan Holm, Andrew Sutton, Connor Aidan Stewart Hunter, Liam Pattel, Yohan Mathew, Ole Teutloff, Abby D'Cruz, Matthew Sharp; and early guidance from Markus Anderljung, Mackenzie Arnold, Gabriel Weil, and Thomas Woodside. I thank everyone here mentioned. I further thank Peter Wills for his extensive mentorship and encouragement. This research was supported by the Centre for the Governance of AI, and the Cambridge Boston Alignment Initiative. This research was not financially supported by the Artificial Intelligence Underwriting Company. *Claude.ai* was used for research, early drafting, and polishing. Errors that remain are my own.

## INTRODUCTION

Insurance is traditionally understood to increase welfare by spreading risk.[2] However, there is a long-running debate regarding another alleged benefit from insurance uptake: insurers serving as effective private regulators disciplined by the market.[3] More precisely, the question is when and how insurance uptake can *reduce harm* (induce closer to optimal levels of care) over some baseline (typically, liability acting alone), *despite* moral hazard. Call this insurers' *net regulatory effect*.

Proponents point to examples of insurers enforcing private safety codes, spreading best practices, auditing company risk management, funding safety R&D, incentivizing the adoption of improved safety technologies, and even lobbying for stricter regulation.[4] Dissenters point to examples of insurers failing to mitigate moral hazard, failing to coordinate amongst themselves, underpricing risk in order to capture more market share, and even encouraging policyholders to avoid liability with box-checking exercises or by hiding information.[5]

The latest instance of this debate surrounds the frontier AI industry.[6] Like other market-based solutions to the AI regulation puzzle,[7] "regulation by insurance" promises to leverage the diversification, experimentation, and powerful information-gathering abilities of markets. It also promises to recalibrate costs and requirements imposed on key actors as markets become more confident in the severity and likelihood of given risks.[8] These are key virtues when trying to manage the high stakes risks of a novel technology developing at the breakneck pace of modern AI,[9] a classic case of the Collingridge dilemma.[10]

---

[2] For a primer, see Michael L. Smith & Stephen A. Kane, *The Law of Large Numbers and the Strength of Insurance*, *in* INSURANCE, RISK MANAGEMENT, AND PUBLIC POLICY: ESSAYS IN MEMORY OF ROBERT I. MEHR 1 (Sandra G. Gustavson & Scott E. Harrington eds., 1994), https://doi.org/10.1007/978-94-011-1378-6_1.

[3] *See e.g.* Tom Baker & Thomas O. Farrish, Liability Insurance and the Regulation of Firearms (Nov. 20, 2004), https://papers.ssrn.com/abstract=621122; Carol A. Heimer, *6. Insuring More, Ensuring Less: The Costs and Benefits of Private Regulation through Insurance*, *in* EMBRACING RISK: THE CHANGING CULTURE OF INSURANCE AND RESPONSIBILITY 116 (Tom Baker & Jonathan Simon eds., 2010), https://www.degruyterbrill.com/document/doi/10.7208/9780226035178-008/pdf?licenseType=restricted; Tom Baker & Rick Swedloff, *Regulation by Liability Insurance: From Auto to Lawyers Professional Liability*, 60 UCLA L. REV. 1412 (2012); Kenneth S. Abraham, *Four Conceptions of Insurance*, 161 U. PA. L. REV. 653 (2012); Omri Ben-Shahar & Kyle D. Logue, *Outsourcing Regulation: How Insurance Reduces Moral Hazard*, 111 MICHIGAN LAW REVIEW 197 (2012); Kenneth S. Abraham & Daniel Schwarcz, *The Limits of Regulation by Insurance*, 98 IND. L.J. 215 (2022).

[4] Ben-Shahar and Logue, *supra* note 3.

[5] Abraham and Schwarcz, *supra* note 3.

[6] Anat Lior, *Insuring AI: The Role of Insurance in Artificial Intelligence Regulation*, 35 HARV. J. L. & TECH. 467 (2021); Cristian Trout, *Liability and Insurance for Catastrophic Losses: The Nuclear Power Precedent and Lessons for AI*, *in* GENERATIVE AI AND LAW WORKSHOP AT THE INTERNATIONAL CONFERENCE ON MACHINE LEARNING (2024); Daniel Schwarcz & Josephine Wolff, The Limits of Regulating AI Safety Through Liability and Insurance: Lessons From Cybersecurity (Aug. 27, 2025), https://papers.ssrn.com/abstract=5411062; Gabriel Weil, *Overcoming Judgment-Proofness: The Law & Economics of Insuring and Mitigating AI Risk* (forthcoming); Lukasz Szpruch et al., Insuring AI: Incentivising Safe and Secure Deployment of AI Workflows (Sept. 19, 2025), https://papers.ssrn.com/abstract=5505759.

This Article contributes to both streams of scholarship. So far, proponents and dissenters of "regulation by insurance" (myself included) have been content to enumerate more and more examples of insurers succeeding or failing to have a net regulatory effect. By collecting together this rich array of case studies and drawing on a broad economics literature, this Article develops a more principled framework for evaluating when and how insurance uptake can have a net regulatory effect.

The framework operates in two steps. First, identify market distortions and biases that insurance uptake might correct. This determines the *potential* for a net regulatory effect. Second, assess whether conditions favor insurers *realizing* that potential through effective moral hazard mitigation. After elaborating each part of the framework, I demonstrate its application to my topical case study, the frontier AI industry.

Part I catalogs the many distortions and biases insurers can address, expanding far beyond the canonical example of judgment-proofness.[11] Part II evaluates the frontier AI industry against this catalog, detecting a number of said biases and distortions at work, especially when it comes to managing catastrophic risks. These distortions include: an underprovisionment of safety R&D due to upfront fixed, costs, and considerable spillovers; an existential race to capture market share incenting unbalanced R&D investment; collective action problems relating to the damage an AI Three Mile Island would inflict on the industry's collective reputation; and the prevalence of overconfidence, availability bias, and the winner's curse in a young industry with many venture-backed startups. Part III identifies the factors that determine whether a potential regulatory effect gets *realized*. This largely turns on when and why insurers succeed or fail at mitigating moral hazard. Factors include the type of risk that's being insured, the type of policyholder being insured, the type of insurers that will be involved, and the structure of the insurance market. I also find that *channeling* liability through a no-fault and or exclusive liability regime synergizes best with a regulation by insurance strategy. Part IV sketches policy implications for the frontier AI industry: I find that realizing insurers' maximum potential here involves inducing the formation of a specialized monopolist insurer (ideally a mutual) that covers a few major AI developers for catastrophic risks typically excluded by insurance policies. This is unlikely to happen by default: policymakers would need to intervene on the market with a carefully designed insurance mandate. Part V concludes and articulates questions for further research.

---

[7] Gillian K. Hadfield & Jack Clark, Regulatory Markets: The Future of AI Governance (Apr. 25, 2023), http://arxiv.org/abs/2304.04914; Trout, *supra* note 6; Mirit Eyal & Yonathan A. Arbel, Racing to Safety: Tax Policy for AI Safety-by-Design (May 8, 2024), https://papers.ssrn.com/abstract=5181207; *AI Security Tax Incentives*, AMERICANS FOR RESPONSIBLE INNOVATION, https://ari.us/policy-bytes/ai-security-tax-incentives/ (last visited Sept. 15, 2025); Philip Moreira Tomei, Rupal Jain & Matija Franklin, AI Governance through Markets (Jan. 29, 2025), http://arxiv.org/abs/2501.17755; Dean W. Ball, A Framework for the Private Governance of Frontier Artificial Intelligence (Apr. 15, 2025), http://arxiv.org/abs/2504.11501; Benjamin Gil Friedman, Shared Residual Liability for Frontier AI Firms (July 5, 2025), https://papers.ssrn.com/abstract=5339887.

[8] A thorough comparison of these market-based solutions cannot be given here, but my argument in favor of regulation by insurance would boil down to this: alternatives seem to be trying to reinvent the wheel; insurance is a mature industry with an ecosystem of tested actors, and whose regulation is more or less a solved problem. Similarly, I cannot provide an in-depth comparison of regulation by insurance with traditional command and control regulatory modes (e.g. licensing regimes or direct supervision).

[9] Noam Kolt, Algorithmic Black Swans (Oct. 14, 2023), https://papers.ssrn.com/abstract=4370566.

[10] David Collingridge, *The Dilemma of Control*, *in* THE SOCIAL CONTROL OF TECHNOLOGY 13 (1980).

[11] Kyle D. Logue, *Solving the Judgment-Proof Problem*, 72 TEX. L. REV. 1375 (1993).

## I. A MODEL OF INSURERS' REGULATORY EFFECT

I model insurer's regulatory effect as a corrective force against distortions to insureds' activity and care levels.[12] A *net* regulatory effect occurs when insurers' efforts to mitigate moral hazard or otherwise reduce losses *more than compensate* for the moral hazard the insurance coverage generates. The *potential* for a net regulatory effect depends on how strong background distortions were before insurance uptake.

Part I of this Article provides an industry-agnostic catalog of such distortions, relying primarily on the theoretical literature. (Readers primarily interested in the AI industry may wish to skip directly to Part II). In general, I find there is much less potential for a net regulatory effect for product accidents and small risks. This is broadly because market forces complement liability as an incentive to take care for these risks. First however, I need to lay some groundwork.

---

[12] Activity levels refer to how much an actor engages in a particular economic activity (e.g. miles driven by car), where care levels refer to the precautionary effort an actor makes while engaging in that activity (e.g. how cautiously one drives, per mile). In what follows I generally gloss over the distinction, raising it only when relevant. I will often focus on care levels, as that is often the more difficult thing for insurers to assess in a policyholder, and therefore harder for them to incentivize. In other words, the claim that insurers correct care levels is the stronger (and hence more interesting) claim: that insurers correct activity levels often more or less follows. For more on the distinction see STEVEN SHAVELL, FOUNDATIONS OF ECONOMIC ANALYSIS OF LAW ch. 2 (2004).

### *A. The Regulation Thesis*

Increasing welfare by spreading risk is the central role of insurance. Call this the *financial efficiency effect* of insurance.[13] I elaborate this effect to set it aside.[14]

Our interest in insurance is as a *regulatory tool*. The claim that insurers can and do play this quasi-regulatory role is what Abraham and Schwarcz call the "regulation thesis."[15] The basic intuition behind this regulation thesis is well articulated by Ben-Shahar and Logue:[16]

> Like a regulator setting standards of conduct and monitoring behavior, insurers have to assess the distribution of harm and determine the desirability of safety measures. And like courts adjudicating liability and awarding damages, insurers have to administer claims, verify harms, and determine the comparative causation of other parties.

This thesis goes beyond the uncontroversial claim that insurance is sometimes purchased for the loss prevention and claims management services that come with it. There, the efficiency gains come from a division of labor:[17] call that the *specialization effect* of insurance.

The regulation thesis is much more controversial: it says that insurance uptake can *reduce harm*, *despite* moral hazard. Or in sharper terms: all else being equal, insurance uptake can bring activity and care levels closer to efficient levels relative to a baseline, i.e. markets, liability, and self-interest operating in the absence of insurance. This is what I call a *net regulatory effect* from insurance.

---

[13] For a primer, see generally Smith and Kane, *supra* note 2; In the same vein, some scholars argue that insurance uptake can spur innovation and competition essentially because smaller companies will struggle to diversify and spread their risk. See e.g. Lior, *supra* note 6 § IV.

[14] Another important function of insurance is to satisfy a preference for risk-aversion (and indirectly, spur economic activity). Given my focus on *firms* (which are typically risk-neutral), I mostly ignore it.

[15] Abraham and Schwarcz, *supra* note 3; cf. Abraham, *supra* note 3 (elaborating a "governance conception" of insurance); cf. Ben-Shahar and Logue, *supra* note 3 (discussing "regulation by insurance").

[16] Ben-Shahar and Logue, *supra* note 3 at 200.

[17] This is the effect that explains the prima facie confusing observation that many large firms still purchase insurance despite their risk being naturally spread over a large operation and their portfolios typically being quite diversified. Álvaro Parra & Ralph A. Winter, *Optimal Insurance Contracts under Moral Hazard*, *in* HANDBOOK OF INSURANCE: VOLUME II 129, § 8 (Georges Dionne ed., 2025), https://doi.org/10.1007/978-3-031-69674-9_6; Tom Baker & Peter Siegelman, *Chapter 7: The Law and Economics of Liability Insurance: A Theoretical and Empirical Review*, § 3.B (2013), https://www.elgaronline.com/edcollchap/edcoll/9781848441187/9781848441187.00015.xml .

That insurance *sometimes* has this effect is undisputed.[18] The question is how common it is, how strong it is, and under what conditions it tends to happen. If we can answer these questions, we can make better policy recommendations regarding insurance, such as when and why an insurance mandate might be advisable. To do so, we first need a clearer model of what this net regulatory effect is.

## *B. The Net Regulatory Effect from First Principles*

### 1. The Idealized Scenario and Efficient Care Level

Consider an idealized scenario: all actors are uninsured, perfectly rational, self-interested, and have perfect information; there are no market failures, all collective action problems are solved, etc.; and the courts are working properly, with the full force of liability acting on agents as the law intends.

It is under these assumptions that scholars derive efficiency results, e.g. proving that under various liability regimes agents will choose the efficient level of care, where the marginal cost equals the marginal social welfare of further precaution.[19]

### 2. Background Distortions

Reality is never so neat, however. To get a more realistic baseline, we need to introduce whatever incentive distortions are present that might thwart efficient outcomes, driving toward excessive or deficient care levels. These distortions include:

- *Judgment-proofness*: Firms lacking sufficient assets to cover potential liabilities, creating moral hazard by capping downside risk. (§I.D.1)
- *Dynamics of competition*: The presence of competition distorting incentives in a wide variety of ways, such as through R&D spillovers, fixed or upfront costs, and first-mover advantages. (§I.D.2)
- *Information asymmetries*: When the party purchasing a product has less information about its risks or accident probabilities than the seller (§I.D.2.a)
- *Collective action problems*: Situations where individually rational behavior leads to collectively suboptimal outcomes. (§I.D.2.a)
- *Financial distress*: When budget constraints or firm survival motives distort incentives. (§I.D.3)
- *Charity hazard*: When the prospect of a government bailout or subsidy caps downside risk. (§I.D.4)
- *Behavioral biases*: Systematic deviations from rational decision-making, such as optimism or overconfidence. (§I.D.5)

---

[18] *See generally* Abraham and Schwarcz, *supra* note 3.

[19] SHAVELL, *supra* note 12 ch. 2.

### 3. Moral Hazard

Now introduce insurance and its moral hazard, the reduction in incentive to take care that occurs when risk is transferred to insurers. The magnitude of moral hazard is primarily determined by insurance policy limits: the greater the coverage, the larger the potential moral hazard.

Moral hazard operates as a countervailing force to any regulatory effect insurers might provide, driving care levels below optimal levels. The net effect on welfare depends on whether insurers' loss prevention efforts – their regulatory effect – can overcome this.

### 4. Regulatory Effect

Finally, introduce what I call insurance's "regulatory effect." This encompasses the positive impact on welfare arising from insurers' efforts to mitigate moral hazard or otherwise reduce losses. For explicitly mitigating moral hazard – monitoring and controlling insureds' behavior – these tools include:[20]

- Exposing insured to some risk through copayments and deductibles.
- Individually risk-pricing premiums
- Monitoring insured behavior
- Refusing to insure, or conditioning coverage on adherence to certain risk management practices and standards
- Voluntary coaching or private safety code enforcement
- Investigating claims and refusing to pay claims when insureds violate policy terms (e.g. through gross negligence)

Thus, to a large extent, moral hazard mitigation and insurers' regulatory effect are two sides of the same coin. Other loss reduction efforts include:

- Conducting safety R&D and disseminating safer technological alternatives
- Lobbying the government for stricter regulation of insured activities

I elaborate on these efforts in Part III.

### 5. Putting it Together: The Net Effects

Figures 1 and 2 illustrate how these concepts fit together, showing how insurers can sometimes produce a net regulatory effect or net moral hazard effect, respectively.

---

[20] Ben-Shahar and Logue, *supra* note 3 § I; *Cf.* Baker and Swedloff, *supra* note 3 § I.

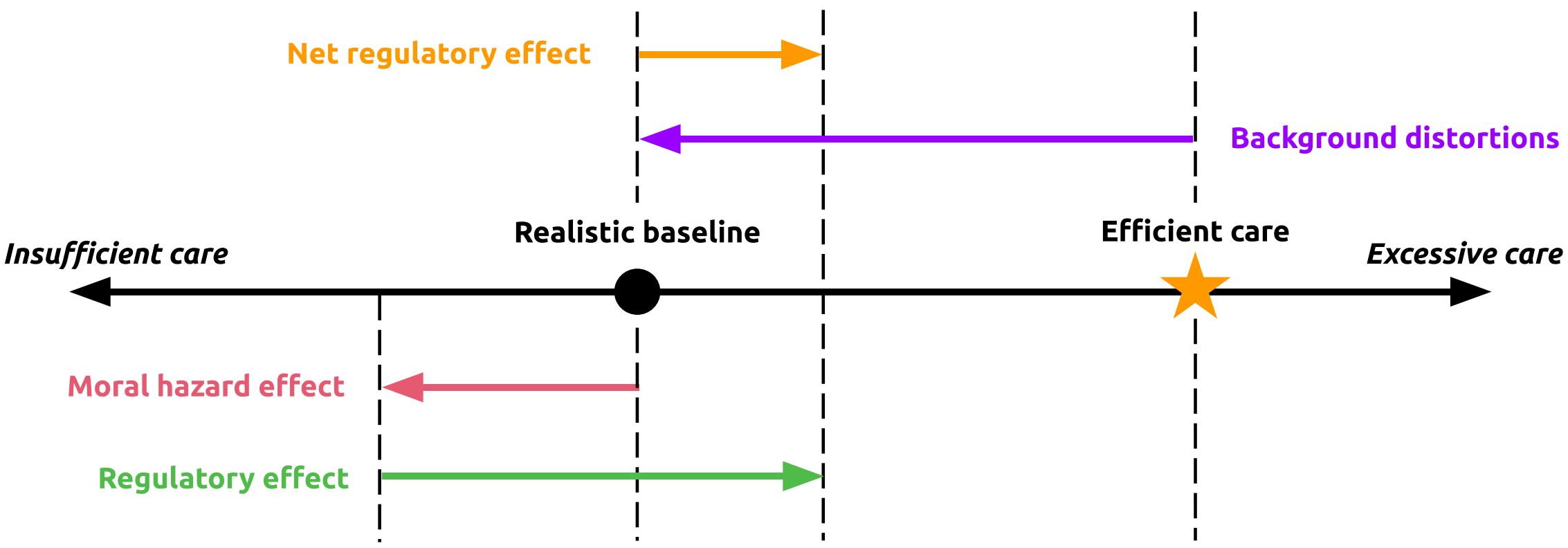

*Figure 1. Model of a net regulatory effect from insurance uptake.*

The welfare gains from the *financial efficiency* and *specialization* effects can also offset the moral hazard effect, but those introduce dimensions beyond the scope of this model. Obviously the model is simple; its purpose is merely to organize and clarify the relation between the concepts in the literature.

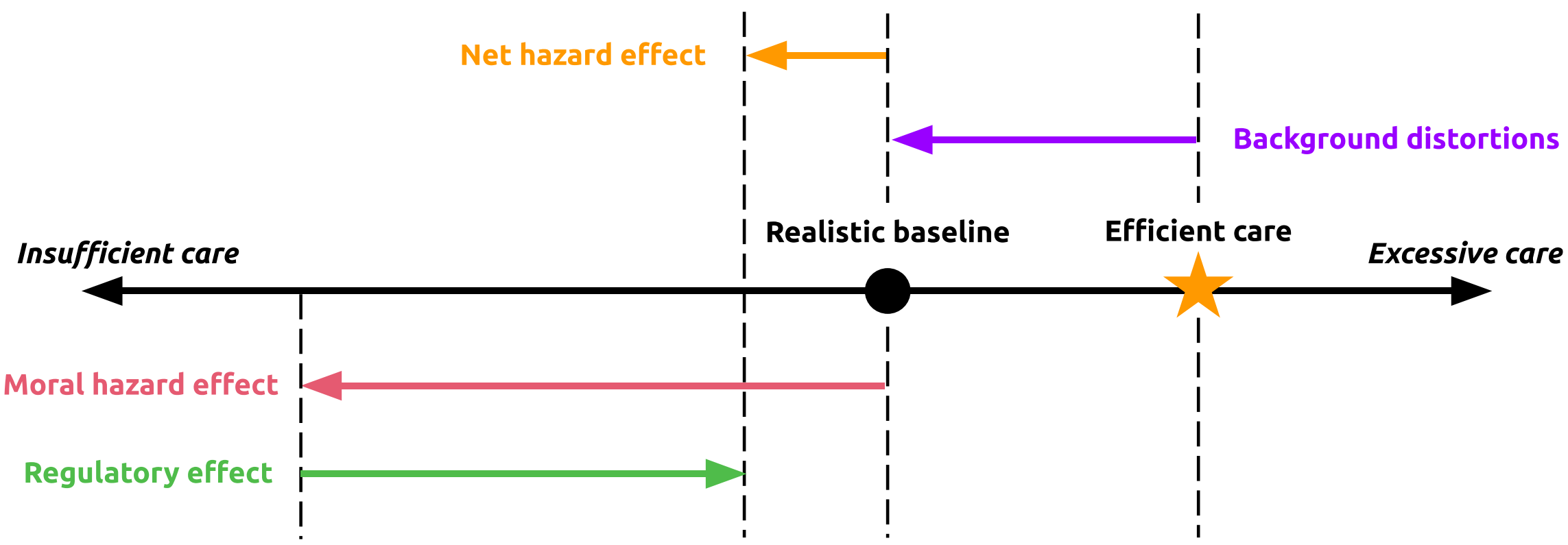

*Figure 2. Model of a net moral hazard effect from insurance uptake.*

Hopefully the model makes clear that a net regulatory effect only occurs when insurers' incentives are *better aligned with welfare maximization* than those of other economic actors. Or in other words, a net regulatory effect is *only possible* when background distortions create inefficiencies that insurers are well-positioned to correct. After all, how are insurers expected to *improve* over the idealized scenario? Thus the presence and magnitude of distortions determine the *potential* for a net regulatory effect.

## 6. Choice of Liability Regime

As the name suggests, *regulation* by insurance is primarily focused on insurance against third-party liability (torts), contractual liability to some extent, and much less so first-party harms. In normative terms: we are primarily concerned with guaranteeing economic actors uphold their duty of care toward *others*. In economic terms: we are primarily concerned with internalizing negative externalities.[21] At any rate, proving insurers can reduce losses through third-party liability policies (as opposed to first-party policies) is the stronger result: moral hazard is typically greater in the former, since, in the latter, an accident's non-pecuniary harms always provide a baseline incentive for self-care.

Thus, however expansive liability is for a given activity, determines how expansive regulation by insurance can be. In that sense, the choice of liability regime is central to the topic of regulation by insurance.

However, my model is liability regime agnostic. We want a way to evaluate the potential for a net regulatory, *given* a liability regime. Earlier I described the idealized scenario as one in which incentives are aligned with efficient outcomes, but we can relax this assumption. We can simply set our "idealized" baseline to however efficient the given liability regime is without insurance, and then introduce all our various distortions to get a realistic baseline.

That is the approach this Article takes: I largely avoid debating which liability regime is best. This isn't to say this Article's findings shouldn't inform discussions of e.g. what AI developers should be liable for. For better or worse, insurance is the *de facto* conduit through which the liability system operates.[22] Thus if it's found that, e.g. AI developer's insurers are very effective at reducing certain kinds of harms, that is *pro tanto* reason to make AI developers more liable for such harms.[23]

Furthermore, sometimes the choice of liability regime does factor into *how effective* regulation by insurance is, and not just its *scope*. For example, I argue in Part III that no-fault liability regimes generally complement regulation by insurance. No-fault regimes reduce the surface area for lawyering one's way out of liability, which insurers have enabled policyholders to do on occasion.[24] For now though, I make no assumptions about the liability regime in force, other than that there is one.

### *C. The Accident Matrix*

Before cataloging distortions, some terminology, and a refresher on accidents and their relation to insurance.

Regulation by insurance can only hope to reduce harms from *accidents* on the part of injurers (or victims), in contrast with various kinds of *intentional attacks* or *misuse*: naturally injurers cannot buy insurance for the latter. Insurers will never replace the police.

---

[21] Today this is seen as the primary purpose of tort law. See SHAVELL, *supra* note 12 ch. 5.5.

[22] By some estimates, 90 per cent of all payments made to tort victims are paid by liability insurers. See *Id.* ch. 5.4.5.

[23] See Ben-Shahar and Logue, *supra* note 3 at 247 for a similar line of reasoning.

[24] Daniel B. Schwarcz, Josephine Wolff & Daniel W Woods, *How Privilege Undermines Cybersecurity*, SSRN JOURNAL (2022), https://www.ssrn.com/abstract=4175523.

I focus on two key dimensions of accidents: whether victims are customers (product accidents)[25] or third parties (non-product accidents),[26] and the severity-frequency of harm (see Table 1). For want of space, I largely focus on accidents caused by *firms*.

In the literature the *severity* of an accident is often distinguished from its *probability*: some care efforts only affect one or the other. This distinction matters for theorists, sometimes leading to qualitatively different results (and not merely minor quantitative differences). However, my analysis is too coarse-grained for such distinctions to matter: I gloss this as a single "severity-frequency" dimension.

| | **product** accident | **non-product** accident |
|---|---|---|
| **small** and **frequent** | Examples:<br>● a consumer is harmed by a power tool that lacks certain safety features<br>In the context of frontier AI:<br>● a minor harm results from a consumer taking faulty advice from an AI model | Examples:<br>● an industrial waste facility regularly leaks a small amount of toxins<br>In the context of frontier AI:<br>● a model reproduces a third party's copyrighted material |
| **large** (>$100 million), **infrequent,** and **acute** (a.k.a. "catastrophic risks") | Examples:<br>● a fatal plane crash due to a manufacturing defect<br>In the context of frontier AI:<br>● firm A uses an AI coding agent from firm B, which introduces a bug into a core technology platform, leading to widespread system shutdowns and delays | Examples:<br>● a core meltdown occurs at a nuclear power plant<br>● a major oil spill occurs on an oil rig<br>In the context of frontier AI:<br>● an AI model, while in testing and not deployed, exfiltrates itself to other servers and commits several major cybercrimes before being caught and shut down. |

*Table 1. Matrix of different accident types studied, with examples.*

I leave out small infrequent accidents because they are, by definition, of least concern. *Prima facie*, activities involving frequent disasters are more interesting.

---

[25] Or so simplest the story goes. I'll complicate this picture later, in §I.D.2.

[26] SHAVELL, *supra* note 12 ch. 3.

However, I set these aside as well since there is rarely much potential for a net regulatory effect here.[27]

Finally, another standard distinction is that of *unilateral* versus *bilateral* accident models.[28] For want of space, my analysis focuses on unilateral accident models.

### *D. What Distortions Can Insurers Correct?*

The rest of this Part systematically identifies distortions which insurers might have the tools and propensity to correct for. In the interest of teasing out different effects, I focus largely on theoretical literature. Cursory explorations of the empirical literature are generally consistent with theory, but a more thorough review awaits further research.

For want of space, I do not attempt to quantify the corrections insurers might provide, opting instead to point out general qualitative patterns to calibrate our instincts on. Ultimately, the working social engineer must rely on their judgment.

#### 1. Judgment-Proofness

An injurer is said to be judgment-proof when they lack sufficient assets to fully compensate victims in the event of an accident. Since the injurer has less at stake than the tort system assumes they do, they are not sufficiently incentivized to take care. Judgment-proofness doesn't only affect individuals, but also large firms:

---

[27] This should become clearer below, but in a nutshell: the frequency of such accidents means there will be less room for time-discounting, thus spurring safety R&D; all else being equal, a self-preserving firm's care levels will increase as the risk of insolvency from liability increases; finally, the combination of frequency and severity means there will be less room for overconfident or optimistic psychological biases to persist among competing firms, regarding these harms. Specifically with regards to frontier AI, we are not yet seeing frequent disasters. If we were, and engineers could not quickly find a way mitigating them, activity levels in the industry would likely collapse and investment would slow down greatly. This may yet happen, but from where we stand today, AI catastrophes are very uncertain, future risks. For the purposes of this Article, such risks are effectively equivalent to *infrequent* disasters. For insurers, "low frequency risks" involving a novel technology is more or less tantamount to "risk of very uncertain probability, and which will remain so for some time."

[28] Unilateral accidents are characterized by situations where only one party (typically the injurer) can affect the probability and severity of harm through their care. Often the potential victim has no meaningful role in reducing accident risk. Bilateral accidents by contrast, involve situations where both parties can influence accident risk. For the core elements of bilateral models, see SHAVELL, *supra* note 12 ch. 2; For explanations of how care complementarities complicate this basic picture, see Miriam Buiten, Alexandre de Streel & Martin Peitz, *The Law and Economics of AI Liability*, 48 COMPUTER LAW & SECURITY REVIEW 105794, § 4.2 (2023); and Dhammika Dharmapala & Sandra A. Hoffmann, *Bilateral Accidents with Intrinsically Interdependent Costs of Precaution*, 34 THE JOURNAL OF LEGAL STUDIES 239 (2005).

scholars have long noted that insolvency is a *de facto* limit on liability.[29] In the case of firms with considerable assets, this is only really relevant for large catastrophes.

The judgment-proof problem is the canonical example of a distortion that insurance mandates can help correct for, and thus yield a net regulatory effect.[30] The potential injurer won't have more assets at risk, but the risk-priced premiums insurers charge them will essentially draw the expected losses forward, spread out over time. To the extent care taken is visible to the insurer and premiums are adjusted accordingly (see Part III), the potential injurer is incentivized to take care in order to reduce their premiums. A mandate is required since it is not in the judgment-proof actor's interest to purchase insurance for liabilities they are essentially immune to.

Empirical studies in the realm of environmental liabilities find strong evidence of the judgment-proof effect at work.[31] Furthermore, mandating insurance can indeed improve outcomes.[32]

The literature on judgment-proofness isn't without its wrinkles. Beard points out that having limited assets can also act as a *subsidy* for (monetary) expenditures on care.[33] As Shavell summarizes: "because expenditures on care would be expenditures of money that might otherwise be paid as liability judgments, care effectively becomes cheaper (even raising the theoretical possibility that a person with low assets would exercise excessive care)."[34] This countervailing effect only applies to care that reduces the *probability* of harm (as opposed to the magnitude).[35] Thus, as Shavell rightly points out, this countervailing effect is less important the lower the probability of harm (i.e. it would not help counter the judgment-proof effect for catastrophic tail risks).

---

[29] *See e.g.* George A. Akerlof et al., *Looting: The Economic Underworld of Bankruptcy for Profit*, 1993 BROOKINGS PAPERS ON ECONOMIC ACTIVITY 1 (1993) and; Taylor Meehan, *Lessons from the Price-Anderson Nuclear Industry Indemnity Act for Future Clean Energy Compensatory Models*, 18 CONN. INS. L.J. 339, 355 (2011).

[30] Abraham and Schwarcz, *supra* note 3 at 271; Logue, *supra* note 11; That said, insurance mandates are not the only remedy: potential injurers could instead be required to hold a certain amount of financial assets instead. Regarding which requirement is superior for handling judgment-proofness, scholars find that theory gives conflicting recommendations based on whether the injurer has more control over the probability or magnitude of the accident. Asset requirements might be better in the former case; liability insurance, in the latter. Respectively, cf. Steven Shavell, *Minimum Asset Requirements and Compulsory Liability Insurance as Solutions to the Judgment-Proof Problem*, 36 THE RAND JOURNAL OF ECONOMICS 63 (2005); and Chulyoung Kim & Paul S. Koh, *Minimum Asset and Liability Insurance Requirements on Judgment-Proof Individuals When Harm Is Endogenous*, 60 HITOTSUBASHI JOURNAL OF ECONOMICS 141 (2019).

[31] Al H. Ringleb & Steven N. Wiggins, *Liability and Large-Scale, Long-Term Hazards*, 98 JOURNAL OF POLITICAL ECONOMY 574 (1990); Judson Boomhower, *Drilling Like There's No Tomorrow: Bankruptcy, Insurance, and Environmental Risk*, 109 AMERICAN ECONOMIC REVIEW 391 (2019).

[32] Boomhower, *supra* note 31.

[33] T. Randolph Beard, *Bankruptcy and Care Choice*, 21 THE RAND JOURNAL OF ECONOMICS 626 (1990).

[34] Shavell, *supra* note 30 n. 13.

While the judgment-proof problem is one of the better studied, it certainly isn't the only distortion insurers can correct for. Among other things, it says nothing of competition between firms, firms' survival motives, key actors' psychologies, public goods or other collective action problems. We turn to these distortions next.

### 2. Competition as a source of distortions

Economists have long known that fierce competition can distort firms' incentives; less widely recognized is that insurers can sometimes help correct these distortions, such as when insurers do better at cooperating on safety research than firms from the industry in question.

A review of the literature at the intersection of industrial organization and accident models reveals that liability plays a relatively lesser role in incenting care for *product* accidents compared to *non-product* accidents, as liability is better complemented (or substituted) by market mechanisms in the former than in the latter. This means there is less room for insurers to play a regulatory role in the case of product accidents.

I consider the case of product accidents first, and then non-product accidents.

#### *a. Product Accidents*

Product accidents are naturally related to the literature on product liability, where product safety can be considered an element of product quality. There are two key questions for understanding liability insurance's potential role here:

1. Does liability meaningfully improve care levels under competitive conditions?
2. If so, do suboptimal care levels persist even with liability, creating space for insurance-driven improvements?

Following Shavell,[36] traditional law and economics models find surprisingly little interaction between market structure and liability policy.[37] As Daughety and Reinganum explain, this policy separability result implies that "antitrust and competition policies can be formulated and implemented without regard to the choice of tort regime in product liability: considerations of market performance are separable from considerations of product performance."[38] Here "product safety" is considered to be a part of "product performance," and consumers are modeled as internalizing expected harm through their willingness to pay. It's based on such arguments and models that Shavell and Polinsky claim market mechanisms can somewhat or altogether substitute for liability in inducing optimal care.[39] These models are simplistic, but provide a useful anchor: we should generally expect liability to play a smaller role in incenting precaution against product accidents than non-product accidents; we shouldn't immediately expect competition to generate a distortion on firms' care levels here.

---

[35] Giuseppe Dari-Mattiacci & Gerrit De Geest, *When Will Judgment Proof Injurers Take Too Much Precaution?*, 26 INTERNATIONAL REVIEW OF LAW AND ECONOMICS 336 (2006).

[36] STEVEN SHAVELL, ECONOMIC ANALYSIS OF ACCIDENT LAW (1987).

[37] Andrew F. Daughety & Jennifer F. Reinganum, Market Structure, Liability, and Product Safety (Mar. 31, 2016), https://papers.ssrn.com/abstract=2772095; in LUIS C. CORCHÓN & MARCO A. MARINI, HANDBOOK OF GAME THEORY AND INDUSTRIAL ORGANIZATION, VOLUME II: APPLICATIONS (2018).

Subsequent scholarship has since extended these simple models to capture a variety of phenomena, making them more realistic. Daughety and Reinganum provide helpful reviews.[40] The phenomena's effects are mixed and complex. I summarize a few of the most relevant results below:

- The less visibility consumers have into the safety of a product, the further care levels tend to be below the socially efficient level.[41] This leaves room for increased liability to improve care levels,[42] including when firms are modeled as signaling the safety of their product through prices.[43] When instead firms can communicate quality (safety) through other channels (e.g. disclosures, third-party verification etc.), increased liability has mixed effects on welfare.[44]
- When safety R&D investments represent relatively fixed and upfront costs, preceding production decisions, increased competition reduces per-firm output and the corresponding returns on said investments. This suppresses safety R&D, and thus care levels.[45]
- Product substitutability introduces non-monotonic effects. High substitutability triggers "business-stealing"[46] dynamics where firms steal market share by competing in safety improvements, potentially leading to excessive care. In a low substitutability context, direct willingness-to-pay effects to dominate: firms act more like monopolies, reducing care levels to pad profits.[47] A threshold for substitutability exists below which deficient care results and above which excessive care occurs.[48] Thus, increased liability can sometimes help and sometimes hinder welfare in this context.

[38] Daughety and Reinganum, *supra* note 37 at 6–7.

[39] A. Mitchell Polinsky & Steven Shavell, *The Uneasy Case for Product Liability*, 123 HARV. L. REV. 1437 (2009).

[40] Andrew F. Daughety & Jennifer F. Reinganum, *Chapter 3: Economic Analysis of Products Liability: Theory* (2013), https://www.elgaronline.com/edcollchap/edcoll/9781848441187/9781848441187.00011.xml ; Daughety and Reinganum, *supra* note 37.

[41] Daughety and Reinganum, *supra* note 40 § 3.

[42] *Id.* § 3.1.

[43] *Id.* § 3.2.1.

[44] *Id.* § 3.2.2.

[45] Daughety and Reinganum, *supra* note 37 at 21.

[46] Note that the spillover effects described below directly dampen this business-stealing effect: the two distortions are antagonists of each other.

- Scholars find firm reputation mechanisms can often substitute for misperceived or missing information about a product's safety, raising care levels closer to efficient levels.[49] However, the effectiveness of such mechanisms diminishes with increased competition, whether through an increase in the number of firms or lower product differentiation. This is true whether reputation is shared between firms[50] or not.[51] In general, it's found that some amount of liability can help raise care levels closer to the efficient level.[52]

In sum, while theorists tend to find some role for liability to bring care levels closer to efficient levels (typically raising them), this is not always the case. The empirical literature largely bears out this mixed picture, finding that liability has non-monotonic effects on firms' incentives to innovate[53] and that increased liability sometimes has little to no effect on product safety.[54] That said, competition can sometimes have a deleterious effect on product safety.[55]

This diminished role of liability relative to market mechanisms and the unclear direction of effects all suggest insurers have limited scope to affect care levels in product accident contexts, as *both* moral hazard and regulatory effects are competing with other market forces.[56]

---

[47] Daughety and Reinganum, *supra* note 37 at 22.

[48] *Id.* at 23.

[49] Baniak Andrzej, Grajzl Peter & Joseph Guse A, *Producer Liability and Competition Policy When Firms Are Bound by a Common Industry Reputation*, 14 THE B.E. JOURNAL OF ECONOMIC ANALYSIS & POLICY 1645 (2014); Yongmin Chen & Xinyu Hua, *Competition, Product Safety, and Product Liability*, 33 THE JOURNAL OF LAW, ECONOMICS, AND ORGANIZATION 237 (2017).

[50] Andrzej, Peter, and A, *supra* note 49.

[51] Chen and Hua, *supra* note 49.

[52] For more on collective reputations, see Jean Tirole, *A Theory of Collective Reputations (with Applications to the Persistence of Corruption and to Firm Quality)*, 63 THE REVIEW OF ECONOMIC STUDIES 1 (1996); Miguel Carriquiry & Bruce A. Babcock, *Reputations, Market Structure, and the Choice of Quality Assurance Systems in the Food Industry*, 89 AMERICAN JOURNAL OF AGRICULTURAL ECONOMICS 12 (2007).

[53] W. Kip Viscusi & Michael J. Moore, *Product Liability, Research and Development, and Innovation*, 101 JOURNAL OF POLITICAL ECONOMY 161 (1993).

[54] Polinsky and Shavell, *supra* note 39.

[55] Gary Lynn & Richard R. Reilly, Effect of Improvisation on Product Cycle Time and Product Success: A Study of New Product Development (NPD) Teams in the United States (2008), https://papers.ssrn.com/abstract=2152625.

[56] Consistent with this conclusion, Hanson and Logue find that consumer-facing first-party insurance against product accidents (e.g. extended warranties) likely produce a net

Of the remaining distortions, there are three that insurers can unambiguously address: helping firms credibly signal product quality (safety) through warranties and certifications; better providing upfront, fixed-cost safety R&D (particularly with large spillovers); and overcoming collective action problems around shared industry reputation. Insurers are well placed to address the first because they are third parties who are financially motivated to verify the safety of the product being insured or certified, giving their public assessments credibility.[57] Insurers are well placed to address the other two distortions when they can better *centralize* and *appropriate* the provision of what are essentially public goods (§III.A).

Auto safety demonstrates these points well: auto insurers successfully coordinated to fund a research institute that develops improved safety technology as well as crashworthiness ratings for vehicles. Insurers were well placed to fill this gap in the market because it's easier for three major auto insurers representing 80% of the market[58] to coordinate than it is for hundreds of millions of drivers,[59] and it's easier for them to appropriate many safety improvements[60] than it is for competing auto manufacturers. Furthermore, insurers are more credible verifiers of safety than the competing auto manufacturers.

That said, such cooperation between insurers is the exception, not the rule (see Part III).

#### *b. Non-product accidents*

Non-product accidents, where victims are 3rd parties to the firms that injure them, present fundamentally different dynamics. Environmental pollution exemplifies this category, where liability plays a critical role in internalizing these otherwise socialized costs.[61] Compared to product accidents then, this accident type immediately offers greater potential for involvement from liability insurers. Does competition reduce care levels below the socially efficient level though, despite liability?

In the simplest models, the answer is no.[62] However, there is a rich literature on the potential benefits of firms cooperating on R&D more generally,[63] balancing hazard effect. Jon D. Hanson & Kyle D. Logue, *First-Party Insurance Externality: An Economic Justification for Enterprise Liability*, 76 CORNELL L. REV. 129 (1990).

[57] Chong Zhang, Man Yu & Jian Chen, *Signaling Quality with Return Insurance: Theory and Empirical Evidence*, 68 MANAGEMENT SCIENCE 5847 (2022).

[58] *Who We Are*, IIHS-HLDI CRASH TESTING AND HIGHWAY SAFETY, https://www.iihs.org/about (last visited Oct. 4, 2025).

[59] Ben-Shahar and Logue, *supra* note 3 at 222.

[60] Such as headlight designs that don't blind oncoming traffic. Matthew L. Brumbelow, *Light Where It Matters: IIHS Headlight Ratings Are Correlated with Nighttime Crash Rates*, 83 JOURNAL OF SAFETY RESEARCH 379 (2022).

[61] Daughety and Reinganum, *supra* note 37 n. 8.

[62] *See e.g.* A. Mitchell Polinsky, *Strict Liability vs. Negligence in a Market Setting*, 70 THE AMERICAN ECONOMIC REVIEW 363 (1980).

overinvestment from business-stealing incentives against underinvestment from a lack of appropriability and spillovers. In the context of non-product accidents, this business-stealing effect comes solely from a reduction in marginal costs, not through improvements to product quality.

The thinly populated literature at the intersection of industrial organization, liability, and environmental economics (where externalities loom large)[64] points to the crucial role of aforementioned spillovers.[65] Furthermore, where safety investments are upfront, more or less fixed costs, we should again expect these investments to fall as competition increases:[66] the marginal benefit of such investments will fall as each firm's output falls.[67]

Empirical evidence for these effects is mixed: while Farber and Martin document positive correlations between industry concentration and investment in pollution control measures,[68] subsequent work contradicts these findings.[69] The sparse empirical literature precludes confident conclusions.[70] In any case, to the extent insurers can better centralize and appropriate safety R&D, insurers can help address these distortions.

The mixed empirical results likely reflect countervailing forces. The wider literature on competition identifies first-mover advantages to R&D investment (in the absence of significant spillovers). This holds for both product quality improvements and marginal cost reduction (the latter being the relevant mechanism for non-product accidents).[71] The welfare effects of this self-reinforcing market dominance effect are ambiguous.[72] This general finding also says nothing of *what kind* of investment is being made: if the returns to *safety* investments are not as great as other investments, these first-mover advantages could *reduce* safety investment both relative to other investments or in the absolute. Thus, the relative returns of different kinds of investments become important.

---

[63] *See e.g*. Claude D'Aspremont & Alexis Jacquemin, *Cooperative and Noncooperative R & D in Duopoly with Spillovers*, 78 THE AMERICAN ECONOMIC REVIEW 1133 (1988); Kotaro Suzumura, *Cooperative and Noncooperative R&D in an Oligopoly with Spillovers*, 82 THE AMERICAN ECONOMIC REVIEW 1307 (1992).

[64] Jiunn-Rong Chiou & Jin-Li Hu, *Environmental Research Joint Ventures under Emission Taxes*, 20 ENVIRONMENTAL AND RESOURCE ECONOMICS 129 (2001).

[65] For more on this intersection of the literature, see generally Roman Inderst & Stefan Thomas, *Competition Policy and the Environment*, 15 ANNUAL REVIEW OF RESOURCE ECONOMICS 199 (2023).

[66] Daughety and Reinganum, *supra* note 37 at 21.

[67] *See also* Jacob Nussim & Avraham D. Tabbach, *A Revised Model of Unilateral Accidents*, 29 INTERNATIONAL REVIEW OF LAW AND ECONOMICS 169 (2009) (discussing "durable precautions").

[68] S. C. Farber & R. E. Martin, *Market Structure and Pollution Control Under Imperfect Surveillance*, 35 THE JOURNAL OF INDUSTRIAL ECONOMICS 147 (1986) NB: "control measures" here refer to e.g. recycling toxic chemicals, treating wastewater, trapping releases etc. These measures are typically referred to as "abatement" in the literature, but Farber and Martin use this last to refer both to such control measures (care levels) and reductions in firm output (activity levels).

[69] Daniel H. Simon & Jeffrey T. Prince, *The Effect of Competition on Toxic Pollution Releases*, 79 JOURNAL OF ENVIRONMENTAL ECONOMICS AND MANAGEMENT 40 (2016).

[70] *See generally* Inderst and Thomas, *supra* note 65.

There is no obvious reason to expect insurers to *systematically* correct for distortions stemming from such first-mover advantages. In fact, if insurers could only modify the behavior of insured firms through actuarially fair premiums, insurance uptake would have no effect on these distortions: firms would face the same cost-benefit tradeoff as before.[73] However, outcomes will change where *insurers* are making the investments in care (e.g. developing risk models, enforcing private safety codes, conducting safety R&D, lobbying for stricter regulations, etc.). This is because insurers focus exclusively on liability reduction: they are indifferent to their policyholders' production costs or product quality, except insofar as it relates to liability exposure. Insurers don't care which policyholder wins in a competition.[74] All things considered then, insurance uptake here certainly can't guarantee efficient outcomes, but can help guarantee a safety *floor*, especially for catastrophic risks. By contrast, minimum asset requirements would have no effect on these first-mover advantage distortions.

Spulber points out one final distortion: inefficiencies arise when a uniform standard of care is imposed on injurers with different cost functions (e.g. an *industry-wide* reasonable person test for determining negligence).[75] The injuring firms will no longer be correctly incentivized to pursue their respective socially optimal care levels (where marginal costs of care equal the marginal reduction in expected damages), which are based on their distinct cost functions. In the case of competing firms, additional inefficiencies arise as "market allocation of output between the two firms will change, reflecting the effect of care on marginal production costs."[76]

---

[71] *See e.g.* Susan Athey & Armin Schmutzler, *Investment and Market Dominance*, 32 THE RAND JOURNAL OF ECONOMICS 1 (2001).

[72] For discussion see Luis M. B. Cabral & Michael H. Riordan, *The Learning Curve, Market Dominance, and Predatory Pricing*, 62 ECONOMETRICA 1115 (1994); Kyle Bagwell, Garey Ramey & Daniel F. Spulber, *Dynamic Retail Price and Investment Competition*, 28 THE RAND JOURNAL OF ECONOMICS 207 (1997).

[73] Modulo some changes in the uncertainty of costs and time discounting.

[74] Especially if the insurance market for the risk in question is highly concentrated, i.e. if a single insurer is insuring a number of competing firms.

[75] DANIEL F. SPULBER, REGULATION AND MARKETS § 14.2.3 (1989).

[76] *Id.* at 406; For a similar and related result involving competing firms under differing budget constraints, see Gérard Mondello & Evens Salies, *Tort law under oligopolistic competition*, § 2.3 (2016), https://sciencespo.hal.science/hal-03459225; *See also* Stéphan

As interesting as this is, this is not a distortion we can expect insurers to remedy with much systematicity. Theoretically, insurers can create a long-run wedge between the losses resulting from a firm's accidents and the premiums that firm pays. However, to correct the distortion here, insurers would need to *systematically overestimate* some firms' risk (those whose socially optimal care levels are above the uniform standard, because precaution is comparatively cheap for them) and *underestimate* other firms' risk (those whose socially optimal care levels are below the uniform standard, because precaution is comparatively costly for them).[77] It's not obvious why insurers would target this efficient outcome.

It's true that insurers offer *menus*[78] of contracts (e.g. reduced premiums for increased precautionary efforts) to attract more customers and better identify "good risk."[79] Firms seeking coverage would then essentially be picking their standard of care based on their marginal costs. However, there is only so far this can realistically go. Setting aside the efforts required of insurers to offer such menus, ultimately insurers only care about the portion of accident costs they must cover, which in this case is determined by the uniform standard of care. Therefore, unless insurers are trying very hard to decrease aggregate losses across all policyholders,[80] insurers are unlikely to offer discounts for care levels well above the uniform standard. Meanwhile, insurers cannot offer policies that explicitly accept care levels far *below* the legal standard, lest the policy be unenforceable or create bad faith liability for the insurer.

In sum, the theoretical literature suggests that in the context of non-product accidents, liability plays an important role in inducing care, and competition can drive care levels below efficient levels when:

1. Safety R&D is an upfront and or fixed cost, and spillovers are large.
2. Firms expect a lower return on safety R&D compared to other investments.
3. When a uniform standard of care is applied to injurers with differing cost functions.

---

Marette, *Minimum Safety Standard, Consumers' Information and Competition*, 32 J REGUL ECON 259 (2007).

[77] If I'm not mistaken, moral hazard needn't result, so long as the firms' premiums remain appropriately sensitive to the firm's care efforts. In other words, the *slope* of the firm's premiums as a function of their care could remain accurate while the *level* is "inaccurate." *Adverse selection* on the other hand would certainly be a problem, though a mandate would eliminate this entirely.

[78] Ben-Shahar and Logue, *supra* note 3 at 231.

[79] David de Meza & David C. Webb, *Advantageous Selection in Insurance Markets*, 32 THE RAND JOURNAL OF ECONOMICS 249 (2001); Amy Finkelstein & Kathleen McGarry, *Multiple Dimensions of Private Information: Evidence from the Long-Term Care Insurance Market*, 96 AMERICAN ECONOMIC REVIEW 938 (2006).

[80] Abraham and Schwarcz, *supra* note 3 § III.B.

With the possible exception of the third, there are proven ways insurance uptake can mitigate these distortions; how often this happens will be discussed in Part III.

### 3. Heterogeneous Insolvency Risk and Firm Survival Motives

In the basic model of judgment-proofness considered earlier, firms only ever stand to lose their current assets in the event of an accident. Evans and Gilpatric point out, this is inaccurate: when a firm faces *insolvency*, the original owners stand not only to lose the firm's current assets, but also discounted future profit flows.[81] This could happen either from the firm shutting down operations, or being taken over by creditors through a bankruptcy process, both forms of "firm death."[82] This creates a firm "survival motive" which fundamentally alters the judgment-proof analysis.

To study this, Evans and Gilpatric introduce a model[83] of firm behavior with heterogeneous insolvency risks: *profit-induced* (from poor performance) versus *liability-induced* (from accident costs). The dominance of one risk type over the other dramatically affects care levels. When liability-induced insolvency dominates, care can *exceed* efficient levels,[84] counteracting judgment-proof effects. When profit-induced insolvency dominates, care falls below *both* efficient levels and pure judgment-proof predictions: the effects compound.[85]

More precisely, they show that where the marginal return on further care expenditures equals the marginal cost from increased profit-induced insolvency risk, depends on which type of risk is more salient for the firm.[86] This parallels the distortions due to first-mover advantages in R&D investment, where the level of investment in *safety* R&D depends on which kind of R&D happens to most cost-effectively secure market advantage. The difference: the distortion here is driven by financial distress. Note that these distortions can interact and compound. Consider e.g. a financially distressed industry where leading firms pursue high-risk strategies to eliminate competitors (e.g. betting the farm on a risky new product).[87]

---

[81] Mary F. Evans & Scott M. Gilpatric, *Abatement, Care, and Compliance by Firms in Financial Distress*, 66 ENVIRON RESOURCE ECON 765, §§ 1–2 (2017).

[82] *Id.* at 770–771.

[83] I'm glossing somewhat. They don't have just one model but two: one in which care taken can only affect the probability of accidents, and one in which care only affects the severity of accidents. While exact quantitative results differ, the qualitative patterns in firm behavior are the same. *Id.* n. 8.

[84] Operationalized here as the level taken by a fully solvent firm (i.e. a firm without capital constraints) under strict liability. This is in keeping with the literature on judgment-proofness, but see also Proposition 3 in Evans and Gilpatric, *supra* note 81.

[85] Their model was built with *non-product* accidents in mind (explicitly environmental pollution). This is why it's reasonable for them to assume that care expenditures will *only* increase profit-induced insolvency risk. For *product* accidents this is a much more objectionable assumption: up to a point, making a product safer is equivalent to increasing its quality, which can make it more profitable (see §I.D.2).

[86] NB: they assume a unimodal (single-peaked) profit distribution, which they recognize could be unrealistic for a "firm engaging in a major new product launch or market entry with the possibilities of success and failure giving rise to a bimodal profit distribution." A bimodal distribution certainly complicates the analysis, often yielding non-monotonic care levels as a function of the firm's asset levels. However, the broader qualitative points remain: a firm's survival motive matters, and when firms primarily face profit-induced insolvency risk they will tend to have lower incentives for precaution than when they primarily face liability-induced insolvency risk. See Evans and Gilpatric, *supra* note 81 at 787.

Empirical studies do find that firms under financial distress (low operating margins) have less safe products or more workplace injuries.[88] This is consistent with the firm survival motive effects noted here, but also a variety of other effects stemming from budget constraints and imperfections in capital markets.

Evans and Gilpatric are skeptical that insurance will be of much help here. Since financial distress is the driver of these distortions, they argue that minimum asset requirements are preferable, even if insurers were to have perfect visibility into care taken.[89] I take issue with their conclusions on three counts:

1. It's not clear how a minimum asset requirement will help: if the firm is *not allowed to operate* with assets below a certain threshold, then raising that threshold *increases* financial distress: it would put *more* firms closer to bankruptcy, all else being equal. They are confusing a firm flush with assets with a firm that has an additional binding constraint.
2. Their claim about insurers being of little aid rests on a shaky premise. They claim "insurance delays the realization of liability costs." This is not quite accurate: insurance temporally *spreads out* liability costs, some of it *drawn forward*, some *pushed back*. Indeed this is the principal function of insurance: to leverage the law of large numbers to smooth out shocks for individual economic actors. Thus large and uncertain future costs become small and very certain costs for the firm. It's not obvious what this would do to the salience of liability-induced insolvency. Large regular premiums could dominate the firm's overall costs. On the other hand, the coverage could make liability costs much more manageable for the firm from a cash flow perspective.

---

[87] This would be an example of the distortions both pushing in the direction of lower care levels. They could of course both push in the direction of excessive care levels. However, it seems less likely they would push in *opposite* directions. That would entail e.g. liability-induced insolvency risk dominating risk due to low profits, but that a market advantage is much more easily secured by investing in non-safety related R&D. Not impossible, but less likely. More research could illuminate how these forces interact.

[88] Nancy L. Rose, *Profitability and Product Quality: Economic Determinants of Airline Safety Performance*, 98 JOURNAL OF POLITICAL ECONOMY 944 (1990); Randall K. Filer & Devra L. Golbe, *Debt, Operating Margin, and Investment In Workplace Safety*, 51 THE JOURNAL OF INDUSTRIAL ECONOMICS 359 (2003).

[89] Evans and Gilpatric, *supra* note 81 at 786.

3. They implicitly assume insurers have no way of reducing losses other than by adjusting premiums they charge firms. As we shall see in Part III, insurers have a number of other tools (e.g. private safety codes, safety R&D, refusing to underwrite, etc).

Will insurers help drive care levels toward the *efficient* level here? As with distortions from first-mover advantages (and for broadly similar reasons), no, we can only expect insurers to help guarantee a safety *floor*. Again, I expect raising this floor will be particularly beneficial in the context of catastrophic risks.

### 4. Charity Hazard

It's widely recognized that the state is society's *de facto* insurer of last resort.[90] This is most relevant regarding catastrophic risks: governments cannot credibly commit to *not* bail out a critical economic sector or *not* provide relief to victims in the event of a major disaster.

However, governments often fail to charge for this implicit social contract, giving rise to a form of moral hazard, sometimes called "charity hazard" or the "Samaritan's dilemma." Hence the too-big-to-fail effect and its role in the 2008 financial crisis.[91] Here then is another potential distortion to care levels.

Occasionally, the government does recognize its role as insurer of last resort. Witness, e.g. the National Flood Insurance Program (NFIP),[92] the Terrorism Risk Insurance Program (TRIP)[93] and the Price-Anderson Act (PAA)[94] for commercial nuclear power. In each, the government explicitly stepped into its role of insurer of last resort, directly insuring policyholders (NFIP) or indemnifying licensees (PAA) or reinsuring primary insurers (TRIP).

Thus, one way governments attempt to mitigate charity hazard is by directly charging risk-priced premiums. For a variety of reasons though, governments struggle to do this. Witness the pricing and financing difficulties of the National

---

[90] *See generally* DAVID A. MOSS, WHEN ALL ELSE FAILS: GOVERNMENT AS THE ULTIMATE RISK MANAGER (2004).

[91] Philip E. Strahan, *Too Big to Fail: Causes, Consequences, and Policy Responses*, 5 ANNUAL REVIEW OF FINANCIAL ECONOMICS 43 (2013).

[92] *See generally* Rawle O King, *National Flood Insurance Program: Background, Challenges, and Financial Status*, WASHINGTON DC: CONGRESSIONAL RESEARCH SERVICE (2011), https://www.fas.org/sgp/crs/misc/RL32972.pdf.

[93] U.S. DEPARTMENT OF THE TREASURY FEDERAL INSURANCE OFFICE, REPORT ON THE EFFECTIVENESS OF THE TERRORISM RISK INSURANCE PROGRAM (2022), https://home.treasury.gov/system/files/311/2022%20Program%20Effectiveness%20Report%20%28FINAL%29.pdf.

[94] United States Nuclear Regulatory Commission, *The Price-Anderson Act: 2021 Report to Congress, Public Liability Insurance and Indemnity Requirements for an Evolving Commercial Nuclear Industry* (2021), https://www.nrc.gov/docs/ML2133/ML21335A064.pdf.

Flood Insurance Program,[95] difficulties which likely would produce moral hazard if it weren't for quirks of human psychology.[96] The TRIP and PAA are even worse: they make no attempt at risk-pricing, charging essentially flat rates instead (and only retrospectively in the case of TRIP).[97] Unsurprisingly, there is evidence of TRIP creating moral hazard (among insurers) when compared to analogous programs in other developed nations.[98] And precisely out of concern for charity hazard, the indemnification program of the PAA was eventually phased out in favor of greater mutualization of risk by industry.[99]

However, from another angle, the TRIP and PAA indemnification programs have arguably been very successful at mitigating charity hazard *by inducing greater private insurance uptake*. These were explicit goals of both programs, and in both cases succeeded.[100] In exchange for an explicit and predictable government backstop, private actors were willing to commit to greater financial responsibility upfront. Here again then, insurance uptake is potentially having a net regulatory effect.

## 5. Catastrophic Risks, Psychology and Paternalism

In the behavioral economics literature, it is widely accepted that individuals tend to underinsure against low-probability, high-consequence risks (a.k.a. catastrophic risks) relative to what classical economic theory would predict.[101]

---

[95] Erwann O. Michel-Kerjan, *Catastrophe Economics: The National Flood Insurance Program*, 24 JOURNAL OF ECONOMIC PERSPECTIVES 165 (2010); King, *supra* note 92.ca

[96] Paul Hudson et al., *Moral Hazard in Natural Disaster Insurance Markets: Empirical Evidence from Germany and the United States*, 93 LAND ECONOMICS 179 (2017); W. J. Wouter Botzen, Howard Kunreuther & Erwann Michel-Kerjan, *Protecting against Disaster Risks: Why Insurance and Prevention May Be Complements*, 59 J RISK UNCERTAIN 151 (2019).

[97] Cristian Trout, *Insuring Uninsurable Risks from AI: Government as Insurer of Last Resort*, *in* GENERATIVE AI AND LAW WORKSHOP AT THE INTERNATIONAL CONFERENCE ON MACHINE LEARNING , 2 (2024).

[98] Erwann Michel-Kerjan & Paul A. Raschky, *The Effects of Government Intervention on the Market for Corporate Terrorism Insurance*, 27 EUROPEAN JOURNAL OF POLITICAL ECONOMY S122 (2011).co

[99] United States Nuclear Regulatory Commission, *supra* note 94 § 3.2.3.

[100] *See generally* FEDERAL INSURANCE OFFICE, *supra* note 93 and; United States Nuclear Regulatory Commission, *supra* note 94. In the case of the PAA, purchase of insurance was made mandatory, but insurers and nuclear power operators made it clear the government would have to provide a backstop in order to make the venture viable. .

[101] *See e.g*. Howard Kunreuther & Mark Pauly, *Neglecting Disaster: Why Don't People Insure Against Large Losses?*, 28 JOURNAL OF RISK AND UNCERTAINTY 5 (2004); Howard Kunreuther & Mark Pauly, *Insurance Decision-Making and Market Behavior*, 1 MIC 63 (2006); Paul Slovic et al., *Preference for Insuring against Probable Small Losses: Insurance Implications*, 44 THE JOURNAL OF RISK AND INSURANCE 237 (1977); Mark J. Browne, Christian Knoller & Andreas Richter, *Behavioral Bias and the Demand for Bicycle and Flood Insurance*, 50 J RISK UNCERTAIN 141 (2015).

Scholars have identified several psychological mechanisms that help explain these irrational outcomes:

- *Coarse chance categories & prospect theory*:[102] In Tversky and Kahneman's cumulative prospect theory, it's noted that people's probability weighting function is not well-behaved near zero.[103] While small probabilities can be overweighted, very small probabilities appear to be either "rounded down" to zero or greatly overweighted. This may be related to coarse chance categories (e.g. "it certainly won't happen" versus "it may happen"). Scholars interested in insurance purchase behavior build upon this framework, arguing that being *worrisome* lowers one's threshold for paying attention to very small risks, leading to higher insurance purchasing.[104]
- *Myopia and overconfidence (or optimism more generally)*:[105] These are distinct biases related to underestimating and under-weighting bad outcomes. Optimistic individuals correctly estimate the average probability of loss for a given risk, but tend to believe they are at a lower probability of loss than this average, thus underweighting the risk. Overconfidence is a special case of optimism, involving a biased perception of one's own skills, prospects, or knowledge.[106] Myopic individuals, on the other hand, underestimate the average probability of loss. Myopia is also related to preferences for short-term benefits.
- *Availability heuristics*:[107] These heuristics affect the assessment of the probability of an event by the "[...] ease which instances or occurrences can be brought to mind."[108] These heuristics can be explained as the overestimation of probabilities for events with vivid impact, and the underestimation of probabilities for events with not such vivid impact.

---

[102] Francisco Pitthan & Kristof De Witte, *Puzzles of Insurance Demand and Its Biases: A Survey on the Role of Behavioural Biases and Financial Literacy on Insurance Demand*, 30 JOURNAL OF BEHAVIORAL AND EXPERIMENTAL FINANCE 100471, § 3.1 (2021).

[103] Amos Tversky & Daniel Kahneman, *Advances in Prospect Theory: Cumulative Representation of Uncertainty*, 5 J RISK UNCERTAINTY 297 (1992).

[104] Christian Schade, Howard Kunreuther & Philipp Koellinger, *Protecting Against Low-Probability Disasters: The Role of Worry*, 25 JOURNAL OF BEHAVIORAL DECISION MAKING 534 (2012); Peter John Robinson & W. J. Wouter Botzen, *The Impact of Regret and Worry on the Threshold Level of Concern for Flood Insurance Demand: Evidence from Dutch Homeowners*, 13 JUDGMENT AND DECISION MAKING 237 (2018).

[105] Pitthan and De Witte, *supra* note 102 § 3.2.

[106] For more on this in the context of insurance purchasing behavior, see Jennifer Coats & Vickie Bajtelsmit, *Optimism, Overconfidence, and Insurance Decisions*, 29 FINANCIAL SERVICES REVIEW 1 (2021).

[107] Pitthan and De Witte, *supra* note 102 § 3.4.

[108] Amos Tversky & Daniel Kahneman, *Availability: A Heuristic for Judging Frequency and Probability*, 5 COGNITIVE PSYCHOLOGY 207 (1973).

While most studies focus on consumer choice or don't differentiate between firms and consumers, research into managerial decision-making confirms that these biases[109] are also at work in firms.[110] While there is evidence that certain groups of consumers over-insure against high-probability, low-consequence risks,[111] the literature doesn't suggest this afflicts firms (though admittedly, the topic appears understudied).

The psychological biases behind these tendencies suggest individuals don't simply under-insure but also fail to take adequate care to manage catastrophic risks.[112] Again, this applies to firms as well, with the caveat that older firms in more mature industries (where catastrophic events have been observed) tend to manage these tail risks better.[113]

Thus the irrationality here suggests significant distortions to care levels for catastrophic risks. This in turn suggests there is room here for insurers to improve loss prevention. Indeed, commentators suggest as much.[114]

There are three key reasons to believe insurers have a comparative institutional advantage for managing tail risks here (see also Appendix A):

1. Actuarial science and risk-modeling, the bread and butter of the insurance industry, is grounded in statistics and slow, deliberative thinking (in contrast with the fast, intuitive thinking that characterizes the psychological biases above). Precise risk-pricing is precisely the insurer's business. (Here my claim is approaching the more widely accepted claim that there is an efficient division of labor between insurers and firms).[115]
2. Insurance is a mature industry. Insurers that have survived market forces and adverse events appear to have e.g. learned heuristics to mitigate the winner's curse,[116] charging higher premiums for risks that are more uncertain.[117]

---

[109] I have already excluded other biases that aren't as emphasized in the research on managerial decision-making. These include narrow framing and mental accounting, decisions based on experience and description, and affection. See generally Pitthan and De Witte, *supra* note 102.

[110] *See e.g*. HOWARD KUNREUTHER & MICHAEL USEEM, MASTERING CATASTROPHIC RISK: HOW COMPANIES ARE COPING WITH DISRUPTION pt. II ch. 3 (2018).

[111] *See e.g*. Daniel Schwarcz, *Insurance Demand Anomalies and Regulation*, 44 JOURNAL OF CONSUMER AFFAIRS 557, 559–560 (2010).

[112] *See e.g*. William Petak, *Mitigation and Insurance*, *in* PAYING THE PRICE: THE STATUS AND ROLE OF INSURANCE AGAINST NATURAL DISASTERS IN THE UNITED STATES (1998), http://www.nap.edu/catalog/5784.

[113] KUNREUTHER AND USEEM, *supra* note 110 ch. 4–5.

[114] *See e.g*. Petak, *supra* note 112 and ; HOWARD C. KUNREUTHER, MARK V. PAULY & STACEY MCMORROW, INSURANCE AND BEHAVIORAL ECONOMICS: IMPROVING DECISIONS IN THE MOST MISUNDERSTOOD INDUSTRY 229–231 (2013).

[115] Baker and Siegelman, *supra* note 17 § 3.B.

[116] In an auction, winning a bid is some evidence that the winner is simply the person who most overestimated the true value of the asset (while other, potentially better-informed

3. Regulators and third-party rating agencies[118] regularly stress-test insurers' portfolios, checking for risk of insolvency. For example, under the European Union's Solvency II requirements, European insurers (including reinsurers critical to the global market) must demonstrate sufficient capital reserves to weather all reasonably foreseeable risks over a one year period with 99.5% confidence. In other words, insurers should only go insolvent (from reasonably foreseeable risks) every 200 years.[119] Requirements like this, or the Risk-Based Capital requirements in the U.S.,[120] force insurers to study and plan for rare catastrophic scenarios. This is expected to trickle down to scrutinizing tail risks among insureds, especially where policies are large or the risk of correlated losses is significant.

Thus, insofar as insurers have the tools to modify firm behavior, we can expect insurers to help correct for firms' biases surrounding tail risk management.

One final note. Empirical studies have found that insurance markets often segment into distinct consumer groups with sharply different purchasing patterns.[121] This appears to apply to catastrophic risk insurance as well.[122] The segmentation stems from preference heterogeneity[123] and cognitive biases:[124] overconfident or less risk-averse individuals forgo insurance, while more anxious or risk-averse individuals purchase it. This pattern can produce advantageous selection rather than

---

bidders, valued the asset less highly). If not adjusted for, this can lead to systematic losses for auction winners.

[117] Jeryl L. Mumpower, *Risk, Ambiguity, Insurance, and the Winner's Curse*, 11 RISK ANALYSIS 519 (1991).

[118] David L. Eckles & Martin Halek, *Insurance Company Financial Strength Ratings*, *in* HANDBOOK OF INSURANCE: VOLUME II 427 (Georges Dionne ed., 2025), https://doi.org/10.1007/978-3-031-69674-9_14.

[119] Rene Doff, *Risk and Solvency Regulation in Europe*, *in* HANDBOOK OF INSURANCE: VOLUME II 483 (Georges Dionne ed., 2025), https://doi.org/10.1007/978-3-031-69674-9_17.

[120] Martin F. Grace, *Regulation of Insurance Markets in the USA*, *in* HANDBOOK OF INSURANCE: VOLUME II 457 (Georges Dionne ed., 2025), https://doi.org/10.1007/978-3-031-69674-9_16.

[121] *See e.g.* David M. Cutler, Amy Finkelstein & Kathleen McGarry, *Preference Heterogeneity and Insurance Markets: Explaining a Puzzle of Insurance*, 98 AMERICAN ECONOMIC REVIEW 157 (2008).

[122] Gary H. McClelland, William D. Schulze & Don L. Coursey, *Insurance for Low-Probability Hazards: A Bimodal Response to Unlikely Events*, *in* MAKING DECISIONS ABOUT LIABILITY AND INSURANCE: A SPECIAL ISSUE OF THE JOURNAL OF RISK AND UNCERTAINTY 95 (Colin Camerer & Howard Kunreuther eds., 1993), https://doi.org/10.1007/978-94-011-2192-7_7; *Cf.* Susan K. Laury, Melayne Morgan McInnes & J. Todd Swarthout, *Insurance Decisions for Low-Probability Losses*, 39 J RISK UNCERTAIN 17 (2009).

[123] Cutler, Finkelstein, and McGarry, *supra* note 121.

[124] Alvaro Sandroni & Francesco Squintani, *Overconfidence, Insurance, and Paternalism*, 97 THE AMERICAN ECONOMIC REVIEW 1994 (2007); Rachel J Huang, Yu-Jane Liu & Larry Y Tzeng, *Hidden Overconfidence and Advantageous Selection*, 35 GENEVA RISK INSUR REV 93 (2010).

adverse selection, undermining a standard justification for insurance mandates. Indeed, if advantageous selection predominates, mandates may simply result in a wealth transfer from risk-averse or worrisome individuals to overconfident or less risk-averse individuals.[125] Whether these dynamics apply to firms remains unclear.

### 6. Managerial Incentives and Short Term Profits

Finally, a wealth of sociological research finds that managers frequently resist safety expenditures when these conflict with profits.[126] This doesn't immediately imply inefficiency, but one generalizable lesson from the literature does suggest it. As Downer summarizes it:[127]

> [O]rganizations, and the individuals staffing them, invariably operate with regard to time frames that are ill suited to managing the kinds of high-consequence and low-probability risks that characterize catastrophic technologies. This is simply to say that — since management positions are only held for a few years at a time, and corporate strategies must cater to investors and quarterly earnings reports — it is difficult to create accountability structures that prioritize the prevention of very rare accidents, no matter how consequential those accidents might be.

Insurers can mitigate these misaligned incentives by drawing future costs forward in the form of premium increases. That said, insurers aren't totally immune from such short-term pressures either.[128] (See §§III.B.2.a and III.B.4). Arguably, the mature regulation surrounding insurance does a fairly effective job of limiting these misaligned incentives.[129]

[125] Cutler, Finkelstein, and McGarry, *supra* note 121; Sandroni and Squintani, *supra* note 124.

[126] SCOTT DOUGLAS SAGAN, THE LIMITS OF SAFETY: ORGANIZATIONS, ACCIDENTS, AND NUCLEAR WEAPONS (1993); Diane Vaughan, *The Dark Side of Organizations: Mistake, Misconduct, and Disaster*, 25 ANNUAL REVIEW OF SOCIOLOGY 271 (1999); CHARLES PERROW, NORMAL ACCIDENTS: LIVING WITH HIGH RISK TECHNOLOGIES - UPDATED EDITION (2011); Charles Perrow, *Cracks in the "Regulatory State,"* 2 SOCIAL CURRENTS 203 (2015); Todd R. LaPorte & Paula M. Consolini, *Working in Practice but Not in Theory: Theoretical Challenges of "High-Reliability Organizations,"* 1 JOURNAL OF PUBLIC ADMINISTRATION RESEARCH AND THEORY: J-PART 19 (1991); Andrew Hopkins, *Why BP Ignored Close Calls at Texas City*, RISK AND REGULATION 4 (2010); JAMES REASON, MANAGING THE RISKS OF ORGANIZATIONAL ACCIDENTS (1997); Susan S. Silbey, *Taming Prometheus: Talk About Safety and Culture*, 35 ANNUAL REVIEW OF SOCIOLOGY 341 (2009); NATIONAL COMMISSION ON THE BP DEEPWATER HORIZON OIL SPILL AND OFFSHORE DRILLING, DEEP WATER: THE GULF OIL DISASTER AND THE FUTURE OF OFFSHORE DRILLING: REPORT TO THE PRESIDENT, JANUARY 2011: THE GULF OIL DISASTER AND THE FUTURE OF OFFSHORE DRILLING (2011); SAGAN.

[127] JOHN DOWNER, RATIONAL ACCIDENTS: RECKONING WITH CATASTROPHIC TECHNOLOGIES (2024).

[128] Christopher Parsons, *Moral Hazard in Liability Insurance*, 28 GENEVA PAP RISK INSUR ISSUES PRACT 448 (2003) (discussing "underwriter hazard," where underwriters are incentivized to write more business in the short-term, and leave the long-term consequences to their successors).

## II. THE POTENTIAL FOR A REGULATORY EFFECT IN THE FRONTIER AI INDUSTRY

Advanced Artificial Intelligence (AI) arrives amid great hope and great fear. Generative AI, AI agents and general-purpose AI promise vast improvements in productivity and welfare, but also threaten widespread disempowerment and great harm.[130] The now hackneyed policy challenge: maximizing the upside, while minimizing the downside, and ruling out socially unacceptable risks altogether.[131] One proposed strategy is regulation by insurance;[132] I take this opportunity to demonstrate my framework in action, evaluating potential for a net regulatory effect in this industry.

I find considerable potential. Fierce competition for market dominance creates temporal mismatches between short-term competitive pressures and long-term safety considerations. Safety research exhibits characteristics of primary research with substantial spillovers and high upfront costs, creating conditions for underinvestment. The industry also faces a collective action problem regarding shared reputational risk: an "AI Three Mile Island" could devastate the entire sector. Psychological biases prevalent in young, venture-backed startups further compound these challenges. Notably, the analysis finds little evidence of traditional judgment-proofness concerns for major AI developers, suggesting other distortions dominate.

The literature on regulating this frontier of general-purpose AI technology is growing rapidly.[133] Like other market-based solutions to the AI regulation puzzle,[134]

---

[129] *See generally* Grace, *supra* note 120 and especially reference to "Risk-Based Capital" requirements. *See also* Eckles and Halek, *supra* note 118.

[130] Yoshua Bengio et al., International AI Safety Report (Jan. 29, 2025), http://arxiv.org/abs/2501.17805; Jan Kulveit et al., Gradual Disempowerment: Systemic Existential Risks from Incremental AI Development (Jan. 29, 2025), http://arxiv.org/abs/2501.16946.

[131] *Cf.* BILL ANDERSON-SAMWAYS ET AL., RESPONSIBLE SCALING: COMPARING GOVERNMENT GUIDANCE AND COMPANY POLICY (2024), https://www.iaps.ai/research/responsible-scaling.

[132] Lior, *supra* note 6; Anat Lior, Innovating Liability: The Virtuous Cycle of Torts, Technology and Liability Insurance (Sept. 11, 2023), https://papers.ssrn.com/abstract=4568702; Trout, *supra* note 6; Weil, *supra* note 6; Schwarcz and Wolff, *supra* note 6.

[133] *See e.g.* Markus Anderljung et al., Frontier AI Regulation: Managing Emerging Risks to Public Safety (Nov. 7, 2023), http://arxiv.org/abs/2307.03718; Stephen Casper et al., *Black-Box Access Is Insufficient for Rigorous AI Audits*2254 (2024), http://arxiv.org/abs/2401.14446; Marie Davidsen Buhl et al., Safety Cases for Frontier AI (Oct. 28, 2024), http://arxiv.org/abs/2410.21572; Peter Wills, Regulatory Supervision of Frontier AI Developers (Mar. 1, 2025), https://papers.ssrn.com/abstract=5122871; Carson

"regulation by insurance" promises to leverage the diversification, experimentation and powerful information gathering abilities of markets. It also promises to efficiently recalibrate the stringency of safety requirements as markets become more confident in the significance or insignificance of risks. These are seen as key virtues when trying to manage the high stakes risks from a novel technology developing at the breakneck pace of modern AI,[135] a classic case of the Collingridge dilemma.[136]

For a more comprehensive catalog of risk from frontier AI, see the MIT AI Risk Repository.[137] For more details on the systemic risks[138] that could arise from the widespread deployment of AI agents, see the work of Hammond et al.[139] For a primer on the risks advanced AI poses to financial systems, see the work of Sutton and Williams,[140] and Jacobs and Sutton.[141] For more on specifically catastrophic risks from AI, see the work of Bengio et al.[142]

---

Ezell, Xavier Roberts-Gaal & Alan Chan, Incident Analysis for AI Agents (Aug. 19, 2025), http://arxiv.org/abs/2508.14231.

[134] Hadfield and Clark, *supra* note 7; Trout, *supra* note 97; Eyal and Arbel, *supra* note 7; AI Security Tax Incentives, *supra* note 7; Tomei, Jain, and Franklin, *supra* note 7; Ball, *supra* note 7; Though cf. Schwarcz and Wolff, *supra* note 6; and Matthew van der Merwe, Ketan Ramakrishnan & Markus Anderljung, *Tort Law and Frontier AI Governance*, LAWFARE (May 24, 2024), https://www.lawfaremedia.org/article/tort-law-and-frontier-ai-governance. I cannot provide a thorough comparison of regulation by insurance with these other market-based solutions to the AI regulation puzzle, but my argument in favor of regulation by insurance would boil down to this: alternatives seem to be trying to reinvent the wheel; insurance is a mature industry with an ecosystem of tested actors, and whose regulation is more or less a solved problem. Similarly, I cannot provide an in-depth comparison of regulation by insurance with traditional command and control regulatory modes (e.g. licensing regimes or direct supervision).

[135] Kolt, *supra* note 9 § V.B, D, E.

[136] Collingridge, *supra* note 10.

[137] Peter Slattery et al., The AI Risk Repository: A Comprehensive Meta-Review, Database, and Taxonomy of Risks From Artificial Intelligence (Apr. 10, 2025), http://arxiv.org/abs/2408.12622; cf. Organisation for Economic Co-operation and Development (OECD), *OECD AI Incidents Monitor (AIM)*, (2024), https://oecd.ai/en/incidents.

[138] For simplicity I treat these as synonymous with "catastrophic risks" in this Article, despite important differences. Most relevant here: without significant tweaks to liability, it's unlikely AI developers would be held liable for various harms arising from multi-agent systemic risks, simply due to difficulties surrounding attributing causation.

[139] Lewis Hammond et al., Multi-Agent Risks from Advanced AI (Feb. 19, 2025), http://arxiv.org/abs/2502.14143.

[140] Andrew Sutton & Sophie Williams, *Submission to the UK Treasury Committee Inquiry into AI in Financial Services*, (2025), https://cdn.governance.ai/Submission_to_the_UK_Treasury_Committee_Inquiry_into_AI_in_FS.pdf.

[141] JULIAN JACOBS & ANDREW SUTTON, RISKS OF AN INTELLIGENT FINANCIAL SYSTEM (2025), https://www.omfif.org/risks-of-an-intelligent-financial-system/.

[142] Bengio et al., *supra* note 130.

## *A. Background: Liability and Insurance for Frontier AI Developers*

Building on previous research,[143] I will explore the potential for a net regulatory effect from insurance uptake by frontier AI developers (e.g. OpenAI). Not argued for here is the claim that AI developers, and not users or operators, are the least cost avoiders.[144] This debate will have to be settled elsewhere, but there appears to be a growing consensus among scholars on this point (modulo a range of caveats).[145]

Elsewhere I've argued that AI developer third-party liability should be strict, limited, and exclusive for a narrow set of non-product catastrophic accidents.[146] Indeed, many of the distortions to care levels I discuss below form the rationale for channeling liability for such risks at developers (see also Part III). However, as noted above, here I make few assumptions about what liability regime is in place, generally assuming background law applies.[147]

Also of relevance is (self-)insurance coverage frontier AI developers already have.[148] Especially relevant policies include:[149]

- **Technology Errors & Omissions (Tech E&O)** – This type of policy might respond if a developer is sued because their negligently built AI agent e.g. provided incorrect information, created a bug in code, or made a financial transaction in breach of contract. Intellectual Property (IP) violations are also sometimes included.

---

[143] Lior, *supra* note 6; Trout, *supra* note 6; Weil, *supra* note 6; Schwarcz and Wolff, *supra* note 6.

[144] I.e. the economic actors who can reduce losses most cost efficiently. For a primer on the topic see Paul Rosenzweig, *Cybersecurity and the Least Cost Avoider*, LAWFARE (Nov. 2013), https://www.lawfaremedia.org/article/cybersecurity-and-least-cost-avoider. For a more theoretical introduction, see SHAVELL, *supra* note 12 ch. 2 §2.11.

[145] David C. Vladeck, *Machines without Principals: Liability Rules and Artificial Intelligence*, 89 WASH. L. REV. 117, 146–147 (2014); Miriam Buiten, Alexandre de Streel & Martin Peitz, *EU Liability Rules for the Age of Artificial Intelligence*, CERRE (2021), https://cerre.eu/publications/eu-liability-rules-age-of-artificial-intelligence-ai/; Buiten, de Streel, and Peitz, *supra* note 28; W. Nicholson Price II & I. Glenn Cohen, Locating Liability for Medical AI (July 21, 2023), https://papers.ssrn.com/abstract=4517740.

[146] Trout, *supra* note 6.

[147] For a primer on AI developers' liability for catastrophic losses under background law, see KETAN RAMAKRISHNAN, GREGORY SMITH & CONOR DOWNEY, U.S. TORT LIABILITY FOR LARGE-SCALE ARTIFICIAL INTELLIGENCE DAMAGES: A PRIMER FOR DEVELOPERS AND POLICYMAKERS (2024), https://www.rand.org/pubs/research_reports/RRA3084-1.html; For a more general primer on AI developer's liability under background law, see Gabriel Weil, Tort Law as a Tool for Mitigating Catastrophic Risk from Artificial Intelligence (Jan. 13, 2024), https://papers.ssrn.com/abstract=4694006.

[148] SCHWARTZ REISMAN INSTITUTE FOR TECHNOLOGY AND SOCIETY, CO-DESIGNING REGULATORY MARKETS 10 (2025), https://srinstitute.utoronto.ca/news/co-designing-regulatory-markets-for-ai.

[149] For a comprehensive overview of how existing insurance policies intersect with AI risk, see Anat Lior, E/Insuring the AI Age: Empirical Insights into Artificial Intelligence Liability Policies (Apr. 1, 2025), https://papers.ssrn.com/abstract=5316376; *Cf.* Szpruch et al., *supra* note 6 at 9–10.

- **Commercial General Liability (CGL)** – This type of policy might respond if a developer is sued because their AI agent, when integrated in a tangible product, caused bodily injury or damage to physical property. Along with Cyber policies, such policies are practically a prerequisite for developers to do business with enterprise clients.
- **Cyber** – It's unusual, but some cyber policies might respond if a developer is sued because their negligently built AI agent e.g. created a security vulnerability or worsened a data breach in some manner.

Very few existing policies will have *affirmative* coverage for AI-specific perils (though this is beginning to change: see §II.B.4). This means there is room for insurers to argue such perils are not in fact silently covered by the more general language in such contracts. That said, where insurance contracts are ambiguous, courts tend to rule in favor of policyholders.[150]

Between AI developers' liability insurance and businesses' first-party insurance, insurance brokers generally expect existing coverage will respond to most or all of the AI perils we have seen so far (e.g. hallucinations, or copyright violations). Thus, liabilities currently arising from small accidents are likely insured for the most part. This could change rapidly if insurers begin writing AI exclusions into their policies *en masse*. Indeed, as AI-related losses mount,[151] we are seeing more exclusions written[152] with a few industry leaders are of "silent AI,"[153] signaling to underwriters to revisit their language, pricing and limits.[154]

---

[150] *See generally* ROBERT H. JERRY & DOUGLAS R. RICHMOND, UNDERSTANDING INSURANCE LAW § 25.a (2025) and especially reference to the contra proferentem principle of insurance contract interpretation.

[151] Marcel Le Gouais, *AI Liability: An Age of Silent Exposures Has Already Begun*, INSURANCE INSIDER (Aug. 6, 2025), https://www.insuranceinsider.com/article/2f5mn8u0mwcdnovlve7e0/lines-of-business/casualty-gl/ai-liability-an-age-of-silent-exposures-has-already-begun.

Catastrophic risks are another story.[155] It's all but guaranteed that existing policies exclude losses related to:

- War-like acts
- Nuclear or radiological harms
- Critical infrastructure failure

In addition, policies will have *per claim* limits[156] (or similar) far below a policy's aggregate limit.[157] Together, these limits and exclusions mean that the economy is largely uninsured against catastrophic risks from advanced AI.

Mandating insurance for catastrophic risks would therefore be the largest shift from the *status quo*, for better or worse. I therefore give these special attention in my analysis.

---

[152] Nancy Germond, *Verisk to Roll Out New General Liability Exclusions for Generative AI Exposures*, INDEPENDENTAGENT.COM (Nov. 1, 2025), https://www.independentagent.com/vu_resource/verisk-to-roll-out-new-general-liability-exclusions-for-generative-ai-exposures/; Geoffrey B. Fehling et al., *The Continued Proliferation of AI Exclusions*, (May 2025), https://www.hunton.com/hunton-insurance-recovery-blog/the-continued-proliferation-of-ai-exclusions.

[153] Swiss Re, *AI – Unintended Insurance Impacts and Lessons from "Silent Cyber" | Swiss Re*, (June 12, 2024), https://www.swissre.com/institute/research/sonar/sonar2024/ai-silent-cyber.html; This is an allusion to "silent cyber." In the 2010s a number of traditional P&C insurers were caught off-guard by exposures to cyber risks they didn't notice. See *End of Silent Cyber in Property Insurance*, (Jan. 19, 2024), https://www.irmi.com/articles/expert-commentary/end-of-silent-cyber-in-property-insurance.

[154] LLOYD'S FUTURESET, GENERATIVE AI: TRANSFORMING THE CYBER LANDSCAPE (2024), https://assets.lloyds.com/media/439566f8-e042-4f98-83e5-b430d358f297/Lloyds_Futureset_GenAI_Transforming_the_cyber_landscape.pdf; MUNICH RE, MIND THE GAP (2024), https://www.munichre.com/content/dam/munichre/contentlounge/website-pieces/documents/MR_AI-Whitepaper-Mind-the-Gap.pdf/_jcr_content/renditions/original./MR_AI-Whitepaper-Mind-the-Gap.pdf; Swiss Re, *supra* note 153; SIMON WOODWARD, NIKHILMON O U & MITALI CHATTERJEE, TECH-TONIC SHIFTS: HOW AI COULD CHANGE INDUSTRY RISK LANDSCAPES (2024), https://www.swissre.com/institute/research/topics-and-risk-dialogues/digital-business-model-and-cyber-risk/ai-and-the-industry-risk-landscape.html.

[155] *Cf.* Schwarcz and Wolff, *supra* note 6 § III.A.5.

[156] Limits for how much the policy will pay out for a single underlying error or wrongful act. For example, if a tech company has a $5 million per claim limit and suffers a data breach affecting 100,000 customers who file 50 separate lawsuits, the insurer's maximum exposure is still $5 million for that single occurrence - not $5 million per lawsuit.

[157] Limits which cap total coverage for the entire policy period across all claims (or wrongful acts, or occurrences).

## *B. What Distortions Plague the Frontier AI Industry?*

For the rest of this Part, I ask: what biases, myopias, collective action problems, market failures and other identified distortions to care levels plague *specifically* the frontier AI industry? In doing so I am evaluating the *potential* for a net regulatory effect in this industry. Where I do find potential, I discuss why an insurance *mandate* in particular may be required to net the largest welfare gain.

I find there is considerable potential for a regulatory effect for insuring against catastrophic non-product accident risks, both because this is where the distortions to care levels are most significant and because these are the sort of distortions I expect insurers can reliably correct for. These distortions include:

- An underprovisionment of safety R&D due to upfront, fixed costs, and considerable spillovers
- An existential race to capture market share incenting unbalanced R&D investment
- Collective action problems relating to the damage an AI Three Mile Island would inflict on the industry's collective reputation
- The prevalence of overconfidence, availability bias, and the winner's curse in a young industry with many startups.

In addition to potential distortions, I identify demand side failures in the market for insurance against catastrophic accidents, highlighting the need for intervention.

I find that, to the extent major frontier AI developers might be judgment-proof, it's unlikely insurers will mitigate this effect by much given the enormous assets these firms control. I also find other distortions to dominate over the judgment-proof effect. In any case, insurers should reliably mitigate the judgment-proofness of smaller frontier AI firms.

I find there is considerably less potential for a regulatory effect regarding product accidents and accidents that are high-frequency low-severity: there is far less evidence to suggest that significant distortions are present (and virtually no evidence of an insurance market failure that would justify intervention).

### 1. Will Frontier AI Companies Be Judgment-Proof?

Recall, an injurer is said to be judgment-proof when they lack sufficient assets to fully compensate victims in the event of an accident, leading to inefficiently low care levels (§I.D.1).

While there is little consensus on the matter,[158] many AI experts believe frontier AI systems could soon cause major catastrophes, a large portion of whose damages

[158] Katja Grace et al., Thousands of AI Authors on the Future of AI (Apr. 30, 2024), http://arxiv.org/abs/2401.02843; Severin Field, *Why Do Experts Disagree on Existential Risk? A Survey of AI Experts*, AI ETHICS (2025), https://doi.org/10.1007/s43681-025-00762-0.

AI developers would plausibly be liable for.[159] Industry has recognized the possibility of catastrophic harms from advanced AI, including e.g. Chemical, Biological, Radiological and Nuclear (CBRN) harms.[160] Legislators in several states have too recognized the possibility of such catastrophic risks from advanced AI, moving to codify them in law.[161] Some experts believe advanced AI might even lead to human extinction.[162]

Literally existential risks are obviously (commercially) uninsurable.[163] However, to the extent that catastrophic yet non-existential risks are correlated with truly existential risks, we should expect mitigating the former to mitigate the latter.

Could such disasters render frontier AI companies insolvent, and thus judgment-proof? This requires making some assumptions about the size of their potential liabilities. Since I want to remain agnostic on those questions, I simplify my task by exploring the very upper bounds of commercial insurability. The largest Property and Casualty (P&C) policies written appear to be in the low billions (e.g. for airlines[164] and nuclear operators[165]). Some companies have multiple such policies, bringing their total coverage into the low tens of billions at the upper end, certainly no greater than $20 billion. This appears to be the limit of commercial insurability. Indeed, all combined, surplus available for P&C lines in the U.S. is around $1.1 trillion.[166] The sum of the limits of all policies will be larger, because insurers never expect all policies to get maxed out at once, but I doubt it's more than one order of magnitude larger.[167]

---

[159] *See generally* Bengio et al., *supra* note 130 and; Weil, *supra* note 147 § I.

[160] FRONTIER MODEL FORUM, FRONTIER MITIGATIONS (2025), https://www.frontiermodelforum.org/technical-reports/frontier-mitigations/.

[161] *See for example* "Critical Harms" in New York Bill A6453A (a.k.a. RAISE act) Alex Bores, *Responsible AI Safety and Education Act (RAISE Act)*, (2025), https://www.nysenate.gov/legislation/bills/2025/A6453/amendment/A; *Cf.* Scott Wiener, *Transparency in Frontier Artificial Intelligence Act (TFAIA)*, (2025), https://leginfo.legislature.ca.gov/faces/billTextClient.xhtml?bill_id=202520260SB53.

[162] Grace et al., *supra* note 158; CAIS, *Statement on AI Risk*, https://www.safe.ai/work/statement-on-ai-risk (last visited June 7, 2024); Yoshua Bengio et al., *Managing Extreme AI Risks amid Rapid Progress*, 384 SCIENCE 842, 3 (2024).

[163] Elsewhere I argue that the government could explicitly step into its role as insurer of last resort and charge risk-priced premiums for coverage against commercially uninsurable disasters. Obviously for literally existential risks, the point of such a scheme is not insurance per se but rather the internalization of negative externalities through what is effectively a Pigouvian tax (the premiums). Those funds could then be redirected toward safety efforts and other socially desirable programs aimed at managing disruptions from advanced AI. See Trout, *supra* note 97.

[164] Aeris Insurance Solutions, *Aircraft Liability - Aeris Insurance Solutions*, HTTPS://AERISINSURANCE.COM/, https://aerisinsurance.com/aircraft-liability/ (last visited July 13, 2025); AIG, US, *Aerospace & Aviation Insurance | AIG US*, https://orgn-aigcom.dmp.aig.com/home/risk-solutions/business/specialty-risks/aerospace-and-aviation (last visited July 13, 2025).

[165] John E. Gudgel, *Insurance and the Public–Private Management of Risk at US Commercial Nuclear Power Plants*, 26 RISK MANAGEMENT AND INSURANCE REVIEW 437, § 5.3.3 (2023).

[166] NATIONAL ASSOCIATION OF INSURANCE COMMISSIONERS, U.S. PROPERTY & CASUALTY AND TITLE INSURANCE INDUSTRIES - 2023 (2024),

Thus one simplified but practical framing asks: would frontier AI firms be judgment-proof in the event they were found liable for ~$20 billion in damages, an amount which insurers could feasibly cover all or most of? This requires knowing something about the finances of AI firms. Below, I offer some basic context and a rough picture of the situation these firms are in at the time of writing.

A brief search reveals that net assets[168] of the S&P 100 firms (Alphabet, Microsoft, Meta, and Amazon) are each measured in the hundreds of billions.[169] Even if these assets were all highly illiquid or sunk,[170] these are all highly profitable firms with revenues that also measure in the hundreds of billions.[171] Finally, their market capitalizations are in the low trillions.[172] Any one of these considerations should be

---

https://content.naic.org/sites/default/files/2023-annual-property-and-casualty-insurance-industries-analysis-report.pdf?utm_source=chatgpt.com.

[167] One way to estimate this is as follows. According to NAIC data, total premiums are approximately ~$0.9 trillion. If we assume the average rate on a line is 2~3% of total coverage, that suggests total coverage for U.S. P&C risks is $30~45 trillion. See *Id*.

[168] Total assets minus liabilities.

[169] *Meta Platforms Total Assets 2010-2025 | META*, https://www.macrotrends.net/stocks/charts/META/meta-platforms/total-assets (last visited July 15, 2025); *Alphabet Total Assets 2010-2025 | GOOGL*, https://www.macrotrends.net/stocks/charts/GOOGL/alphabet/total-assets (last visited July 15, 2025); *Microsoft Total Assets 2010-2025 | MSFT*, https://www.macrotrends.net/stocks/charts/MSFT/microsoft/total-assets (last visited July 15, 2025); *Amazon Total Assets 2010-2025 | AMZN*, https://www.macrotrends.net/stocks/charts/AMZN/amazon/total-assets (last visited July 15, 2025).

[170] A highly unrealistic assumption, especially for the likes of Google, Microsoft and Amazon that all have plenty of physical capital (e.g. datacenters).

[171] *Meta Platforms Revenue 2010-2025 | META*, https://www.macrotrends.net/stocks/charts/META/meta-platforms/revenue (last visited July 15, 2025); *Microsoft Revenue 2010-2025 | MSFT*, 2, https://www.macrotrends.net/stocks/charts/MSFT/microsoft/revenue (last visited July 15, 2025); *Alphabet Revenue 2010-2025 | GOOGL*, https://www.macrotrends.net/stocks/charts/GOOGL/alphabet/revenue (last visited July 15, 2025); *Amazon Revenue 2010-2025 | AMZN*, https://www.macrotrends.net/stocks/charts/AMZN/amazon/revenue (last visited July 15, 2025).

[172] *Meta Platforms (Facebook) (META) - Market Capitalization*, https://companiesmarketcap.com/meta-platforms/marketcap/ (last visited July 15, 2025); *Alphabet (Google) (GOOG) - Market Capitalization*, https://companiesmarketcap.com/alphabet-google/marketcap/ (last visited July 15, 2025); *Microsoft (MSFT) - Market Capitalization*, https://companiesmarketcap.com/microsoft/marketcap/ (last visited July 15, 2025); *Amazon (AMZN) - Market Capitalization*, https://companiesmarketcap.com/amazon/marketcap/ (last visited July 15, 2025).

more or less dispositive: it seems extremely unlikely any of these firms would be judgment-proof in the face of an additional $20 billion liability.

Information is scarce on the net assets of the smaller private startups (e.g. Imbue, Mistral, Cohere, Scale AI, and Anthropic), but it's fair to say most of their assets are either intangible (e.g. algorithms, branding, tacit knowledge and other IP) or in the form of human capital. Compared to physical assets (such as data centers), intangible assets are often less liquid or largely sunk in the event of a firm's defaulting.[173] Meanwhile, human capital flight is very real in this industry, since non-compete clauses are void in California, where key portions of the workforce are employed.[174] At the time of writing, the annualized revenues of these startups range from the low millions (Imbue, <$10 million)[175] to the low billions (Scale AI, ~$0.9 billion;[176] Anthropic, ~$2B).[177] Others sit in the middle (Mistral, ~$30 million;[178] Cohere, ~$100 million).[179] These startups are not profitable as of yet, and losses are projected to grow for those training and running cutting-edge chatbots. However, their revenues are growing aggressively (especially Anthropic and Scale AI) and they've

---

[173] *See generally* JONATHAN HASKEL & STIAN WESTLAKE, RESTARTING THE FUTURE: HOW TO FIX THE INTANGIBLE ECONOMY ch. 1 (2022).

[174] *Attorney General Bonta Issues Consumer Alert Reminding California Workers of Their Rights*, STATE OF CALIFORNIA - DEPARTMENT OF JUSTICE - OFFICE OF THE ATTORNEY GENERAL (Oct. 15, 2024), https://oag.ca.gov/news/press-releases/attorney-general-bonta-issues-consumer-alert-reminding-california-workers-their; Employees have used this to great effect. Witness e.g. the very credible threat of exodus that OpenAI employees made when the board of its non-profit arm tried to oust Sam Altman as CEO Ty Roush, *More Than 700 OpenAI Employees Threaten To Quit—And Join Microsoft—Unless Board Resigns*, FORBES, Nov. 20, 2023, https://www.forbes.com/sites/tylerroush/2023/11/20/more-than-500-openai-employees-threaten-to-quit-over-sam-altmans-removal/.

[175] *How Much Did Imbue Raise? Funding & Key Investors | Clay*, https://www.clay.com/dossier/imbue-funding (last visited July 15, 2025); *Imbue Company Overview, Contact Details & Competitors | LeadIQ*, https://leadiq.com/c/imbue/5ffde0a2b31988eee12bcfba (last visited July 15, 2025).

[176] *8 Scale AI Statistics (2025): Revenue, Valuation, Funding, Competitors*, TAPTWICE DIGITAL (Apr. 24, 2025), https://taptwicedigital.com/stats/taptwicedigital.com/stats/scale-ai; Anna Tong, Kenrick Cai & Krystal Hu, *Exclusive: Google, Scale AI's Largest Customer, Plans Split after Meta Deal, Sources Say*, REUTERS, June 13, 2025, https://www.reuters.com/business/google-scale-ais-largest-customer-plans-split-after-meta-deal-sources-say-2025-06-13/.

[177] PYMNTS, *Report: Anthropic's Annualized Revenue Reaches $1.4 Billion*, PYMNTS.COM (Mar. 11, 2025), https://www.pymnts.com/artificial-intelligence-2/2025/report-anthropics-annualized-revenue-reaches-1-4-billion/.

[178] Supantha Mukherjee, *French Startup Mistral Launches Chatbot for Companies, Triples Revenue in 100 Days*, REUTERS, May 7, 2025, https://www.reuters.com/technology/french-startup-mistral-launches-chatbot-companies-triples-revenue-100-days-2025-05-07/.

[179] Echo Wang, *AI Firm Cohere Doubles Annualized Revenue to $100 Million on Enterprise Focus*, REUTERS, May 15, 2025, https://www.reuters.com/business/ai-firm-cohere-doubles-annualized-revenue-100-million-enterprise-focus-2025-05-15/.

achieved valuations ranging from the low billions (Imbue ~$1 billion,[180] Mistral ~$6 billion,[181] Scale, ~$13 billion)[182] to mid tens of billions (Anthropic, ~$60 billion).[183] In sum, it seems very plausible that, at least for the next few years, a sudden ~$20 billion liability would render these startups insolvent and force them to file for bankruptcy. At that point victims would likely receive only partial compensation. In other words, these startups are likely judgment-proof for such catastrophes. The possible exception here is Anthropic, which, of them all, is the most likely to be able to simply raise more money if its investors aren't spooked.

What of OpenAI, the star of the current frontier AI boom? In many ways, its situation is just a more extreme version of the other startups (most closely resembling Anthropic). Currently its assets are likely similar in nature to the other startups: mostly intangible or human capital. The growth of its annual revenue has been incredibly aggressive, multiplying by roughly 18x from 200 million in 2022[184] to 3.7B in 2024, and OpenAI projects its annual revenue to triple again in 2025 to 13B.[185] However, its losses have also been staggering, with one report estimating it lost 5B in 2024 and that OpenAI's own projections imply it could lose as much as 14B in 2026.[186] Nevertheless, it reached a valuation of $300 billion in March 2025, one of the largest ever for a private tech startup.[187] Clearly many investors are willing to open their pocketbooks for OpenAI, despite its voracious burn rate. So would OpenAI be judgment-proof in the face of an unexpected $20 billion liability? They're unlikely to have sufficient cash immediately on hand to compensate victims but, like Anthropic, they may be able to simply raise more capital. However, it's unclear how securitized OpenAI's backers' investments are, and it's certainly not clear how investors would react to such a liability: they could decide to invest further, or instead withdraw, collecting collateral or attempting to sell what shares they can. If a fire sale began, OpenAI could quickly be forced to file for bankruptcy, at which point victims would again be left to shoulder undue losses.

---

[180] How Much Did Imbue Raise?, *supra* note 175.

[181] Mukherjee, *supra* note 178.

[182] Tong, Cai, and Hu, *supra* note 176.

[183] PYMNTS, *supra* note 177.

[184] Mike Isaac & Erin Griffith, *OpenAI Is Growing Fast and Burning Through Piles of Money*, THE NEW YORK TIMES, Sept. 27, 2024, https://www.nytimes.com/2024/09/27/technology/openai-chatgpt-investors-funding.html; Nathan Latka, *OpenAI Revenue of $3.7B: How It's Defining the Future of AI*, LATKA (Nov. 8, 2024), https://getlatka.com/blog/openai-revenue/ NB: annual revenue is not the same as annualized revenue. The latter is simply the monthly revenue multiplied by 12.

[185] The Economist, *Will OpenAI Ever Make Real Money?*, THE ECONOMIST, May 2025, https://www.economist.com/business/2025/05/15/will-openai-ever-make-real-money.

[186] Cory Weinberg, *OpenAI Projections Imply Losses Tripling to $14 Billion in 2026*, THE INFORMATION, Oct. 10, 2024, https://www.theinformation.com/articles/openai-projections-imply-losses-tripling-to-14-billion-in-2026.

[187] Cade Metz, *OpenAI Completes Deal That Values Company at $300 Billion*, THE NEW YORK TIMES, Mar. 31, 2025, https://www.nytimes.com/2025/03/31/technology/openai-valuation-300-billion.html.

Finally, I cover Elon Musk's frontier AI startup, xAI. Valued at $80 billion as of March 2025, with an annualized revenue[188] of $100 million as of Dec 2024, it's a special case among the startups for two reasons:

1. Notoriously, it is directly backed by the world's richest man, whose net worth currently sits around $374 billion.[189]
2. It controls at least one major physical capital asset, the supercomputer Colossus. It's unclear if it owns or is merely renting the 200,000 cutting-edge NVIDIA chips powering the supercomputer. If it owns them, those alone are worth roughly 4B. If they eventually purchase the total 1 million chips they intend to, that would bring the value of the asset closer to $27 billion.[190]

Despite its low revenue then, xAI seems considerably less judgment-proof in the face of a sudden $20 billion liability.

Considerable uncertainty surrounds all the startups due to their volatile nature. We're currently witnessing an AI boom; we may yet see a bust, possibly precipitated by one or more major AI-related disasters (§II.B.4). One thing is sure regarding the startups: outcomes are likely high variance. They will likely end up being very judgment-proof or not at all, and the liability threshold for which outcome we get is mercurial.

On the whole then, it's very implausible the S&P 100 firms will be judgment-proof in the face of a $20 billion liability, but entirely plausible that many of the startups will be. However, I conclude judgment-proofness is *not* where the most potential for a regulatory effect lies (especially in light of following sections). This conclusion came as a surprise. It runs somewhat counter to arguments from Trout[191] and Weil,[192] which present judgment-proofness as the principal distortion plaguing the industry, and make policy recommendations accordingly. Of course judgment-proofness *is* reduced by adding $20 billion of coverage against liability (whether that coverage kicks in after a comparatively small deductible or after all the company's assets have been exhausted, as Weil would prefer).[193] $20 billion just isn't so large a number relative to the assets available to most of the key players though.

---

[188] NB: *annual* revenue is not the same as *annualized* revenue. The latter is simply the monthly revenue multiplied by 12. Thus, for growing companies, it's typically much higher than that year's annual revenue.

[189] John Hyatt, *Elon Musk Scores $33 Billion In Self-Serving Merger*, FORBES, Apr. 2, 2025, https://www.forbes.com/sites/johnhyatt/2025/04/02/the-xai-x-deal-is-a-33-billion-windfall-for-elon-musk/.

[190] Grace Kay & Ellen Thomas, *Elon Musk's xAI Is Spending at Least $400 Million Building Its Supercomputer in Memphis. It's Short on Electricity.*, BUSINESS INSIDER, Apr. 1, 2025, https://www.businessinsider.com/elon-musk-xai-data-center-colossus-power-memphis-2025-4.

[191] Cristian Trout, *Liability and Insurance for Catastrophic Losses: The Nuclear Power Precedent and Lessons for AI*, *in* GENERATIVE AI AND LAW WORKSHOP AT THE INTERNATIONAL CONFERENCE ON MACHINE LEARNING (2024); Trout, *supra* note 97.

[192] Weil, *supra* note 6.

More importantly, there's good reason to believe there are other equally if not *more* significant distortions at work. So far in the theoretical literature, judgment-proofness has been the canonical example of an incentive distortion insurance uptake can help correct, but it is not because it is canonical that it should dominate our attention. After all, the model it's based on is quite simplistic: it assumes the economic actor in question only cares about its current, pecuniary assets. It completely ignores future profit flows, privileges it might lose by declaring bankruptcy, and non-pecuniary goods.

This can hardly capture the full story, especially in an industry as complex and high-stakes as this. As an intuition pump, consider: do you expect the CEOs of OpenAI or xAI, Sam Altman and Elon Musk, respectively, to act much differently if their companies were required to carry $20 billion in cash at all times, in case of a major accident? Make it $80 billion if you like. If your gut, like mine, says "no," then there is a lot more going on besides simple judgment-proofness. I turn to these forces next.

### 2. An (Existential) Race for Market Share

The frontier AI industry is famously in an aggressive race to capture as much share of a growing market as possible.[194] For OpenAI to have reached a valuation of $300 billion (~75 times its last annual revenue); xAI, $80 billion (~800 times its last annual revenue); and Anthropic, $60 billion (~30 times its last annual revenue), clearly the industry's backers expect this market to be very large.

Furthermore, the industry is pouring tens of billions of dollars into R&D each year, and that number is only increasing.[195] To take one illustrative example, OpenAI is estimated to have spent over $5 billion last year on R&D, over half of its total budget or roughly 150% of its revenue.[196] Those ratios are expected to remain roughly

---

[193] This is intended to minimize moral hazard. As I hope to make clear though, distortions other than moral hazard and judgment-proofness should be given more consideration. For example, if the firm survival motive is much more significant than other distortions, the choice of insulating companies from financial distress or not should take precedence.

[194] *See e.g*. Nico Grant & Cade Metz, *A New Chat Bot Is a 'Code Red' for Google's Search Business*, THE NEW YORK TIMES, Dec. 21, 2022, https://www.nytimes.com/2022/12/21/technology/ai-chatgpt-google-search.html.

[195] *See generally* Nestor Maslej et al., Artificial Intelligence Index Report 2025 (July 2, 2025), http://arxiv.org/abs/2504.07139.

[196] Weinberg, *supra* note 186. R&D costs consist of training compute, research compute, employee salaries, and data acquisition.

the same for some time, as the cost of training new frontier AI models continues to rise.[197]

The perceived stakes here are also clearly enormous. Whatever one believes about "Artificial General Intelligence" (AGI), it's undeniable that *industry leaders* frame the pursuit of AGI in explicitly existential terms.[198] These stakes are echoed by prominent narratives from AI insiders,[199] as well as leading experts. The "godfather of AI," Geoffery Hinton, has predicted there is "10% to 20%" chance that AGI would lead to human extinction within the three decades.[200] Yoshua Bengio, a colleague of

---

[197] *See generally* Epoch AI, *Machine Learning Trends*, EPOCH AI (Apr. 11, 2023), https://epoch.ai/trends#investment.

[198] See comments by Dario Amodei, CEO of Anthropic Dario Amodei, *Machines of Loving Grace*, (2024), https://www.darioamodei.com/essay/machines-of-loving-grace (characterizes AGI as "a country of geniuses in a datacenter"); Dario Amodei, *Statement from Dario Amodei on the Paris AI Action Summit*, (2025), https://www.anthropic.com/news/paris-ai-summit (predicting AGI will usher "the largest change to the global labor market in human history"); Liron Shapira [@liron], *Dario Amodei's P(Doom) Is 10–25%. CEO and Co-Founder of @AnthropicAI. Https://T.Co/sQyHCsasjS*, TWITTER (2023), https://x.com/liron/status/1710520914444718459 (estimating there is a 10-25% chance that AGI causes a catastrophe "on the scale of [...] human civilization."); See comments by Demis Hassabis, CEO of Google DeepMind Jess Kinghorn, *DeepMind CEO Makes Big Brain Claims, Saying AGI Could Be Here within "five to 10 Years" and Cause Humanity to Experience Widespread Change That's "10 Times Bigger than the Industrial Revolution, and Maybe 10 Times Faster,"* PC GAMER, Aug. 5, 2025, https://www.pcgamer.com/software/ai/deepmind-ceo-makes-big-brain-claims-saying-agi-could-be-here-in-the-next-five-to-10-years-and-that-humanity-will-see-a-change-10-times-bigger-than-the-industrial-revolution-and-maybe-10-times-faster/ (describing the disruption from AGI as likely "10 times bigger than the Industrial Revolution, and maybe 10 times faster"); Billy Perrigo, *Google DeepMind CEO Demis Hassabis on AGI and AI in the Military*, TIME, Apr. 2025, https://time.com/7280740/demis-hassabis-interview/ (claiming society post-AGI will require "a 'new political philosophy' to organize society" beyond liberal democracy); See comments by Elon Musk, CEO of xAI *Elon Musk Says Artificial Intelligence Is like "Summoning the Demon" - CBS News*, Oct. 2014, https://www.cbsnews.com/news/elon-musk-artificial-intelligence-is-like-summoning-the-demon/; Ryan Browne, *Elon Musk Says Global Race for A.I. Will Be the Most Likely Cause of World War III*, CNBC, Sept. 2017, https://www.cnbc.com/2017/09/04/elon-musk-says-global-race-for-ai-will-be-most-likely-cause-of-ww3.html; See comments by Sam Altman, CEO of OpenAI Sam Altman, *The Intelligence Age*, (Sept. 23, 2024), https://ia.samaltman.com/ (predicting AGI will enable "astounding triumphs – fixing the climate, establishing a space colony, and the discovery of all of physics"); Sarah Jackson, *The CEO of the Company behind AI Chatbot ChatGPT Says the Worst-Case Scenario for Artificial Intelligence Is "Lights out for All of Us,"* BUSINESS INSIDER, 2023, https://www.businessinsider.com/chatgpt-openai-ceo-worst-case-ai-lights-out-for-all-2023-1 (warning that the worst case scenario for AGI is "lights out" for humanity).

[199] Leopold Aschenbrenner, *Situational Awareness: The Decade Ahead*, (2024), https://situational-awareness.ai/; Daniel Kokotajlo et al., *AI 2027*, (2025), https://ai-2027.com/.

[200] Dan Milmo, *'Godfather of AI' Shortens Odds of the Technology Wiping out Humanity over next 30 Years*, THE GUARDIAN, Dec. 27, 2024, https://www.theguardian.com/technology/2024/dec/27/godfather-of-ai-raises-odds-of-the-technology-wiping-out-humanity-over-next-30-years.

Hinton's, warns that superintelligent AGI "would open a Pandora's box," potentially creating "existential risks"[201] or enabling "a world government dictatorship"[202] if one company achieves a decisive advantage and leverages economic dominance into political control.

Whether these claims are accurate is somewhat beside the point. What matters for competitive dynamics is that key decision-makers appear to genuinely believe them. We should assume these are basically rational actors who believe winning the race to AGI is tantamount to deciding the fate of humanity. These beliefs cannot be ignored when analyzing industry behavior.

All this suggests there are significant first-mover advantage dynamics at work, driving R&D investment higher than it would be in the absence of fierce competition. As a reminder, this can lead to either an over- or underinvestment in *safety* R&D, depending on the returns of safety R&D relative to other kinds of R&D (see §§I.D.2 and 3).

What kind of R&D is the industry prioritizing? I make the case that frontier AI companies are neglecting the forward-looking safety research required to investigate and mitigate potential catastrophic tail risks.

The frontier AI industry is built upon deep learning, a machine learning subfield that leverages abundant computing power to train enormous neural networks on very large datasets.[203] Large Language Models (LLMs), the technology powering the most advanced AI agents, represent some of the largest neural networks ever trained.[204] Their scale reflects a simple economic calculus: increasing data, compute, and inference time[205] reliably improves capabilities across knowledge domains.[206] This reliable return on investment explains the exponentially increasing R&D directed toward training new models. [207]

---

[201] Yoshua Bengio, *Implications of Artificial General Intelligence on National and International Security*, YOSHUA BENGIO (Oct. 30, 2024), https://yoshuabengio.org/2024/10/30/implications-of-artificial-general-intelligence-on-national-and-international-security/.

[202] Susan D'Agostino, *'AI Godfather' Yoshua Bengio: We Need a Humanity Defense Organization*, BULLETIN OF THE ATOMIC SCIENTISTS (Oct. 17, 2023), https://thebulletin.org/2023/10/ai-godfather-yoshua-bengio-we-need-a-humanity-defense-organization/.

[203] *See generally* Bengio et al., *supra* note 130 ch. 1.

[204] *Data on Notable AI Models*, EPOCH AI (June 19, 2024), https://epoch.ai/data/notable-ai-models.

[205] *Learning to Reason with LLMs*, https://openai.com/index/learning-to-reason-with-llms/ (last visited July 15, 2025).

[206] Deep Ganguli et al., *Predictability and Surprise in Large Generative Models*, *in* 2022 ACM CONFERENCE ON FAIRNESS, ACCOUNTABILITY, AND TRANSPARENCY 1747 (2022), http://arxiv.org/abs/2202.07785.

[207] Epoch AI, *supra* note 197; Maslej et al., *supra* note 195 ch. 4 § 2.

Despite the success of scaling, certain problems persist. LLMs are notoriously prone[208] to "hallucinating," i.e. presenting fabricated claims as fact or following imaginary instructions.[209] Such hallucinations likely cause frequent but currently small-scale accidents (chatbots hallucinating false company policies, for example).[210] These accidents remain small-scale likely because unreliability slows adoption in high-stakes contexts like core financial sector activities;[211] the firm that solves hallucinations would gain a substantial competitive advantage. Thus AI firms undoubtedly invest considerable R&D effort here (though as progress stalls,[212] investments likely default back to scaling).

This illustrates a general point: safety investments are difficult to disentangle from product quality improvements (cf. §I.D.2). This is good news: for many accident types, fierce competition should induce considerable safety investment. These will be the relatively frequent accidents that impose customer losses or otherwise hamper adoption (i.e. typical product accidents). Mitigating them generally confers a competitive edge.

Unfortunately this does not apply to efforts aimed at mitigating the catastrophic risks that many experts warn of.[213] Precisely because tail risks are rare and highly uncertain, investments in mitigating them do not clearly translate into a substantial or reliable competitive advantage. Temporality compounds the problem. Though such investments might serve firms' long-term interests, they are bound by the short-term constraints of the race for market dominance. This applies especially to startups facing intense pressure to achieve aggressive revenue increases.[214]

As evidence of neglect for catastrophic risks, I hold up the growing number of stories about leading firms rushing dangerous capability evaluations[215] or only doing them after releasing the models in question.[216] These evaluations are precisely the early warning system intended to alert society to the increased risk of catastrophe.[217] Witness too, OpenAI disbanding their "superalignment" team, which was its only team dedicated to solving technical challenges surrounding the safety and controllability of future superhuman AI systems.[218]

---

[208] Simon Hughes, Minseok Bae & Miaoran Li, *Vectara Hallucination Leaderboard*, (2023), https://github.com/vectara/hallucination-leaderboard; Giwon Hong et al., The Hallucinations Leaderboard -- An Open Effort to Measure Hallucinations in Large Language Models (Apr. 17, 2024), http://arxiv.org/abs/2404.05904.

[209] Bengio et al., *supra* note 130 § 2.2.1.

[210] Marisa Garcia, *What Air Canada Lost In 'Remarkable' Lying AI Chatbot Case*, FORBES, Feb. 19, 2024, https://www.forbes.com/sites/marisagarcia/2024/02/19/what-air-canada-lost-in-remarkable-lying-ai-chatbot-case/; Cade Metz & Karen Weise, *A.I. Is Getting More Powerful, but Its Hallucinations Are Getting Worse*, THE NEW YORK TIMES, May 5, 2025, https://www.nytimes.com/2025/05/05/technology/ai-hallucinations-chatgpt-google.html.

[211] *See e.g*. Sutton and Williams, *supra* note 140 at 7.

[212] Metz and Weise, *supra* note 210.

[213] *See e.g*. Bengio et al., *supra* note 130 § 2.2.3.

[214] Managerial incentives likely prove misaligned as well, inefficiently prioritizing short-term revenue growth over long-term safety (see §I.D.6).

[215] Pranshu Verma, Nitasha Tiku & Cat Zakrzewski, *OpenAI Promised to Make Its AI Safe. Employees Say It 'Failed' Its First Test.*, THE WASHINGTON POST, July 12, 2024, https://www.washingtonpost.com/technology/2024/07/12/openai-ai-safety-regulation-gpt4/.

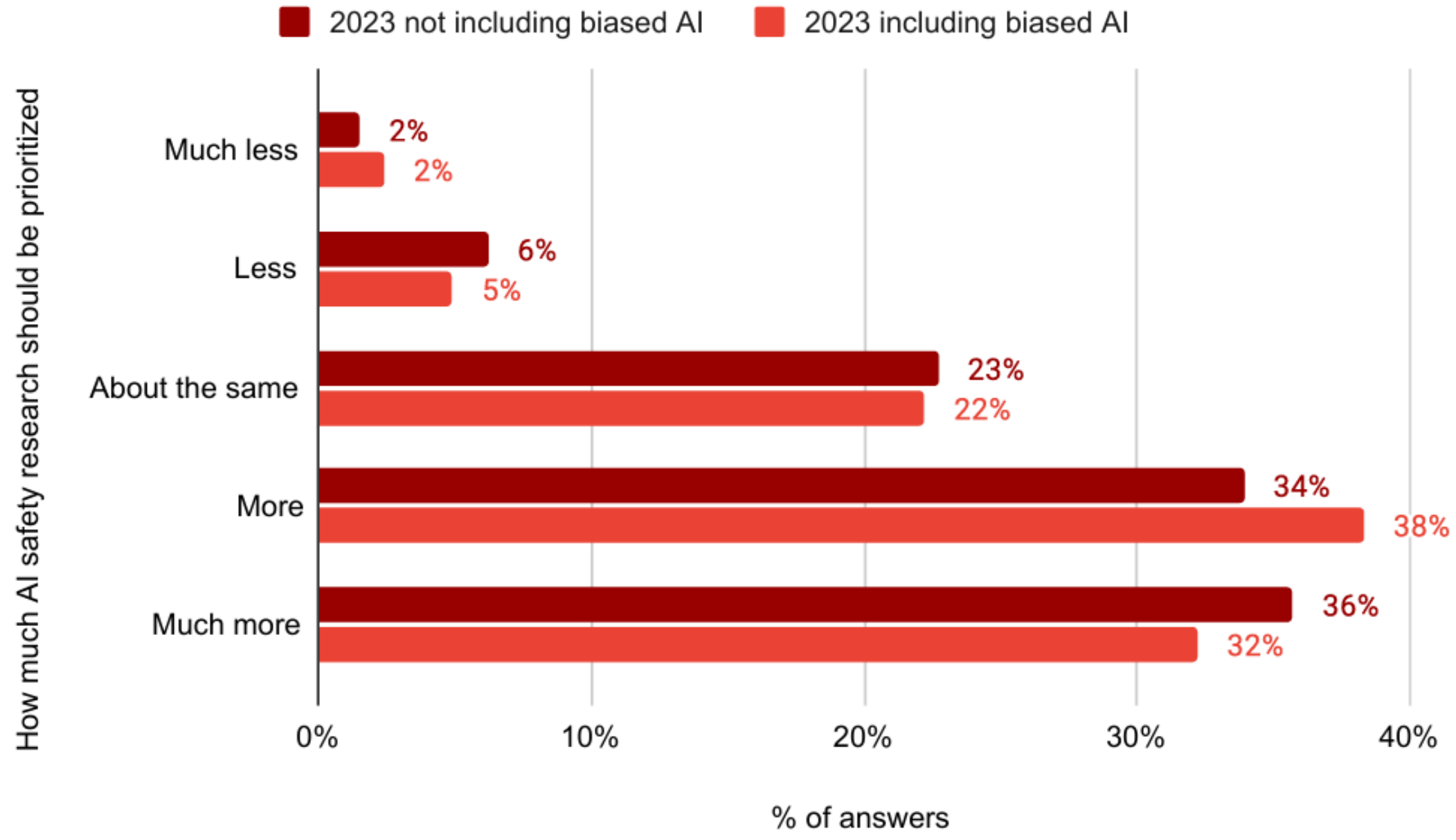

*Figure 3. Two framings of the question "How much should AI safety research be prioritized?" one including and one not including biased AI as an example.*[219]

Finally, in a survey of machine learning experts, 70% of respondents (68% the preceding year) answered that safety research should be prioritized "more" or "much more," suggesting systemic underinvestment.[220] Tellingly, when examples of such research included more high-frequency low-consequence accidents (such as unfair or biased AI decision-making), the aggregate "more" or "much more" response didn't change, but the emphasis on "much more" declined (Figure 3). This suggests it is catastrophic risks that are particularly neglected.

---

[216] Jennifer Elias, Hayden Field & Jonathan Vanian, *AI Research Takes a Backseat to Profits as Silicon Valley Prioritizes Products over Safety, Experts Say*, CNBC, May 2025, https://www.cnbc.com/2025/05/14/meta-google-openai-artificial-intelligence-safety.html.

[217] Toby Shevlane et al., Model Evaluation for Extreme Risks (Sept. 22, 2023), http://arxiv.org/abs/2305.15324.

[218] Will Knight, *OpenAI's Long-Term AI Risk Team Has Disbanded*, WIRED, https://www.wired.com/story/openai-superalignment-team-disbanded/ (last visited July 15, 2025).

[219] Figure 20 from Grace et al., *supra* note 158.

[220] Grace et al., *supra* note 158 at 17 Among the examples of such research was "research on long-term existential risks from AI systems."

In sum, the competing AI firms largely appear to be taking a "we'll deal with it when we get there" approach to catastrophic risks from advanced AI. This hardly seems socially efficient: competition is driving a race to the bottom.

Mandatory insurance for catastrophic risks could help address this market failure. By requiring frontier AI firms carry such insurance, policymakers would outsource part of the regulatory challenge to the market: insurers will enforce minimum safety mitigations for catastrophic risks that likely raise the floor. This is because insurers don't face the same trade-offs between different R&D streams that the competing firms do (§I.D.2), and the supply of such insurance could be highly concentrated (§IV.B). However, outcomes will depend heavily on the liability regime in force (§III.B.4) and mandate design (§IV).

One final distortion merits attention. The race for market dominance produces distortions similar to those stemming from firm survival motives facing heterogeneous insolvency risks (see §§I.D.2 and I.D.3). Though these companies command large revenue streams, many are startups operating at enormous losses, burning through their "runway" very fast (§II.B.1). This rapid growth phase is often critical for startups: failure to make inroads against competitors or grow revenue as promised can cause startups to stall and lose investor backing. In this context, stalling likely means eventual shutdown or a forced sale to owners willing to pursue more aggressive growth. Thus, despite high valuations, the owners and managers of the startups likely feel pressure not only from the race for market dominance but also survival imperatives (§I.D.3). For many, the most salient survival risk is likely falling behind competitors and not tail liability risks, which, despite outsized welfare impacts, are by definition rare and uncertain. Firm survival motives are therefore likely *compounding* competitive distortions, driving care further below efficient levels with regard to tail risks.

### 3. The Spillovers and Fixed, Upfront Costs of AI Safety Research

We saw that spillovers combined with fixed and upfront costs can lead to inefficiently low R&D investments in competitive environments (§I.D.2). AI safety research exhibits the key features that make such failures likely.

Economists consistently find that spillovers occur more frequently for intangible assets (e.g. algorithms) than other asset types.[221] AI safety research – like all AI R&D – is predominantly intangible. Moreover, primary research typically generates substantial spillovers and represents upfront, fixed costs, which explains the standard rationale for public research grants. Is AI safety research primary research?

Not all of it: much of it involves making marginal improvements to established techniques. Reinforcement Learning from Human Feedback (RLHF)[222] and its descendants are good examples here. Firms have extensively refined these methods and they remain central to contemporary frontier systems.[223]

---

[221] HASKEL AND WESTLAKE, *supra* note 173 ch. 1.

[222] Paul F Christiano et al., *Deep Reinforcement Learning from Human Preferences*, 30 *in* ADVANCES IN NEURAL INFORMATION PROCESSING SYSTEMS (2017),

However, many experts question whether such incremental refinements will suffice as AI systems grow more capable. Many argue we must entirely rethink how we build powerful general AI systems, calling instead for systems that are e.g. "safe-by-design" or have formally-verifiable safety guarantees.[224] Such approaches will require significant exploration and genuine scientific breakthroughs, rather than incremental optimization.

The mix of incremental versus primary research we will ultimately require remains unclear. What *is* clear is that the engineering fields that feed into AI safety are protean. Consider: the science for controlling useful but untrustworthy AI systems is less than a year old, beginning with the work of Greenblatt et al.;[225] the science of evaluating AI systems for dangerous capabilities is no more than a year old, beginning with the work of Kinniment et al.;[226] mechanistic interpretability – "neuroscience for AI" – arguably didn't begin in earnest until 5 years ago with Cammarata et al.'s investigation into neural network circuits;[227] and the Deep Learning revolution itself, the breakthrough that has enabled this entire industry, doesn't date back much more than a decade, beginning with the work of LeCun et al.[228] In other words, there is clearly fertile ground for primary research.

We should strongly suspect spillovers and fixed, upfront costs are leading to an undersupply of such exploratory safety research. Public grants remain essential for addressing such market failures, but an insurance mandate might also be an effective, if unconventional, policy response. A mandate could concentrate safety R&D efforts – potentially into a single entity (§IV.B) – providing an elegant solution to multiple distortions simultaneously. Spillovers are minimized (or eliminated), and the returns to fixed upfront costs are maximized for the actor incurring them. If the

---

https://proceedings.neurips.cc/paper_files/paper/2017/hash/d5e2c0adad503c91f91df240d0cd4e49-Abstract.html.

[223] Yuntao Bai et al., Constitutional AI: Harmlessness from AI Feedback (Dec. 15, 2022), http://arxiv.org/abs/2212.08073; Melody Y. Guan et al., Deliberative Alignment: Reasoning Enables Safer Language Models (Jan. 8, 2025), http://arxiv.org/abs/2412.16339.

[224] *See e.g*. David "davidad" Dalrymple et al., Towards Guaranteed Safe AI: A Framework for Ensuring Robust and Reliable AI Systems (July 8, 2024), http://arxiv.org/abs/2405.06624.

[225] Ryan Greenblatt et al., AI Control: Improving Safety Despite Intentional Subversion (July 23, 2024), http://arxiv.org/abs/2312.06942.

[226] Megan Kinniment et al., Evaluating Language-Model Agents on Realistic Autonomous Tasks (Jan. 4, 2024), http://arxiv.org/abs/2312.11671.

[227] Nick Cammarata et al., *Thread: Circuits*, 5 DISTILL e24 (2020).

[228] Yann LeCun, Yoshua Bengio & Geoffrey Hinton, *Deep Learning*, 521 NATURE 436 (2015).

insurer is a joint underwriting association[229] or a mutual, then by forcing buy-in to the insurance regime, policymakers would essentially be forcing buy-in to a joint venture on safety R&D (between insurers or the AI firms themselves, respectively) aimed at providing what is essentially a public good.

This approach has limitations: insurers cannot be expected to fund entirely new, speculative techniques for building general AI systems, so long as such techniques have no bearing on their policyholders' existing products. Nonetheless, a number of promising research agendas would likely see a boost.[230] Part III discusses in greater depth why insurers would engage in such R&D efforts.

#### 4. Industry Reputation: Implications of an AI Three Mile Island and the Unilateralist Curse

We saw that shared industry reputation can depress care below efficient levels (§I.D.2.a). The mechanism is straightforward: customers observe industry or firm reputation more readily than individual product quality. The public good character of a shared industry reputation enables free-riding: firms adopting fewer precautions benefit from competitors' higher standards.

How rich are customers' insights into frontier AI products? The day-to-day safety and reliability of competing consumer-facing products appear fairly transparent: hallucination benchmarks are widely available,[231] and public reviewers extensively document issues like code generation reliability. There are exceptions: most users likely lack awareness of data collection and usage practices that may impose diffuse long-term harms.[232] There are also a growing number of "AI psychosis" cases, in which the delusions of unsuspecting users are slowly reinforced.[233]

The situation is quite different for business customers, where extensive deployment of an AI system might have far-reaching and difficult-to-foresee consequences. Commentators emphasize AI systems' complexity, training data opacity, and rapid evolution, arguing these factors make comprehensive risk evaluation extremely difficult for business customers.[234]

---

[229] These are special risk-pooling associations, in which multiple insurers collectively underwrite niche, high-risk, or generally low profitability risk. Whether voluntary or by force, these special entities can only be formed with the blessing of insurance regulators, and are typically only so created to ensure coverage availability where the standard market has failed.

[230] Controlling untrustworthy AI agents, dangerous capability evaluations, and possibly mechanistic interpretability, to name a few.

[231] Hughes, Bae, and Li, *supra* note 208.

[232] Bengio et al., *supra* note 130 § 2.3.5.

[233] Rachel Fieldhouse, *Can AI Chatbots Trigger Psychosis? What the Science Says*, 646 NATURE 18 (2025).

[234] *See e.g.* Price II and Cohen, *supra* note 145; Buiten, de Streel, and Peitz, *supra* note 28 § 5.1.3.

Insurance markets can address this market failure by enabling developers to signal product quality credibly, with insurers effectively acting as independent quality assurance.[235] Indeed, a market for this is emerging in the frontier AI space, with indemnification commitments from AI vendors themselves,[236] as well as emerging insurance products from startups like Armilla AI,[237] the Artificial Intelligence Underwriting Company,[238] Testudo,[239] AiShelter,[240] as well as traditional players like Chubb[241] and Munich RE.[242] This is a "regulatory market" in action.[243] It's too early to tell what the effects on care levels will be, but given that there is both demand and supply for such products, it's much harder to claim we're in the presence of a significant market failure. Intervention is harder to justify, bringing into question legislative proposals such as California's SB 813.[244]

So far these insurance products only cover small to medium product accidents though. Could a shared industry reputation distort care levels for rare, severe events? No such accident has yet occurred, so empirical evidence is lacking. The risks experts warn of are on the scale (or larger) of catastrophic accidents we have observed in other industries though: can we learn something from these analogies?

---

[235] Zhang, Yu, and Chen, *supra* note 57.

[236] Hossein Nowbar & Brad Smith, *Microsoft Announces New Copilot Copyright Commitment for Customers*, MICROSOFT ON THE ISSUES (Sept. 7, 2023), https://blogs.microsoft.com/on-the-issues/2023/09/07/copilot-copyright-commitment-ai-legal-concerns/; Kyle Wiggers, *OpenAI Promises to Defend Business Customers against Copyright Claims*, TECHCRUNCH (Nov. 6, 2023), https://techcrunch.com/2023/11/06/openai-promises-to-defend-business-customers-against-copyright-claims/; Chris Stokel-Walker, *Adobe Is so Confident Its Firefly Generative AI Won't Breach Copyright That It'll Cover Your Legal Bills*, FAST COMPANY (June 8, 2023), https://www.fastcompany.com/90906560/adobe-feels-so-confident-its-firefly-generative-ai-wont-breach-copyright-itll-cover-your-legal-bills.

[237] *Armilla: Insurance for AI Risks*, https://www.armilla.ai/ (last visited July 16, 2025).

[238] *AIUC | AI Agent Standard & Insurance*, https://aiuc.com/ (last visited Sept. 10, 2025).

[239] *Testudo | Generative AI Insurance & Underwriting Company*, TESTUDO, https://www.testudo.co (last visited Sept. 10, 2025).

[240] AiShelter LLC, *Artificial Intelligence Liability Insurance | AiShelter*, ARTIFICIAL INTELLIGENCE LIABILITY INSURANCE | AISHELTER, https://www.aishelter.com (last visited Sept. 10, 2025).

[241] Chubb, *Chubb Google Cloud Risk Protection Program*, CHUBB GOOGLE CLOUD RISK PROTECTION PROGRAM, https://www.chubb.com/content/dam/chubb-sites/chubb-com/us-en/business-insurance/products/cyber/documents/chubb-google-risk-protection-program.pdf (last visited Sept. 10, 2025).

[242] *aiSure™ More AI Opportunity. Less AI Risk | Munich Re*, https://www.munichre.com/en/solutions/for-industry-clients/insure-ai.html (last visited July 16, 2025).

[243] Hadfield and Clark, *supra* note 7.

[244] Jerry McNerney, *Multistakeholder Regulatory Organizations*, (2025), https://leginfo.legislature.ca.gov/faces/billTextClient.xhtml?bill_id=202520260SB813.

Drops in demand for commercial aviation following fatal accidents and the 9/11 terrorist attacks offer relevant case studies.[245] However, for want of space, I focus on the case of major accidents in the commercial nuclear power industry. After both the 1979 Three Mile Island accident in the U.S. and the 2011 Fukushima Daiichi accident in Japan, each country's respective nuclear industry was met with major public and regulatory backlash, severely damaging the industry's profitability and slowing expansion (or in Japan's case, leading to the indefinite shutdown of nearly all reactors for several years).[246] Indeed, these accidents, along with the 1986 Chernobyl accident, were so damaging to the industry's reputation that they fueled successful anti-nuclear movements beyond the borders of where the accident took place (e.g. in Germany and Switzerland).[247]

In the U.S., the industry quickly realized after Three Mile Island (TMI) they were "hostages of each other," their collective reputation representing a public good amongst them; to protect it, they would have to police each other.[248] For example, when the owner of the Three Mile Island plants filed suit against its manufacturer, damning information about the operator's practices and their manufacturer's quality control began coming to light; the case was abruptly settled out of court *under pressure from other nuclear power plant owners* to avoid further public fallout.[249] After a lengthy and prolonged effort to establish a trusted and credible industry regulatory body, as well as reform the professional culture of both managers and operators, the U.S. commercial nuclear power industry stands today as an unusually effective case of industry self-regulation, substantially raising the floor (and bar) for safety across the industry.[250]

It's important to emphasize the role of government regulation here. Federal regulators provide an important check on the nuclear industry, and their power to e.g. revoke licenses is still an important recourse when self-regulatory pressures are insufficient to change the behavior of a delinquent firm.[251] It's argued that a perennial *threat* of *increased* regulation is a crucial ingredient for *maintaining* the self-regulatory regime.[252]

---

[245] Haoming Liu & Jinli Zeng, *Airline Passenger Fatality and the Demand for Air Travel*, 39 APPLIED ECONOMICS 1773 (2007).

[246] Andrew Glen Benson, Three Essays on the Political Economy of Nuclear Power (2021) (UC Irvine), https://escholarship.org/uc/item/4q6110jf.

[247] *See generally* BENJAMIN K. SOVACOOL & SCOTT VICTOR VALENTINE, THE NATIONAL POLITICS OF NUCLEAR POWER: ECONOMICS, SECURITY, AND GOVERNANCE (2012); and J. SAMUEL WALKER, THREE MILE ISLAND: A NUCLEAR CRISIS IN HISTORICAL PERSPECTIVE ch. 10 (2004); For quantitative evidence of the regulatory backlash to Three Mile Island, see e.g. Benson, *supra* note 246 ch. 3; For a quantitative analysis of the different factors leading to the increased cost and delay in nuclear power build out, see e.g. *Id.* ch. 1.

[248] *See generally* JOSEPH V. REES, HOSTAGES OF EACH OTHER: THE TRANSFORMATION OF NUCLEAR SAFETY SINCE THREE MILE ISLAND (1996), https://press.uchicago.edu/ucp/books/book/chicago/H/bo3618989.html.

[249] Susan Q. Stranahan, *Opinion | The Real Three Mile Island Story Leaks Out*, THE WASHINGTON POST, Feb. 20, 1983, https://www.washingtonpost.com/archive/opinions/1983/02/20/the-real-three-mile-island-story-leaks-out/75d044b4-6bab-42f2-8738-d4292c9d8ed0/.

[250] REES, *supra* note 248; Neil Gunningham & Joseph Rees, *Industry Self-Regulation: An Institutional Perspective*, 19 LAW & POLICY 363 (1997); Gudgel, *supra* note 165.

This self-regulatory regime revolves around membership in the Institute of Nuclear Power Operations (INPO), which is closely linked to the industry's mutual insurer, Nuclear Electric Insurance Limited (NEIL). Purchasing a NEIL policy requires INPO membership.[253] INPO runs inspections of plants, conducts research, collects best practices, as well as delivers report cards and recommendations to operators. INPO mostly relies on peer pressure to advance its mission of promoting power plant safety,[254] but ever since its attachment to NEIL, its inspections now inform NEIL's underwriting and premiums as well.[255]

INPO and NEIL were established after TMI, but skeptics will point out that TMI happened *despite* operators *already* being insured for third-party liability risks by American Nuclear Insurers (ANI). Judging the counterfactual (how different TMI would have been, or how many TMIs would have transpired *without* ANI) is quite difficult, and the evidence suggests ANI makes meaningful effort in monitoring and controlling their policyholders,[256] but skeptics have a point. Looked at another way, this suggests the relative efficacy of *mutuals* (like NEIL) over *stock insurers* (like ANI), especially in handling collective action problems (see §III.B.3).

As mentioned, many experts believe the frontier AI industry faces TMI-like risks: rare, catastrophic, and likely to be highly mediatized against a backdrop of general public pessimism about AI.[257] What often goes unnoted is that these risks are largely *collective*, borne by *the industry as a whole*, in the form of shared reputation. All this suggests they face a collective action problem and will need to make collective efforts similar to those made by the U.S. nuclear industry to improve safety, satisfy regulators, and ultimately, earn the public's trust.

---

[251] For example, the INPO orchestrated the firing of a COO and CEO of a nuclear plant owner, by leaking damning information with the Nuclear Regulatory Commission. See REES, *supra* note 248 at 110–118.

[252] Gunningham and Rees, *supra* note 250 at 391.

[253] Gudgel, *supra* note 165 at 458.

[254] REES, *supra* note 248 ch. 6.

[255] Gudgel, *supra* note 165 § 5.3.3.

[256] John E. Gudgel, *Insurance as a Private Sector Regulator and Promoter of Security and Safety: Case Studies in Governing Emerging Technological Risk From Commercial Nuclear Power to Health Care Sector Cybersecurity*, 158 (2022), https://hdl.handle.net/1920/13083.

[257] Colleen McClain Pasquini Brian Kennedy, Jeffrey Gottfried, Monica Anderson and Giancarlo, *How the U.S. Public and AI Experts View Artificial Intelligence*, PEW RESEARCH CENTER (Apr. 3, 2025), https://www.pewresearch.org/internet/2025/04/03/how-the-us-public-and-ai-experts-view-artificial-intelligence/.

There are likely two collective action problems at work. First, where firms are rational and purely self-interested, we have a classic free-rider problem. The carelessness of one AI developer could devastate the entire industry, and in an industry as competitive as this, the incentive to cut corners on safety while externalizing most of the costs onto your competitors will be strong: we should expect significantly lower than efficient levels of care. To be sure, it's producer welfare that will be hit first, but there is no reason to believe AI developers won't pass costs onto consumers. In the extreme, the entire industry could severely contract or collapse, triggering a new "AI winter." Consumers would likely lose out on the benefits of AI for a generation as society developed a different way to manage the technology's risks and rebuild trust in it.[258]

Second, even if firms were purely altruistic,[259] we still have a case of the unilateralist's curse.[260] That is, a situation in which an increase in the number of uncoordinated altruistic agents with imperfect knowledge increases the odds of a mistaken *harmful* action being taken, *despite* best intentions. Imagine an idealistic startup founder who doesn't care they're operating at a loss, sees themselves as bestowing upon humanity the greatest invention since fire, believes they have solved the problem of aligning superhuman AI to human values, brushes off the warnings from the few outside voices that have any visibility into what they're doing, and unilaterally decides to unleash their creation upon the world. Increase the number of such actors, and the odds that a catastrophic error gets made go up.

Thus firms need some way of policing each other, and or sharing information to avoid catastrophic mistakes. We might be seeing the germ of this in the Frontier Model Forum,[261] but this body has a very long way to go if it wants to become an analog of INPO. Naturally, building trust between competitors, minding anti-trust law, signaling credible commitments, navigating differing worldviews and other overhead costs all pose serious challenges to spontaneous coordination. These are the symptoms of a market failure.

The regime developed for commercial nuclear power was reactive, but there is no reason the frontier AI industry can't be proactive. Governments that want to protect the long-term growth of this industry should encourage such efforts, lest they be forced to stymie the industry under public pressure after a disaster.

Making the industry collectively liable for residual losses from a catastrophe would help make this collective action problem more explicit,[262] but this may not be sufficient to overcome it.[263] A well-designed insurance mandate – especially one that encourages mutualization[264] (§IV.B) – could effectively force participation in an adaptable self-regulatory regime that has the government's blessing, shielding the industry from anti-trust in matters of safety practices. The insurer becomes the centralized body for industry self-regulation. Thus a mandate might not only increase *welfare* in expectation, but also *profits* for frontier AI firms: by avoiding an AI Three Mile Island, they can preserve the public's trust and protect themselves against a new "AI winter." Solving this coordination problem could well be a win-win, for the industry and the public.

---

[258] Should that be desired.

[259] I.e. strove to maximize total social welfare.

[260] Nick Bostrom, Thomas Douglas & Anders Sandberg, *The Unilateralist's Curse and the Case for a Principle of Conformity*, 30 SOCIAL EPISTEMOLOGY 350 (2016).

[261] *Frontier Model Forum*, FRONTIER MODEL FORUM, https://www.frontiermodelforum.org/ (last visited July 16, 2025).

[262] Friedman, *supra* note 7 § IV.B.2.

### 5. Is the Frontier AI Industry Too Big to Fail?

We saw that when an industry or firm is critical to the broader economy, governments can't credibly commit to *not* bail them out were disaster to strike (§I.D.4). This creates charity hazard.

At least two elements are necessary for charity hazard to materialize in the frontier AI industry. First and foremost, the catastrophic risks from advanced AI must be sufficient in magnitude to trigger an emergency response from the government. Second, the frontier AI industry must be perceived to be sufficiently critical to the broader economy or national security that the government will be willing (or forced) to prop it up in the event of such a shock.

Until more of the catastrophic risks experts warn of can be ruled out, the first element is certainly satisfied, given that threats include:[265]

- CBRN risks, especially through the intervening agency of a criminal actor
- AI agents causing critical infrastructure failure or mass casualties, whether at the behest of a criminal actor or because law-abiding actors lost control over the AI agents.

The second element seems more and more likely to be satisfied as time goes on. The industry seems to be increasingly viewed, by administrations of both parties, as critical to the future of national security and the economy.[266] Many commentators no doubt argue that the industry has been successfully lobbying the government in this respect.

[263] For example, when the Price-Anderson Act was amended in 1975, legislators added a mechanism quite like shared residual liability. All US nuclear power plant operators are now required to participate in a secondary layer of insurance that covers excess losses from major incidents, essentially forcing operators nationwide to insure each other. However, pro-rated premiums are collected only retrospectively, ex post – much as shared residual liability will be assessed ex post. Sure enough, nuclear power operators didn't create a mutual or effective self-regulatory body until they experienced a major catastrophe (Three Mile Island) and started paying premiums upfront. See United States Nuclear Regulatory Commission, *supra* note 94 § 1.1.2.

[264] In a mutual, policyholders can vote on the board of the mutual. Thus, if all frontier AI firms joined the same mutual, this would be the closest feasible regime to allowing competing firms to vote on each other's behavior, a potential solution to the unilateralist curse as well as the free-rider problem. *Cf.* Friedman, *supra* note 7 § IV.B.2.d.

[265] FRONTIER MODEL FORUM, *supra* note 160; Alex Bores, *supra* note 161; Bengio et al., *supra* note 130 § 2.1.4, 2.2.3.

Can insurance uptake make a dent in this growing charity hazard? This is subtle. To some extent, this refers back to the question of whether insurance uptake can make AI firms substantially less judgment-proof in the face of $20 billion liabilities (the upper limit of commercial insurability). However, to keep things clear, judgment-proofness refers only to the distortion in incentives caused by *insolvency,* limiting what compensation victims receive *through courts*.

Charity hazard needn't come in the form of bailouts to insolvent firms. It could also come in the form of government-provided emergency containment services or emergency relief to victims: these are costs that the injuring firms are externalizing. It could also come in the form of favorable loans or contracts from the government, aimed at helping a critical industry rebound faster. These would amount to subsidies.

Further research should investigate how often such charity occurs in the event of disasters where there are clearly responsible parties. A cursory look into the *Deepwater Horizon* oil spill suggests such charity was quite limited.[267] However, even if the government was ultimately paid back, the injuring parties unquestionably relied on the government's emergency response services and funds. Some of this work can be outsourced to private insurers, as was done in the nuclear power industry through the Price-Anderson Act. Under that regime, private insurers are an integral part of the government's emergency response plan, standing ready to provide immediate financial relief and deploy dedicated emergency claim personnel in the event of a disaster.[268]

Finally, note that to the extent one believes an AI Three Mile Island is likely to trigger a *regulatory backlash* (as opposed to a *bailout*), one must believe charity hazard is less likely: these two distortions to incentives are mutually exclusive to an extent.

---

[266] *See e.g*. Kevin Frazier, *Vance Outlines an America First, America Only AI Agenda*, LAWFARE (2025), https://www.lawfaremedia.org/article/vance-outlines-an-america-first--america-only-ai-agenda; Ezra Klein, *Opinion | The Government Knows A.G.I. Is Coming*, THE NEW YORK TIMES, Mar. 4, 2025, https://www.nytimes.com/2025/03/04/opinion/ezra-klein-podcast-ben-buchanan.html.

[267] U. S. Government Accountability Office, *Deepwater Horizon Oil Spill: Preliminary Assessment of Federal Financial Risks and Cost Reimbursement and Notification Policies and Procedures | U.S. GAO*, (Nov. 12, 2010), https://www.gao.gov/products/gao-11-90r; Jonathan L Ramseur, *Deepwater Horizon Oil Spill: Recent Activities and Ongoing Developments*, WASHINGTON, DC, USA: CONGRESSIONAL RESEARCH SERVICE (2014); National Oceanic and Atmospheric Administration, *10 Years of NOAA's Work After the Deepwater Horizon Oil Spill: A Timeline | NOAA Fisheries*, NOAA (Apr. 24, 2025), https://www.fisheries.noaa.gov/national/habitat-conservation/10-years-noaas-work-after-deepwater-horizon-oil-spill-timeline.

[268] United States Nuclear Regulatory Commission, *supra* note 94 § 2.3.15.nrc

### 6. Psychology: an Immature Industry and Overconfident, Optimistic Founders

Efficiency results are only as neat as their assumptions about human psychology. We saw in §I.D.5 that insurers can sometimes help correct certain systemic biases in human decision-making with regard to catastrophic risks, biases that are very prevalent in the frontier AI industry.

For starters, this industry is unquestionably young: its oldest players are a few decades old.[269] Meanwhile, most keystone insurance companies were established over a century ago.[270] In any case, standards in the frontier AI industry are still very much in flux, and no firm has more than a few years of experience with this technology. The industry has not experienced any major accident, and is still mapping out the technology's vulnerabilities. To be sure, the industry is developing guidelines to manage such risks[271] but so far it's unclear how much of this is simply good PR.[272]

Granted, the lack of data on accidents means insurers also face serious challenges in estimating risks: that cuts both ways. The difference, however, is that insurers *can have their gaze trained* on catastrophic risks, risks that will otherwise be paid insufficient attention due to a variety of psychological biases (§I.D.5).

The prevalence of startups in the industry is also noteworthy. Tech startup founders exhibit famously high risk-tolerance, optimism, overconfidence, and seem to seek risk more than the general population.[273] This is at least in part due to how startups are financed. Hits-based venture capital heavily incentivizes pursuing high-risk, high-reward opportunities with extremely rapid growth targets, hence Silicon Valley's "move fast and break things" mantra. The sheer youth of startups also means their risk management practices are likely underdeveloped, relying more on intuitive thinking than deliberative reasoning,[274] and having yet to draw on the wealth of risk-management expertise developed in other industries.[275] Lack of repeated play means they are also more susceptible to a version of the winner's curse:[276] the competitor

---

[269] Microsoft, Alphabet, and Meta to some extent.

[270] E.g. Lloyd's of London or Swiss Re.

[271] *See e.g.* FRONTIER MODEL FORUM, *supra* note 160.

[272] Consider e.g. Billy Perrigo, *Exclusive: OpenAI Lobbied E.U. to Water Down AI Regulation*, TIME, June 2023, https://time.com/6288245/openai-eu-lobbying-ai-act/.

[273] *See generally* Thomas Astebro et al., *Seeking the Roots of Entrepreneurship: Insights from Behavioral Economics*, 28 JOURNAL OF ECONOMIC PERSPECTIVES 49 (2014); and Sari Pekkala Kerr, William R. Kerr & Tina Xu, *Personality Traits of Entrepreneurs: A Review of Recent Literature*, 14 ENT 279 (2018).

[274] KUNREUTHER AND USEEM, *supra* note 110 ch. 4–5.

[275] Simon Mylius, Systematic Hazard Analysis for Frontier AI Using STPA (June 2, 2025), http://arxiv.org/abs/2506.01782 (see reference to "unstructured hazard analysis").

[276] In an auction, winning a bid is some evidence that the winner is simply the person who most overestimated the true value of the asset (while other, potentially better-informed

willing to build the most powerful AI agents the fastest (in pursuit of market dominance), is also likely the actor who most underestimates the risks. For high-frequency risks, players can quickly calibrate; not so for low-frequency catastrophic risks.

By contrast, major insurers have been disciplined by the market to seek "good risk" and otherwise estimate risks accurately. They have mature risk management practices and are repeat players familiar with catastrophic risks. Their aversion to ambiguity and uncertainty[277] is consistent with a learned heuristic that mitigates the winner's curse.[278]

Thus introducing insurers as stakeholders in the management of catastrophic risks from frontier AI should improve decision-making, correcting deficient care levels. However, a mandate will likely be required to realize this improvement, given that less risk-averse, overconfident and optimistic actors are precisely the kinds of actors who purchase less insurance than they should.[279]

## III. OVERCOMING MORAL HAZARD AND REALIZING THE POTENTIAL

Part I described a model of insurers' net regulatory effect and net hazard effect. According to the model, the outcome we get depends on:

1. How distorted the incentives to take care are in the base case. This determines the *potential* for a net regulatory effect.
2. How much effort insurers make to mitigate *moral hazard* (or otherwise reduce losses), and how effective those efforts are. This determines how much of that potential gets *realized*.

Here in Part III, we're concerned with this second question: assessing how much (if any) of that *potential* gets realized. This largely turns on how much moral hazard is generated in the first place, as well as when and why insurers fail or succeed to mitigate it. My analysis examines four key factors: the type of policyholder, the type of risk being insured, the type of insurer, and the structure of the insurance market.

It's found that niche insurers, and especially mutuals, are particularly good at keeping moral hazard in check. It's also found that when policy limits are large and risks are catastrophic, insurers will be induced to make greater *ex ante* effort, since they cannot afford to "wait and see" and slowly collect actuarial data. In some situations, fierce competition between insurers can spur greater effort, while in

---

bidders, valued the asset less highly). If not adjusted for, this can lead to systematic losses for auction winners.

[277] Howard Kunreuther, Robin Hogarth & Jacqueline Meszaros, *Insurer Ambiguity and Market Failure*, 7 J RISK UNCERTAINTY 71 (1993).

[278] Mumpower, *supra* note 117.

[279] Coats and Bajtelsmit, *supra* note 106.

others, a concentrated market allows insurers to avoid spillovers and capture more of the returns on their loss reduction efforts. In general, due to transaction costs, insurers make much greater effort at reducing losses with large commercial clients than with consumers and small businesses.

Regarding liability insurance in particular, it's found to function more effectively with greater legal stability: without this, insurers cannot price their products or invest in loss reduction measures with confidence. For realizing a net regulatory effect, *clearer* and *stricter* liability assignment is possibly even more important: this makes *legal maneuvering* less viable as a strategy for reducing losses (in contrast with genuine harm reduction efforts). Discouraging the former while encouraging the latter is crucial for avoiding the worst failures of regulation by insurance, such as insurers helping insureds bury damning information, engage in box-checking exercises, or possibly resist adopting safety improvements that would alter the standard of care in ways unfavorable for insurers or insureds.

Case studies of how insurers handle other catastrophic risks indicate insurers are generally capable of pricing risks with known unknowns, but struggle with unknown unknowns. That said, it's unclear what social institution does better. The upshot is that horizon-scanning exercises that simply try to *notice* and *distinguish* new risks[280] can offer a particularly high social return on investment.

Readers primarily interested in the frontier AI industry may wish to skip directly to Part IV, where policy implications are synthesized.

### *A. Efforts to Mitigate Moral Hazard or Otherwise Reduce Losses*

Regulation by insurance can fail. It's not hard to find examples of insurers failing to mitigate moral hazard, or otherwise increasing harm by helping policyholders skirt liability.[281] Some scholars argue insurers sometimes *deliberately* attempt to

---

[280] *Cf.* Daniel Carpenter et al., Dark Speculation: Combining Qualitative and Quantitative Understanding in Frontier AI Risk Analysis (Nov. 26, 2025), http://arxiv.org/abs/2511.21838.

[281] *See e.g*. Alma Cohen & Rajeev Dehejia, *The Effect of Automobile Insurance and Accident Liability Laws on Traffic Fatalities*, 47 THE JOURNAL OF LAW AND ECONOMICS 357 (2004) (finding strong evidence of moral hazard arising from auto insurance uptake) ; Tom Baker & Sean J. Griffith, *The Missing Monitor in Corporate Governance: The Directors' & (and) Officers' Liability Insurer*, 95 GEO. L.J. 1795 (2006) (demonstrating that D&O insurers often fail to take measures that can meaningfully limit the risk of corporate malfeasance); Shauhin Talesh, *Legal Intermediaries: How Insurance Companies Construct the Meaning of Compliance with Antidiscrimination Laws*, 37 LAW & POLICY 209 (2015) (finding that insurers' risk management efforts focus predominantly on coaching employers to avoid being sued rather than on promoting the underlying goals of liability, such as fostering a healthy workplace environment); Marcos Antonio Mendoza, *The Limits of Insurance as Governance: Professional Liability Coverage for Civil Rights Claims against Public School Districts*, 38 QUINNIPIAC L. REV. 375 (2019); Shauhin A. Talesh & Bryan Cunningham, *The Technologization of Insurance: An Empirical Analysis of Big Data an Artificial Intelligence's Impact on Cybersecurity and Privacy*, 2021 UTAH L. REV. 967 (2021)

*increase* risks in order to expand long-term demand for their products,[282] though others dispute this.[283] After all, most of insurers' efforts are already geared toward risk-reduction, as we shall see.

The basic problem of moral hazard is given its modern treatment in the seminal work of Stiglitz.[284] Essentially, the incentives of insureds become distorted when insurers cannot perfectly observe the true probabilities of accidents or the efforts made by insureds.

Insurers do collect data in order to better predict the frequency of accidents, often more effectively than insureds. Pricing risk is, after all, the core business of insurance. *Actuarial* risk-modeling is the norm, relatively simple statistical analysis of accident data that assumes historical patterns will be predictive of future ones. However, insurers do sometimes engage in *causal* risk-modeling, such as physics-based models of hurricane damage,[285] catastrophe models for terrorism risk,[286] network analysis for cyber risks,[287] epidemiological models of disease spread,[288] and supply chain resilience models for business interruption insurance.[289]

---

("Although reliance on technology and data are increasingly transforming the way insurers advertise, underwrite, and price insurance, the actual impact on insurer behavior seems to have remained minimal and is largely symbolic."); Schwarcz and Wolff, *supra* note 6 § III.4 (finding that insurers help firms reduce liability, while increasing risk of harm, by hiring lawyers who bury damning information).

[282] Ronen Avraham & Ariel Porat, *The Dark Side of Insurance*, 19 REVIEW OF LAW & ECONOMICS 13 (2023).

[283] *Cf.* Abraham and Schwarcz, *supra* note 3 § III.B.2.

[284] Richard J. Arnott & Joseph E. Stiglitz, The Basic Analytics of Moral Hazard (1988), https://www.nber.org/papers/w2484; *See also* Mark V. Pauly, *The Economics of Moral Hazard: Comment*, 58 THE AMERICAN ECONOMIC REVIEW 531 (1968); *and* Steven Shavell, *On Moral Hazard and Insurance*, 93 THE QUARTERLY JOURNAL OF ECONOMICS 541 (1979).

[285] PATRICIA GROSSI, HOWARD KUNREUTHER & CHANDU C. PATEL, CATASTROPHE MODELING: A NEW APPROACH TO MANAGING RISK ch. 3, 4 (2005).

[286] *Id.* ch. 10.

[287] *See e.g.* RISK MANAGEMENT SOLUTIONS, INC., MANAGING CYBER INSURANCE ACCUMULATION RISK 2016 (2016), https://www.jbs.cam.ac.uk/centres/risk/publications/technology-and-space/cyber-risk-outlook/managing-cyber-insurance-accumulation-risk-2016/; Lloyd's Futureset, *Lloyd's Systemic Risk Scenario Reveals Global Economy Exposed to $3.5trn from Major Cyber Attack*, (Oct. 18, 2023), https://www.lloyds.com/about-lloyds/media-centre/press-releases/lloyds-systemic-risk-scenario-reveals-global-economy-exposed-to-3.5trn-from-major-cyber-attack.

[288] *See e.g.* Runhuan Feng & Jose Garrido, *Actuarial Applications of Epidemiological Models*, 15 NORTH AMERICAN ACTUARIAL JOURNAL 112 (2011); Charles Iwebuke Nkeki & Emeka Henry Iroh, *On Epidemiological and Actuarial Analyses of Health Insurance Models for Communicable Diseases*, 29 NORTH AMERICAN ACTUARIAL JOURNAL 422 (2025).

[289] QUANTIFYING BUSINESS INTERRUPTION: RISK PROPAGATION IN COMPLEX SUPPLY CHAINS | SWISS RE (2024), https://www.swissre.com/institute/research/topics-and-risk-dialogues/economy-and-insurance-outlook/complex-supply-chains.html.

Insurers are incentivized to monitor and influence an insured's behavior to avoid adverse selection and ensure losses don't exceed premiums collected. Why insurers would provide risk-mitigation services more broadly is less obvious: why reduce that which creates demand for their product? Ben-Shahar and Logue go over the various rationales in more detail,[290] but here I summarize: any risk reduction achieved after policy issuance provides a net benefit to the insurer; risk reduction can make premiums more affordable, allowing insurers to expand the potential customer base or steal market share; often risk-management services are a key value proposition for large corporate policyholders;[291] insurers can directly profit from predictable changes in aggregate losses, most easily accomplished through targeted risk reduction;[292] risk-reduction willingness serves as a screening device to identify profitable customers, a.k.a. "good risk."[293] On top of these, *mutuals* have a straightforward incentive for cost-effective loss reduction: it is in the interest of their policyholders, who own the mutual.

Ben-Shahar and Logue provide a thorough description of the tools insurers use to mitigate moral hazard or otherwise reduce losses. As a reminder, for explicitly mitigating moral hazard – monitoring and controlling insureds' behavior – these tools include:[294]

- Leaving insureds exposed to some of the risk through copayments and deductibles.[295]
- Individually risk-pricing premiums (e.g. experience rating in auto insurance,[296] or the Engineering Rating Factor in nuclear power plant insurance[297]).
- Monitoring insured behavior (e.g. the use of telematics in auto,[298] or power plant inspection in nuclear[299])

---

[290] Ben-Shahar and Logue, *supra* note 3 at 203–205.

[291] Baker and Siegelman, *supra* note 17 § 3.B.

[292] Abraham and Schwarcz, *supra* note 3 § III.B.2.

[293] Ben-Shahar and Logue, *supra* note 3 at 204.

[294] *Id.* § I; *See also* Baker and Swedloff, *supra* note 3 § I.

[295] Abraham and Schwarcz argue that copayments and deductibles should really be understood as the absence of insurance rather than a tool for mitigating moral hazard from the insurance purchased. Abraham and Schwarcz, *supra* note 3 § IV.B; While they have a point, this is misleading: even e.g. small deductibles or co-payment mechanisms can have an outsized impact on correcting incentives, due to threshold effects, varying demand elasticities at different price points, and small frictions halting a race to the bottom between policyholders. In brief, insurance consumption and precautionary behavior do not respond linearly to insurance coverage. See Pauly for a clear exposition of the theory Pauly, *supra* note 284; For empirical verification, see e.g. Aviva Aron-Dine, Liran Einav & Amy Finkelstein, *The RAND Health Insurance Experiment, Three Decades Later*, 27 JOURNAL OF ECONOMIC PERSPECTIVES 197 (2013).

[296] Ben-Shahar and Logue, *supra* note 3 § I.A.1.

[297] Gudgel, *supra* note 256 at 158.

- Refusing to insure, or conditioning coverage on adherence to certain risk management practices and standards.[300]
- Voluntary coaching[301] (e.g. cyber insurers offering cybersecurity consulting) or private safety code enforcement (e.g. property insurers developing building code ratings and lobbying localities to enforce higher standards).[302]
- Investigating claims and refusing to pay claims when insureds violate policy terms (e.g. through gross negligence).[303]

Thus, to a large extent, moral hazard mitigation and insurers' regulatory effect are two sides of the same coin. Other loss reduction efforts include:

- Conducting safety R&D and disseminating safer technological alternatives (e.g. auto insurers developing crashworthiness tests for automobiles long before the National Highway Traffic Safety Administration)[304]
- Lobbying the government for stricter regulation of insured activities (e.g. when insurers lobbied for making airbags mandatory for new automobiles)[305]

## *B. Factors that Determine Insurer Effort and Effectiveness*

The initial severity of moral hazard, and the efforts to combat it or otherwise reduce losses, vary widely in their scope and effectiveness. Here I cover some of the key factors that determine outcomes.

### 1. Policyholder Type

Hanson and Logue find that insurers rarely make much effort when providing first-party insurance to consumers for small-frequent product accidents.[306] There are two reasons for this. First, the number and heterogeneity of policyholders and insured activities are large, driving up transaction costs. Second, consumers are simply ill-equipped to generate and collect the information insurers would need to make risk classifications. By contrast, manufacturers are well equipped to generate

---

[298] Ben-Shahar and Logue, *supra* note 3 at 208; *See also* Yizhou Jin & Shoshana Vasserman, Buying Data from Consumers: The Impact of Monitoring Programs in U.S. Auto Insurance (July 2021), https://www.nber.org/papers/w29096.

[299] Gudgel, *supra* note 165 § 5.3.

[300] Ben-Shahar and Logue, *supra* note 3 § I.A.3.

[301] *Id.* § I.A.4.

[302] *Id.* § I.A.5.

[303] *Id.* § I.B. Note, for want of space I largely ignore claims management as a regulatory tool, mostly because I'm interested in ex ante loss reduction efforts.

[304] *Id.* § I.A.6. and p. 222.

[305] Robert Kneuper & Bruce Yandle, *Auto Insurers and the Air Bag*, 61 THE JOURNAL OF RISK AND INSURANCE 107 (1994).

[306] Hanson and Logue, *supra* note 56 at 145–153.

and collect such information. They are also fewer, with more homogeneous risk-bearing preferences.

Whether insurance generates *any* moral hazard effect in the first place can depend heavily on psychology. For example, individuals who purchase more insurance against natural disasters tend to take greater precaution, and the underlying causal factors appear to be psychological traits of the insurance consumer, such as risk-aversion and worry.[307] For such insurance lines, insurers likely spend more effort identifying this "good risk" and extracting rents than realigning policyholder incentives. That said, under these conditions, some loss reduction efforts are also more profitable for insurers, such as safety R&D, safety coaching, and private safety codes. This is because "inducement costs"[308] for getting these policyholders to take greater care should be lower than with unbiased, risk-neutral actors.[309]

Conversely, overconfident or optimistic individuals likely take too few precautions by default (§I.D.5). There is some evidence that, e.g. well-designed training programs can reduce overconfidence (and thus reduce losses).[310] Theoretically, inducement costs (essentially a form of agency cost)[311] could be lower or higher depending on whether overconfidence or optimism dominates (respectively), and how much the risk depends on the policyholder's actions.[312] Obviously if the policyholder is genuinely *risk-happy,* inducement costs will be higher.

Overall, few studies appear powerful enough to determine whether insurance uptake tends to have a strong *counterfactual* impact on loss reduction for either subgroup of policyholders (i.e. the risk-averse or worrisome, and the overconfident or optimistic). One recent study in the U.S. did find that safer drivers tended to self-select for greater oversight from insurers in exchange for potential premium discounts, and that when they did so, they became 30% safer still.[313]

---

[307] Hudson et al., *supra* note 96; Botzen, Kunreuther, and Michel-Kerjan, *supra* note 96; Jantsje M. Mol, W. J. Wouter Botzen & Julia E. Blasch, *Risk Reduction in Compulsory Disaster Insurance: Experimental Evidence on Moral Hazard and Financial Incentives*, 84 JOURNAL OF BEHAVIORAL AND EXPERIMENTAL ECONOMICS 101500 (2020).

[308] E.g. premium discounts, or marketing for taking extra precautionary measures.

[309] Hanson and Logue, *supra* note 56 § D.I.b.

[310] Alvaro Sandroni & Francesco Squintani, *A Survey on Overconfidence, Insurance and Self-Assessment Training Programs* (2004).

[311] Michael C. Jensen & William H. Meckling, *Theory of the Firm: Managerial Behavior, Agency Costs and Ownership Structure*, 3 JOURNAL OF FINANCIAL ECONOMICS 305 (1976).

[312] Leonidas Enrique De La Rosa, *Overconfidence and Moral Hazard*, 73 GAMES AND ECONOMIC BEHAVIOR 429 (2011); *See also* Luis Santos-Pinto & Leonidas Enrique de la Rosa, *Overconfidence in Labor Markets*, *in* HANDBOOK OF LABOR, HUMAN RESOURCES AND POPULATION ECONOMICS 1 (2020), https://link.springer.com/rwe/10.1007/978-3-319-57365-6_117-1.

[313] Jin and Vasserman, *supra* note 298.

### 2. Risk Type

There are myriad ways in which characteristics of a risk (or set of risks) influence insurers' ability and efforts to reduce losses. Below I go over a few of the most important.

#### *a. Liability Risk*

In the context of liability insurance, it's critical to emphasize the wedge between *harm* and *liability*: one can reduce one's liability (e.g. by destroying records of one's negligence) without reducing accidents or harms. As such, liability insurers, like injurers, can always minimize liability risk in one of two ways: legal maneuvering or improved harm reduction. When the former is more cost-effective than the latter, we can expect insurers to engage in it, as sometimes seen in, e.g. Cyber insurance[314] and Employment Practice Liability Insurance.[315] This is the most severe failure mode for regulation by insurance, and one reason why clearer and strict(er) forms of liability synergize well with the regulation by insurance strategy: by reducing the surface area for lawyering, they make genuine harm reduction the more appealing investment for insurers. Incidentally, channeling liability can also induce further effort on the part of insurers by protecting *them* from liability: scholars note that *insurers* can be held liable when they undertake to provide serious risk-reduction services to their policyholders, which can discourage insurers from making such efforts.[316]

Setting aside such pernicious cases, how else does liability insurance compare with other lines? The extensive literature comparing States with no-fault automobile accident compensation schemes versus those with traditional tort schemes finds no conclusive evidence of increased accident rates.[317] That suggests that insurers, regardless if they are offering first-party bodily injury insurance or third-party liability insurance, are no more or less successful at mitigating moral hazard.[318]

The virtual collapse of the U.S. liability insurance market in the 70s and 80s spawned a large literature investigating its causes.[319] I highlight two insights that emerged. First, the supply of liability insurance depends on a stable legal landscape, or at least a landscape that only changes slowly and more or less predictably. This stability is not only critical for insurers to be able to correctly price the coverage, but also for determining which investments in loss reduction to make.

---

[314] Schwarcz and Wolff, *supra* note 6 § III.4.

[315] Talesh, *supra* note 281.

[316] Kyle D. Logue, *Encouraging Insurers to Regulate: The Role (If Any) for Tort Law*, 5 UC IRVINE L. REV. 1355 (2015).

[317] JAMES M. ANDERSON, PAUL HEATON & STEPHEN J. CARROLL, THE U.S. EXPERIENCE WITH NO-FAULT AUTOMOBILE INSURANCE: A RETROSPECTIVE 79–82 (2010), https://www.rand.org/pubs/monographs/MG860.html.

[318] However, Parsons points out that auto accidents are not a representative case. Typically the relationship between injurers and potential victims is highly asymmetrical: a few actors have a disproportionate impact on a disproportionate number of other actors. Parsons, *supra* note 128 at 457. This is the case for insurance against e.g. product liability or environmental liabilities. Further research could probe how this asymmetry influences moral hazard mitigation in third-party liability coverage vs first-party coverage.

[319] Mark A. Geistfeld, *Chapter 11: Products Liability*, § 13 (2009), https://www.elgaronline.com/display/book/9781782547457/b1_chapter11.xml.

Second, insurers struggle to profit from (let alone mitigate) long-tail risks. These are risks with a significant time lag between the initial occurrence of exposure or harm, and manifestation of injury, together with claim filing. Such risks are particularly acute for *occurrence* based policies,[320] which were common in the mid-20th century[321]. Asbestos poisoning is the most famous example: exposures in the 50s and 60s led to very costly mass tort suits decades later.[322] The temporal separation of long-tail risks presents significant difficulties that insurers struggle to manage with their traditional toolkit. Deductibles and co-payments are moot if the original policyholder no longer exists (e.g. if the insured firm was liquidated).[323] Most notably, experience rating has limited effectiveness because claims can emerge years after the inciting incident(s), breaking the feedback loop between behavior change and premium consequences.[324] That's assuming the relevant policyholder can even be accurately identified, let alone the relevant bad behavior.[325] Some have suggested that underwriters may also be incentivized to underprice policies involving long-tail risk, in order to write more business in the present and leave the consequences to their successors.[326] This is essentially a form of temporal judgment-proofing, or "looting," a problem well-documented among firms with long-tail liabilities.[327] So here is a risk type for which the judgment-proof problem is especially acute but difficult for insurers to remedy. There is great *potential* for a net regulatory effect, but insurers struggle to *realize* it.

---

[320] Occurrence-based policies cover claims arising from incidents that occurred during the policy period, regardless of when the claim is made. The insurer can thus remain liable for decades-old exposures. By contrast, claims-made policies cover only claims made during the policy period (or extended reporting period). Once the policy expires, no new claims are covered, even for past exposures.

[321] Kenneth S. Abraham, *The Long-Tail Liability Revolution: Creating the New World of Tort and Insurance Law*, 6 U. PA. J.L. & PUB. AFF. 347, 27 (2020).

[322] Jeffrey W. Stempel, *Assessing the Coverage Carnage: Asbestos Liability and Insurance after Three Decades of Dispute*, 12 CONN. INS. L.J. 349 (2005).

[323] Timothy Swanson & Robin Mason, *Long Tail Risks and Endogenous Liabilities: Regulating Looting*, 23 THE GENEVA PAPERS ON RISK AND INSURANCE. ISSUES AND PRACTICE 182 (1998).

[324] Ringleb and Wiggins, *supra* note 31 n. 5; *Cf.* Boomhower, *supra* note 31.

[325] Abraham, *supra* note 321 § I.C.1.

[326] Parsons, *supra* note 128 § 7.

[327] Ringleb and Wiggins, *supra* note 31; Akerlof et al., *supra* note 29; Swanson and Mason, *supra* note 323.

### b. *Third-party Moral Hazard*

Some risks pose acute *third-party* moral hazard (a.k.a. *external* moral hazard). Kidnap insurance provides a vivid illustration. Economic models and common sense suggest that carrying such insurance could make one a *greater* target of such a crime, as kidnappers will be more confident they can extract a greater ransom. There is some anecdotal evidence suggesting bans of said insurance may have decreased kidnapping rates.[328] On the other hand, in key markets where kidnapping insurance is allowed, insurers have made significant efforts to reduce this third-party moral hazard.[329] Other examples of third-party moral hazard include insurance against ransomware,[330] and imperfect repair markets (e.g. auto repair services being overpriced since carrying insurance makes customers less price-sensitive).[331]

Related to third-party moral hazard are the coordination problems surrounding security investments. Security externalities mean that when one actor hardens their security, they may make other actors more attractive targets. Conversely, interdependencies between security systems mean that uncoordinated investment in security will likely be far below efficient levels.[332] So far, insurers have struggled to overcome these coordination problems in e.g. cyber, but there is much talk of greater private and public partnerships to better standardize and pool incident reporting.[333]

In sum, insurers face significant additional hurdles when trying to reduce losses that involve intentional misuse or criminal activity.

### c. *Dynamic Risk*

A shifting legal environment is not the only uncertainty insurers must contend with when attempting to price premiums and determine which efforts to invest in.

---

[328] Gideon Parchomovsky & Peter Siegelman, *Third-Party Moral Hazard and the Problem of Insurance Externalities*, 51 THE JOURNAL OF LEGAL STUDIES 93, § 2.3 (2022).

[329] Anja Shortland, *Governing Kidnap for Ransom: Lloyd's as a "Private Regime,"* 30 GOVERNANCE 283 (2017).

[330] Kyle D. Logue & Adam B. Shniderman, *The Case for Banning (and Mandating) Ransomware Insurance*, SSRN JOURNAL (2021), https://www.ssrn.com/abstract=3907373.

[331] Martin Nell, Andreas Richter & Jörg Schiller, *When Prices Hardly Matter: Incomplete Insurance Contracts and Markets for Repair Goods*, 53 EUROPEAN ECONOMIC REVIEW 343 (2009).

[332] Erwann Michel-Kerjan & Burkhard Pedell, *How Does the Corporate World Cope with Mega-Terrorism? Puzzling Evidence from Terrorism Insurance Markets*, 18 JOURNAL OF APPLIED CORPORATE FINANCE 61, 5 (2006); Martin Eling, *Cyber Risk and Cyber Insurance*, *in* HANDBOOK OF INSURANCE: VOLUME I 199, 217 (Georges Dionne ed., 2025), https://doi.org/10.1007/978-3-031-69561-2_7.

[333] Eling, *supra* note 332 at 217; Bruce Schneier & Josephine Wolff, *Building a Cyber Insurance Backstop Is Harder Than It Sounds*, LAWFARE (2024), https://www.lawfaremedia.org/article/building-a-cyber-insurance-backstop-is-harder-than-it-sounds; Some loss data is beginning to be pooled by two consortiums. See *Cyber Industry Loss Index*, PERILS, https://www.perils.org/products/cyber (last visited Oct. 7, 2025); and *About*, CYBERACUVIEW, https://cyberacuview.com/about-us/ (last visited Oct. 7, 2025).

Large changes in underlying technology (e.g. in cyber security)[334] or science (e.g. the discovery of asbestos's harmful health effects) or broad behavioral shifts (e.g. in the tactics of cyber attackers[335] or terrorists)[336] make it difficult for insurers to predict what kinds of loss prevention efforts will be cost-effective.

#### *d. Magnitude, Concentration and Frequency of Losses*

The magnitude, concentration and frequency of losses are particularly salient in determining the effort insurers make.

Risk manifests differently depending on whether exposure is concentrated in a few large policies or distributed across many small ones. A single large policy will typically trigger more intensive loss prevention efforts from insurers than equivalent aggregate exposure spread across numerous small policies, due to lower transaction costs.[337] While smaller policies receive standardized coverage with minimal individual attention, large corporate policyholders will tend to receive more comprehensive risk management services: detailed safety audits, customized loss control programs, and ongoing technical consultation, etc. Indeed, as noted before, that is often a key selling point for such policyholders.[338]

Separate from the concentration of risk across policies is simply the size of policy limits. According to conversations with industry insiders, policies with limits over the ~$100 million mark receive a disproportionate amount of attention from insurers. This makes sense: such policies represent not only key customers in an insurer's portfolio, but also potential sources of increased variance in losses over a period.

Limiting variance in losses is also partly what drives insurers to add exclusions and sub-limits for catastrophic risks.[339] However, when insurers are exposed to such risks, they make significant effort to mitigate them.[340] Indeed, when exposure to catastrophic risk threatens firm insolvency, these efforts by insurers may well reflect firm survival motives at work (§I.D.3). The stricter solvency requirements imposed on insurers compound this, effectively amplifying the survival motive at higher levels of liquidity: as capital levels deplete below certain thresholds, regulators increasingly interfere with business operations, seizing control as a last resort.[341]

---

[334] See reference to the "dynamic nature of cyber risks" Eling, *supra* note 332 § 2.5.

[335] Schwarcz and Wolff, *supra* note 6 at 28.

[336] Michel-Kerjan and Pedell, *supra* note 332 at 5.

[337] Paul Hudson & Annegret H. Thieken, *The Presence of Moral Hazard Regarding Flood Insurance and German Private Businesses*, 112 NAT HAZARDS 1295, 1310 (2022) This finding is also corroborated by own conversations with insurance industry insiders. *See also* Schwarcz and Wolff, *supra* note 6 n. 135.

[338] Baker and Siegelman, *supra* note 17 § 3.B.

[339] *Cf*. Schwarcz and Wolff, *supra* note 6 § II.B.5.

[340] *See e.g*. Gudgel, *supra* note 165 § 5.3.2.

[341] Grace, *supra* note 120 at 472.

Consistent with these conclusions (and those of §I.D.1, §I.D.2.b and §§I.D.5-6) are studies involving environmental liability insurance: insurance uptake for high-frequency low-severity environmental pollution appears to sometimes have a net *hazard* effect,[342] or at best a weak net regulatory effect;[343] uptake appears to have a net *regulatory* effect when it comes to catastrophic pollution risks (e.g. oil spills or waste storage tank leaks).[344]

Finally, Mol finds that, under experimental conditions, moral hazard among individuals is comparatively weaker for very low probability risks (compared to higher probability risks).[345] This may be for the same reason individuals underinsure against said risks: the risk fails to cross an attention threshold. This is particularly relevant for catastrophic risks.

#### e. *Predictability and Novelty*

When actuarial data is sparse or unreliable – whether because risks are genuinely new or because historical loss patterns provide poor guidance for future exposures – insurers cannot rely on traditional actuarial approaches to risk assessment. Instead, they must invest in (typically more costly) causal risk modeling that attempts to understand underlying loss mechanisms. This shift toward causal analysis typically occurs for catastrophic risks where the potential for severe losses justifies the additional modeling investment, even when data limitations make such efforts inherently speculative.[346] Such efforts are most prominent in insurance against natural catastrophes,[347] but can also be seen in lines exposed to nuclear power,[348] catastrophic cyber,[349] terror,[350] pandemic risks,[351] and supply chain disruptions.[352]

---

[342] Shiyi Chen et al., *New Evidence of Moral Hazard: Environmental Liability Insurance and Firms' Environmental Performance*, 89 JOURNAL OF RISK AND INSURANCE 581 (2022).

[343] Beibei Shi et al., *The Impact of Insurance on Pollution Emissions: Evidence from China's Environmental Pollution Liability Insurance*, 121 ECONOMIC MODELLING 106229 (2023).

[344] PAUL K. FREEMAN & HOWARD KUNREUTHER, MANAGING ENVIRONMENTAL RISK THROUGH INSURANCE 25 (1997); John Merrifield, *A General Equilibrium Analysis of the Insurance Bonding Approach to Pollution Threats*, 40 ECOLOGICAL ECONOMICS 103 (2002); Haitao Yin, Howard Kunreuther & Matthew W. White, *Risk-Based Pricing and Risk-Reducing Effort: Does the Private Insurance Market Reduce Environmental Accidents?*, 54 THE JOURNAL OF LAW AND ECONOMICS 325 (2011); Haitao Yin, Alex Pfaff & Howard Kunreuther, *Can Environmental Insurance Succeed Where Other Strategies Fail? The Case of Underground Storage Tanks*, 31 RISK ANALYSIS 12 (2011).

[345] Mol, Botzen, and Blasch, *supra* note 307.

[346] *See generally* GROSSI, KUNREUTHER, AND PATEL, *supra* note 285; while these methods were pioneered in Excess and Surplus lines, there is growing acceptance by regulators to allow their use in the admitted market. See e.g. *Commissioner Lara Invites Public Input on Final Phase of New Wildfire Modeling Regulation*, https://www.insurance.ca.gov/0400-news/0100-press-releases/2024/release037-2024.cfm (last visited July 17, 2025); and KARA BAYSINGER ET AL., CALIFORNIA'S PROPOSED REGULATIONS TO ENABLE CATASTROPHE MODELING AND EXPAND WILDFIRE PROTECTION IN HIGH-RISK AREAS (2024), https://www.willkie.com/-/media/files/publications/2024/08/californiasproposedregulationstoenablecatastrophemodelingandexpandwildfireprotectioninhighriskareas.pdf.

The novelty of a risk also influences insurer behavior. All else being equal, initial pricing for such risks tends to be conservative, with high premiums reflecting both genuine uncertainty and ambiguity in what is being covered.[353]

The novelty, frequency and magnitude of risks also interact with market structure in subtle ways. A novel risk is a novel market, offering a growth opportunity for insurers. The preferred strategy for entering a new market is to offer small or limited coverage without making enormous effort in modeling the risk *ex ante*, and instead waiting for loss data to come in. The first coverage will likely be offered by major players through Excess and Surplus lines.[354] If the market is competitive, insurers might underprice their coverage not only to capture more market share but also to gather more loss data faster in the hopes of turning that into a competitive edge.[355]

Such a "wait and see" approach isn't feasible for novel catastrophic risks, typically prompting insurers to simply exit the market. Witness, e.g. the lack of affordable terrorism coverage post 9/11,[356] and the total lack of nuclear power plant coverage in the early days of the industry.[357] Both necessitated government intervention to strong-arm insurers into offering coverage, but once they did, insurers made *ex ante* efforts to model the risks.

---

[347] GROSSI, KUNREUTHER, AND PATEL, *supra* note 285 ch. 3, 4.

[348] *See e.g.* Rudy Mustafa, Insuring Nuclear Risk (July 2017).

[349] *See e.g.* RISK MANAGEMENT SOLUTIONS, INC., *supra* note 287; Lloyd's Futureset, *supra* note 287.

[350] GROSSI, KUNREUTHER, AND PATEL, *supra* note 285 ch. 10.

[351] *See e.g.* Feng and Garrido, *supra* note 288; Nkeki and Iroh, *supra* note 288.

[352] QUANTIFYING BUSINESS INTERRUPTION, *supra* note 289.

[353] Kunreuther, Hogarth, and Meszaros, *supra* note 277; For an example, see e.g. the (still emerging) cyber market Eling, *supra* note 332.

[354] The E&S market is typically the one that first provides coverage very novel or hard to place risks. For more see Patrick L. Brockett, *An Economic Overview of the Market for Excess & Surplus Lines Insurance*, 9 JOURNAL OF INSURANCE REGULATION (1990).

[355] Regarding competition and the underwriting cycle, see Baker and Siegelman, *supra* note 17 § 6.D; Regarding competition in a new market see Xiaoying Xie, Charles Lee & Martin Eling, *Cyber Insurance Offering and Performance: An Analysis of the U.S. Cyber Insurance Market*, 45 GENEVA PAP RISK INSUR ISSUES PRACT 690 (2020).

[356] R. Glenn Hubbard, Bruce Deal & Peter Hess, *The Economic Effects of Federal Participation in Terrorism Risk*, 8 RISK MANAGEMENT AND INSURANCE REVIEW 177 (2005); Michel-Kerjan and Pedell, *supra* note 332 at 6–7; Howard Kunreuther & Erwann Michel-Kerjan, *Looking Beyond TRIA: A Clinical Examination of Potential Terrorism Loss Sharing*, w12069, 4 (2006).

[357] United States Nuclear Regulatory Commission, *supra* note 94 § 1.1.

Cyber insurance is illustrative of both behaviors. Insurers initially wrote limited policies with many exclusions, before slowly expanding coverage. Even today,[358] the vast majority of cyber lines either have small limits or exclude various forms of cyber catastrophes, despite clear demand for increased coverage.[359] Nevertheless, large policies are being written by a small number of major insurers[360] who are investing in sophisticated modeling and loss prevention services specifically for cyber.[361]

In order to assess insurers' ability to accurately price novel (or extremely rare) catastrophic risks, I go over four case studies in Appendix A. The upshot is as follows. Insurers appear to price risks with *known* unknowns fairly well. Insurance lines involving such risks are profitable, and the supply of coverage has been stable or increasing. Coverage for nuclear power and coverage for pandemics in the form of life insurance are good examples here. Aggregate pricing for cyber also appears to be going fine. However, risks that involve *unknown* unknowns are another story. These are the risks that have completely caught insurers off guard, leading to ruinous losses, and sudden contractions in supply or a spike in premiums (or both). Terrorism coverage and coverage for pandemics in the form of *business interruption* insurance are good examples here. That said, it's not clear what social institution is better at pricing in risks with unknown unknowns: neither markets nor governments foresaw 9/11 or the extended lockdowns of COVID-19.

If this is right, then transforming *unknown* unknowns into *known* unknowns is pivotal for effective risk management. The post-9/11 terrorism insurance market and the rapid recalibration of cyber premiums following pandemic-related losses demonstrate how merely elevating a risk category into conscious consideration – even with highly uncertain models – substantially improves insurers' ability to price and manage exposure. To take another example: because of the high salience of Y2K disaster stories in the 90s, insurers took extensive action to exclude such risks and prompt policyholders to remedy their technology.[362]

### 3. Insurer Type

There are two dimensions regarding the type of insurer offering coverage that scholars agree have a large impact on moral hazard mitigation and loss reduction

[358] For reference, despite the first cyber policies dating back several decades, the cyber market is still considered immature. Uncertainty and experimentation with policy language continues. See Eling, *supra* note 332.

[359] *Id.* at 212.

[360] Martin Eling, *Cyber Risk and Cyber Insurance*, *in* HANDBOOK OF INSURANCE: VOLUME I 199, 212 (Georges Dionne ed., 2025), https://doi.org/10.1007/978-3-031-69561-2_7; AM BEST, US CYBER: HOT PRICING COOLS OFF, RAPID GROWTH STALLS (2024), https://web.ambest.com/docs/default-source/events/best's-marketing-segment-report---us-cyber---hot-pricing-cools-off-rapid-growth-stalls.pdf?sfvrsn=cabe2a5f_1.

[361] Though cf. Schwarcz and Wolff, *supra* note 6 § II.B.3.

[362] Tony Pyne, *The Exclusion of Y2K Related Losses from Aviation Insurance Policies - Practicalities, Politics, and Legalities*, 65 J. AIR L. & COM. 769 (1999).

efforts: ownership structure (mutual versus stock) and market focus (niche versus general).

Scholars agree that mutuals – insurance companies owned by their policyholders[363] – tend to generate less moral hazard and demonstrate superior loss prevention performance compared to investor-owned companies (a.k.a. stock insurers). Indeed, the most prominent success stories of insurers acting as quasi-regulators nearly all involve mutuals.[364] The explanations are quite straightforward: more closely aligned incentives between policyholders and the insurer mean reduced agency costs and greater returns on safety R&D for the insurer;[365] peer pressure can serve as a very efficient monitoring and enforcement mechanism;[366] reduced information asymmetries mean more accurate risk-pricing and less room for fraud.[367]

Even the skeptics of regulation by insurance agree that niche insurers can also be quite effective at reducing losses.[368] These specialized insurers tend to develop domain expertise and maintain close industry relationships. Mutuals are naturally specialized, typically focusing on the industry their policyholders are a part of, but stock insurers can also specialize. American Nuclear Insurers (ANI), a stock insurer, demonstrates this well: exclusively offering coverage for the nuclear power industry, they have nuclear engineers as part of their permanent staff, regularly run inspections on power plants, and risk-price their premiums based on an Engineering Rating Factor determined during inspections.[369]

Finally, theory and common sense predict that the forms of insurance which generate the worst moral hazard (and zero regulatory effect) are national compensation funds, such as the Oil Spill Liability Trust Fund, whose primary income source is a fixed per-barrel tax on the oil industry.[370] Such funds aim only to ensure compensation is available to victims, at the cost of unmitigated moral hazard. Such a scheme is only reasonable when paired with stringent regulation.

---

[363] For more on how the governance of mutuals differs from stock insurers, see Narjess Boubakri & Pascale Valéry, *Corporate Governance Issues from the Insurance Industry: Updated Review*, *in* HANDBOOK OF INSURANCE: VOLUME II 439, § 3 (Georges Dionne ed., 2025), https://doi.org/10.1007/978-3-031-69674-9_15.

[364] *See e.g.* Henry Hansmann, *The Organization of Insurance Companies: Mutual versus Stock*, 1 THE JOURNAL OF LAW, ECONOMICS, AND ORGANIZATION 125, § 4 (1985); REES, *supra* note 248; George M. Cohen, *Legal Malpractice Insurance and Loss Prevention: A Comparative Analysis of Economic Institutions*, 4 CONN. INS. L.J. 305, 340 (1997); FREEMAN AND KUNREUTHER, *supra* note 344 at 22–25; KUNREUTHER, PAULY, AND MCMORROW, *supra* note 114 at 229–231; John Rappaport, *How Private Insurers Regulate Public Police*, 130 HARV. L. REV. 1539 (2016); Tom Baker & Rick Swedloff, *Mutually Assured Protection among Large U.S. Law Firms*, 24 CONN. INS. L.J. 1 (2017).

[365] Hansmann, *supra* note 364; Boubakri and Valéry, *supra* note 363 § 3.

[366] *See e.g.* REES, *supra* note 248 ch. 6; and Cohen, *supra* note 364 at 340.

[367] Hansmann, *supra* note 364 at 127; Bruce D. Smith & Michael Stutzer, *A Theory of Mutual Formation and Moral Hazard with Evidence from the History of the Insurance Industry*, 8 THE REVIEW OF FINANCIAL STUDIES 545 (1995).

[368] Abraham and Schwarcz, *supra* note 3 § I.B.3.

[369] Gudgel, *supra* note 256 § VII; *See also, generally* REES, *supra* note 248.

#### 4. Market Structure

The structure of the insurance market significantly shapes insurers' efforts to reduce losses. It should be no surprise that the same market distortions that affect other firms (§I.D.2) can sometimes affect insurers' efforts to mitigate losses. For example, as Abraham and Schwarcz point out, in competitive insurance markets, spillover effects can dampen loss prevention efforts. After an insurer invests in effective risk-mitigation strategies for its policyholders, competing insurers can often appropriate those benefits by offering lower premiums to those same policyholders.[371] That's assuming the policyholder doesn't simply forego future coverage, having reduced their underlying risk. (A mandate eliminates this last possibility, helping alleviate insurers' fears of being "too successful" at reducing risk). Furthermore, competing insurers capture fewer of the positive externalities from their loss reduction efforts, passing them on to competitors instead (e.g. fire prevention investments which also protect neighboring property insured by a competitor). Empirical studies are consistent with theory here (showing e.g. that monopolist insurers invest more in loss reduction measures).[372] Again, the extensive loss prevention and risk-pricing efforts of ANI (which has a monopoly on certain mandated coverage) are also consistent with this explanation.[373] Similarly, scholars argue that the market concentration of kidnap insurance has been critical for enabling effective mitigation of third-party moral hazard.[374]

One major caveat is in order. Theoretically, a monopolistic *liability* insurer might make *fewer* investments in loss prevention if the liability regime is *negligence*-based. As noted above, liability insurers can always minimize losses in one of two ways: helping insureds skirt liability with legal maneuvering or genuine harm reduction (increased care). Ideally, Judge Learned Hand's formula is the standard of care negligence invokes: the marginal cost of caution should equal the marginal reduction in harm.[375] However, in practice, it's widely understood to boil down to comparing behavior against industry best practices.[376] Thus, under a negligence standard, liability can be kept tamped down by keeping industry standards low. Therefore, a monopolistic insurer will have much less interest in e.g. safety R&D that could *raise* that standard.[377] Under negligence then, insurers are incentivized to ensure all policyholders *meet* industry standards: insurers will help raise the floor, but not necessarily the bar. When the efficient care level is actually much higher than said standards, one is potentially leaving significant welfare on the table by not adopting strict liability, where damages are always internalized. Once again then, a strict liability regime synergizes well with regulation by insurance. That said, in practice, what insurers will actually do likely depends mostly on *aggregate demand* (under a mandate, insurers are happy to let risk go to near zero), *aggregate risk levels* (if very high, they still might be happy to raise industry standards), *risk volatility* (where improved standards can reduce this, insurers will be amenable) and *the predictability of changes in risk levels* (gradually increasing standards could lead to aggregate risk falling at a predictable, i.e. profitable rate for insurers).[378]

[370] *The Oil Spill Liability Trust Fund (OSLTF)*, https://www.uscg.mil/Mariners/National-Pollution-Funds-Center/About_NPFC/OSLTF/ (last visited June 29, 2025).

[371] Abraham and Schwarcz, *supra* note 3 § III.B.

[372] Gebhard Kirchgässner, *On the Efficiency of a Public Insurance Monopoly: The Case of Housing Insurance in Switzerland*, *in* PUBLIC ECONOMICS AND PUBLIC CHOICE: CONTRIBUTIONS IN HONOR OF CHARLES B. BLANKART 221 (Pio Baake & Rainald Borck eds., 2007), https://doi.org/10.1007/978-3-540-72782-8_13; *See also* Schwarcz and Wolff, *supra* note 6 n. 135.

[373] Gudgel, *supra* note 165 § 5.3.2.

[374] Shortland, *supra* note 329.

[375] SPULBER, *supra* note 75 at 405–406; SHAVELL, *supra* note 12 ch. 2 fn. 23.

The insurance business cycle further complicates the relationship between competition and loss prevention. During "soft markets," when insurers compete aggressively for market share, the pressure to underprice coverage might reduce incentives for rigorous loss prevention. Conversely, "hard markets" characterized by restricted capacity and higher prices may provide insurers with both the resources and leverage to implement more robust loss prevention requirements.[379] A more concentrated market should see less variance in insurer behavior over these cycles, as evidenced by insurance for commercial nuclear power.[380]

## IV. IMPLICATIONS FOR FRONTIER AI

Using the framework from Part I, Part II assessed the frontier AI industry, finding great potential for a net regulatory effect; here in Part IV, I lean on the framework from Part III to evaluate what policy interventions could help ensure that potential gets realized. This also serves as "proof in the pudding" for the frameworks developed in Part I and III: the depth of analysis they provide enables much greater nuance in policy recommendations.

---

[376] SPULBER, *supra* note 75 at 406; SHAVELL, *supra* note 12 ch. 2 p. 10; RAMAKRISHNAN, SMITH, AND DOWNEY, *supra* note 147 ch. 3.

[377] In fact, where market forces would otherwise push policyholders to *raise* the industry standard, a monopolistic insurer might *oppose* such efforts, since it would only increase the liabilities of their other policyholders.

[378] Abraham and Schwarcz, *supra* note 3 § III.B.

[379] Baker and Siegelman, *supra* note 17 § 6.D; Scott E. Harrington & Patricia M. Danzon, *Price Cutting in Liability Insurance Markets*, 67 THE JOURNAL OF BUSINESS 511 (1994).

[380] United States Nuclear Regulatory Commission, *supra* note 94 at xx.

I sketch two contrasting interventions: a poorly designed insurance mandate, and a better one. I choose mandates because they are the simplest mechanism for overcoming the demand side failures identified in Part I, stemming most notably from a coordination failure between competitors (§§II.B.2-4) and misaligned incentives (e.g. §II.B.1).

However, supply side failures are also an obstacle: by default, meaningful coverage is not and will not be available for the catastrophic risks from this novel, fast-moving technology, with virtually no relevant case law or loss data. That is exactly the sort of risk insurers hate to underwrite. Indeed, many would call it "uninsurable,"[381] though it should be noted that "insurability" is highly subjective, depending heavily on insurer appetite.[382]

If insurers don't simply refuse they will do some mix of demanding truly exorbitant premia (>10% of the policy limit),[383] and placing many exclusions on it so as to water it down significantly.[384] However, as I hope to illustrate in the following scenarios, these supply side issues provide both challenges and opportunities for using mandate design to shape the market.

In all scenarios, with demand propped up by a mandate, insurers should be encouraged to invest more in loss prevention services (§III.B.4), and offer more coverage due to the increased pool of insureds. Furthermore, regulation by insurance will always be much more effective with clearer and stricter assignment of liability (§§III.B.2.a and III.B.4). Every other choice in the design of the mandate can greatly change the outcome: these are what I wish to highlight.

## *A. The Worse Intervention: A Broad, Shallow and Permissive Mandate*

In this intervention, the mandate is *broad* in that many developers and downstream deployers have to carry liability insurance for a wide variety of harms, including many small and frequent product accidents. If it includes catastrophic risks, it only does so incidentally (not making clear that these are risks insurers must explicitly cover and price for). The mandate is *shallow* in that the coverage required is not too great (e.g. policy limits of ~$100 million), and allows insurers to write in

---

[381] A risk is considered more insurable the more of the following criteria are met: there are many independent and identically distributed exposure units; losses are unintentional and accidental; losses are definite as to time, place, amount, and causes; the chance of loss is calculable; premiums are affordable. For more see Robert W Klein, *A Regulator's Introduction to the Insurance Industry*, NAIC, 9 (1999), https://content.naic.org/sites/default/files/inline-files/prod_serv_marketreg_rii_zb.pdf.

[382] Baruch Berliner, *Large Risks and Limits of Insurability*, 10 THE GENEVA PAPERS ON RISK AND INSURANCE 313 (1985).

[383] For reference, typical cyber premia range from 0.05% to 0.5% of policy limits (occasionally another order of magnitude higher for very complex large exposures), and while typical Tech E&O premia range from 0.3% to 3%.

[384] Schwarcz and Wolff, *supra* note 6 at 50.

many exclusions. Finally, the mandate is *permissive* in terms of which insurers qualify for satisfying the mandate. What will happen?

What is *unlikely* is a supply side failure. Several major Excess and Surplus (E&S) insurers[385] would likely compete to offer coverage. The ambiguity surrounding such novel risks will push premia up. However, competing insurers looking to capture market share in a new market and collect valuable loss data will push premia down. If insurers aren't required to meet any financial strength rating score to satisfy the mandate (e.g. an AM Best rating of A or above),[386] a race to the bottom could occur, leading to further underpricing.[387] Competing insurers will likely only make some *ex ante* effort to price the risk accurately, mostly relying on a "wait and see" strategy (§III.B.2.e).

The fact that the mandate is not targeted will lead to coverage being spread over many different policyholders and increase transaction costs, translating into weaker loss reduction efforts from insurers (see §III.B.1 and §III.B.2.d respectively).

Such a mandate would prompt a major reshuffling of existing coverage. Greatly accelerating a process which appears already underway (§II.A), many insurers would likely move to explicitly exclude AI risks, and then offer coverage as an add-on or standalone product. This should create greater predictability and clarity for businesses.

However, the largest AI firms (e.g. Google DeepMind or Microsoft) will simply self-insure through their pure captives,[388] subsidiary insurance companies that only insure their parent company.[389] To permit the use of pure captives (or simply hold a minimum amount of financial assets) would be to sacrifice an *enormous* amount of potential from regulation by insurance here. Such captives would do little or nothing to correct for the distortions of psychological biases (§II.B.6), the existential race for market share (§II.B.2), and the underprovisionment of public goods such as safety research or industry reputation (§§II.B.3-4).[390]

---

[385] Brockett, *supra* note 354.

[386] David L. Eckles & Martin Halek, *Insurance Company Financial Strength Ratings*, *in* HANDBOOK OF INSURANCE: VOLUME II 427 (Georges Dionne ed., 2025), https://doi.org/10.1007/978-3-031-69674-9_14.

[387] Though I don't expect this effect to be great: low rating insurers might even self-select *out* of such a novel market, as they attempt to rehabilitate their rating.

[388] Steve Evans, *Google Parent Alphabet Turns to Cat Bonds for Earthquake Insurance - Artemis.Bm*, ARTEMIS (Nov. 9, 2020), https://www.artemis.bm/news/google-parent-alphabet-turns-to-cat-bonds-for-earthquake-insurance/; *Washington Doubles down on "unlawful" Captives Following Microsoft Settlement*, CAPTIVE INTERNATIONAL (Dec. 12, 2018), https://www.captiveinternational.com/services/accounting-and-tax/washington-doubles-down-on-unlawful-captives-following-microsoft-settlement-2626.

[389] It's very common for major corporations to have a captive (often to fulfill a mandate, such workers compensation). Coverage from captives offer several advantages for their parent companies over market coverage. These include: cost savings from reduced overhead (rate and form filing requirements are reduced for captives), cost savings from avoiding markups, tailored coverage, reduced information sharing requirements, and direct access to reinsurance markets. As subsidiaries, captives are strongly aligned with their parent companies. For more see *Insurance Topics | Captive Insurance Companies | NAIC*, https://content.naic.org/insurance-topics/captive (last visited July 1, 2025).

If it is permitted to exclude catastrophic risks, such as harms involving critical infrastructure failure, or Chemical Biological Radiological and Nuclear (CBRN) harms, then it's all but guaranteed they will be excluded, as noted before. The exclusion of catastrophic risks would be even more lost potential, since distortions to care levels are most pronounced for precisely these risks (see generally §II.B).

Making catastrophic risks *not* excludable would certainly be an improvement. However, under such a mandate they would be included in a broad portfolio of risks under a policy with a relatively small limit. This will prompt insurers to dedicate *some* resources to pricing these risks, but it's unlikely they will form a dedicated institution whose *sole purpose* is to price and mitigate these risks.[391]

In sum, this first intervention would do little if anything to counterfactually increase total welfare, while leaving much on the table.

## *B. The Better Intervention: A Narrow, Deep and Restrictive Mandate*

In this intervention, the mandate is *narrow* in that only a few foundation model developers – those at the very frontier[392] – have to carry liability insurance for a very narrow set of harms, namely, specific catastrophic risks. The mandate is *deep* in that the coverage required is large (e.g. policy limits of ~$2 billion or greater), and cannot exclude various threat modalities insurers typically exclude (e.g. critical infrastructure failure, CBRN,[393] or systemic risks from widespread AI agent deployment[394]). Finally, the mandate is *restrictive* in terms of which insurers qualify for satisfying the mandate: no pure captives and only insurers with e.g. an AM Best rating of A or above. What will happen?

For starters, we can be sure such a mandate would have a much larger counterfactual impact, since existing policies don't come anywhere close to fulfilling such a mandate, both in their limits and their exclusions (§II.A).

[390] Moreover, the lack of a rating requirement for the insurer strikes here again, possibly much harder. Because captives don't serve the wider market, they needn't care about their rating nearly as much, limiting the market disciplining ratings can have. If its parent company wishes it to take on much more risk than the market thinks wise, it will likely acquiesce and shrug at the rating downgrade it receives for doing so. See generally Eckles and Halek, *supra* note 118; Regulatory solvency requirements will, of course, still apply, but these raise the floor not the ceiling. See generally Grace, *supra* note 120 Thus, allowing for captives greatly undercuts the core policy objectives of regulation by insurance.

[391] As one insider put it, at best insurers might form a dedicated working group. At worst, it will fall as an additional task on someone's already crowded desk.

[392] Possibly proportional to their activity levels.

[393] Bengio et al., *supra* note 130 § 2.1.4; FRONTIER MODEL FORUM, *supra* note 160.

[394] Hammond et al., *supra* note 139 § 3.4 and p. 44.

Based on conversations with insurers, if the mandated coverage wasn't more than a few 100 million or so, we might still see a few major E&S insurers offer coverage, albeit at very steep prices (~6-10% of the policy limit). Much past $500 million though, and we are likely to see a supply side market failure.

Again, if pure captives were permitted, the largest AI firms would simply self-insure. That would leave the rest of the industry out in the cold, severely undermining competition and chilling innovation. By barring captives, we not only protect competition and innovation, but unlock the potential for regulation by insurance to correct many of the distortions identified in §II.B. Assuming the supply side failures can be fixed, that is. Completely halting AI development for lack of insurance availability is a pyrrhic victory in the quest to deliver safe and secure AI.

Two very sensible solutions present themselves, one a carrot, the other a stick. Policymakers should wield both. Option one, the carrot, is to encourage the frontier AI industry to form a mutual[395] (much as nuclear power operators eventually did with Nuclear Electric Insurance Limited). Option two, the stick, is to threaten to form a Joint Underwriting Association (JUA)[396] and force the insurance industry to offer coverage (much as American Nuclear Insurers was formed).

Forming a mutual is the carrot for two reasons. First, because it would lead to substantially lower premia: mutuals are non-profits that return surpluses to policyholders.[397] Second, because it should result in greater investments in cost-effective loss reduction measures the industry can collectively profit from (see §III.B.3). These investments are precisely the increases in care levels and safety R&D we want to see. Forming a JUA is the stick because it would lead to substantially higher premia, and likely less investment in loss reduction efforts (but second only to those of a mutual).

By wielding both carrot and stick, policymakers can set up favorable conditions for the AI frontier industry to cooperate on safety and security. A mutual is likely the best outcome, both for welfare more broadly and the industry in particular. However, if the industry fails to coordinate, a JUA is still an excellent second best.

Either way, because of the mandate design – barring pure captives and requiring substantial coverage – the resulting insurer, whether a JUA or mutual, would almost certainly have a monopoly on this niche market. This is also expected to help reduce spillovers in safety R&D, concentrate private regulatory power, and thus encourage greater loss prevention (§III.B.4).

---

[395] *Cf.* Friedman, *supra* note 7 § IV.B.2.d.

[396] These are special risk-pooling associations, in which multiple insurers collectively underwrite niche, high-risk, or generally low profitability risk. Whether voluntary or by force, these special entities can only be formed with the blessing of insurance regulators, and are typically only so created to ensure coverage availability where the standard market has failed.

[397] Hansmann, *supra* note 364 § 4.

Whether a JUA or a mutual, this insurer will be an institution dedicated to pricing and mitigating the particular risks policymakers name, better focusing attention and expertise (§III.B.3). If a mutual, there's a better chance it will benefit from expertise currently concentrated among AI developers themselves: engineers from frontier AI firms could occasionally be "loaned" to the mutual, similar to programs in the nuclear industry.[398] In any case, such a dedicated institution would also likely be much better suited to integrate with the government's emergency response plans than a generalist insurance company, should the government wish to so integrate it (§II.B.5).[399]

Because the mandate explicitly targets catastrophic risks and will force the insurer to get significant skin in the game, the "wait and see" strategy won't be cost-effective, prompting the insurer to invest significantly more in *ex ante* causal risk-modeling and risk mitigation efforts (§III.B.2.e). Compounding this is the concentration of coverage onto a few exposure units (a few AI developers), and reduced transaction costs (§III.B.2.d).

Premia will be undoubtedly high, though nothing the major frontier AI developers can't afford.[400] Channeling liability (e.g. by making it exclusive, as was done for nuclear) could also help concentrate insurance capacity and thus reduce premia.[401] Obviously, setting a limit on liability would also reduce premia, at the potential cost of moral hazard. If the limit is high enough though, this hazard should be quite limited, since safety efforts on the part of insurers and their policyholders are likely inelastic beyond a certain damage level.

By requiring qualifying insurers meet a high financial strength rating (as e.g. ANI does),[402] policymakers can ensure the resulting insurer takes its solvency requirements seriously and is resistant to any pressure from the AI industry to underprice the risk. Finally, the mandate could require insurers (perhaps in partnership with government) to engage in systematic horizon-scanning and threat identification, much like provisions in the Terrorism Risk Insurance Act (TRIA).[403] This promises to be a particularly high leverage intervention, prompting insurers to transform unknown unknowns into known unknowns, critical for managing novel catastrophic risk (§III.B.2.e).

---

[398] REES, *supra* note 248 at 56–60.

[399] Many policymakers will understandably be interested in the swift restitution of victims. This is of particular concern in the event of a disaster, when critical aid needs to be distributed rapidly. This job falls on the government by default, as society's *de facto* insurer of last resort (§I.D.4), but can be partially outsourced (§II.B.5).

[400] If the mandate were to also fall on smaller developers, required coverage could potentially scale with activity levels to make it more affordable.

[401] United States Nuclear Regulatory Commission, *supra* note 94 § 2.3.8.

[402] *Membership – American Nuclear Insurers*, https://www.amnucins.com/membership/ (last visited July 19, 2025).

[403] Carpenter et al., *supra* note 280 § 1.

In sum, this second intervention should do much more for welfare in expectation, while minimizing downsides.

## C. *Objections and Rebuttals*

In my remaining space, I address a few well-warranted concerns about the regulation by insurance strategy for frontier AI.

### 1. Pricing the Risk of Frontier AI Systems

Drawing on the track record of cyber insurers, Schwarcz and Wolff describe just how challenging it will be for insurers to price company-specific risk in the frontier AI industry.[404] The risks from AI are unquestionably novel and dynamic, posing serious challenges for insurers. Note though that a government agency will also struggle: after all, this is the crux of the AI policy challenge. Part of the problem is that virtually all relevant expertise is concentrated in industry (as opposed to insurers or government). Above (§IV.B), I argued that mutuals might help rectify this by leaning on in-house expertise from the leading AI firms, but the challenge is undeniable.

When insurers can't price the company-specific *technological* risk, insurers lose a critical tool for handling moral hazard and encouraging better risk-management practices. However, premium adjustments are not insurers' only tools: often, for example, coverage is simply *conditioned* on adopting better practices (e.g. cyber insurers requiring Multi-Factor Authentication).[405] One could imagine insurers requiring AI vendors adopt e.g. certain prompt injection protections before offering coverage.[406]

Furthermore, Schwarcz and Wolff recognize that policies will still likely be accurately risk-priced, if only in a crude fashion: premiums will depend heavily on e.g. the scale of the company's operations, the sector its products are being deployed in, and other such readily verifiable characteristics.[407] These proxies for risk, however coarse-grained, are effective at keeping loss ratios in check. In other words, risk-pricing here may struggle to induce efficient *care* levels, but it can at least induce efficient *activity* levels, transmitting liability's *deterrence* effect if not its effect on *care*. Given many of the distortions identified in §II.B revolve around competition driving not only care levels too low, but plausibly *activity* levels too high,[408] there is good reason to believe insurance uptake here could still substantially improve welfare by simply drawing forward crude approximations of future catastrophic costs that frontier AI firms are currently incentivized (or psychologically disposed) to downplay.

---

[404] Schwarcz and Wolff, *supra* note 6 § III.A.2.

[405] *Cyber Insurance in the Fight Against Ransomware | AJG United States*, GALLAGHER, https://www.ajg.com/news-and-insights/cyber-insurance-fight-against-ransomware/ (last visited Oct. 7, 2025).

[406] Such as those described in Edoardo Debenedetti et al., Defeating Prompt Injections by Design (June 24, 2025), http://arxiv.org/abs/2503.18813.

[407] Schwarcz and Wolff, *supra* note 6 § II.B.2.

[408] E.g. the existential race for market share is likely leading competitors to deploy products faster and more widely than socially optimal.

### 2. The Insurance Lobby

One might worry that with a mandate in place, insurers will work very hard to limit any further expansion of liability for their policyholders by winning favorable precedents in case law and lobbying the government. That is, after all, one way of limiting risk for themselves and their policyholders.

I highlight three observations in response. First, as Peck points out, "[w]ithin bar associations, lawyers representing insurance companies are balanced against those representing injured plaintiffs."[409] Second, Parsons points out that in many jurisdictions (the U.S. included), courts take into account a defendant's insurance coverage when determining damages. This "deep pockets" effect tends to increase damages awarded.[410] Third and finally, an empirical study of the major tort reforms of the 70s and 80s finds surprisingly little evidence of the insurance lobby having a large influence.[411]

### 3. Crowding Out Ex Ante Regulation

Finally, as with any market-based solution to the AI regulation puzzle, there is a danger that effective private regulation will crowd out public regulation, stifling the growth of critical state capacity in matters of AI.[412] A complete analysis of the topic requires its own paper; here I content myself with a few comments.

While crowding out public institutions is a very warranted concern, sometimes private regulation is a complement to public regulation. For example: the perennial threat of government intervention can sustain more effective private regulation;[413] explicit public private partnerships can better aggregate incident data, identify best practices,[414] and encourage private actors to shoulder more risk (§I.D.4). Furthermore, there are cases of insurers setting a standard or developing a safety technology, and then lobbying the government to mandate said standard or technology more widely (as in the case of the airbag).[415] In other words, sometimes insurers act as effective laboratories for future regulation.

---

[409] Cornelius J. Peck, *The Role of the Courts and Legislatures in the Reform of Tort Law*, 48 MINN. L. REV. 265, 283 (1963).

[410] Parsons, *supra* note 128 § 6; *Cf.* KENNETH S. ABRAHAM, THE LIABILITY CENTURY: INSURANCE AND TORT LAW FROM THE PROGRESSIVE ERA TO 9/11 4 (2008).

[411] Yiling Deng & George Zanjani, *What Drives Tort Reform Legislation? An Analysis of State Decisions to Restrict Liability Torts*, 85 JOURNAL OF RISK AND INSURANCE 959 (2016).

[412] For an overview of the literature on such concerns, see Russell W Mills, *The Interaction of Private and Public Regulatory Governance: The Case of Association-Led Voluntary Aviation Safety Programs*, 35 POLICY AND SOCIETY 43, § 2 (2016); For articulations of this concern in the frontier AI space, see e.g. Anton Leicht, *Don't Outsource AI Governance Just Yet*, THE DISPATCH (July 2, 2025), https://thedispatch.com/article/artificial-intelligence-private-governance-california/.

[413] *See e.g*. Gunningham and Rees, *supra* note 250 at 391.

## V. Conclusion and Further Research

After it was noted insurers sometimes act as private regulators, a scholarship studying the phenomenon developed. "Regulation by insurance" is now a debated public policy strategy. However, as of yet no first principles account has been given regarding when and why such a strategy might be appropriate. This Article attempts to do so, advancing a two-step framework. First, it evaluates how many and how severe certain market distortions are for the economic activity in question. This tells us the *potential* for a *net regulatory effect*. Second, it determines if the conditions are conducive for insurers to *realize* that potential.

It's found that potential for a net regulatory effect tends to lie in the management of catastrophic non-product accidents, where incentives are most distorted and psychological biases are strongest. Interventions (e.g. mandates) are often required to realize this potential though. This should be no surprise: there is a long history of market failure and government intervention for managing catastrophic risks.[416] Regulation by insurance is also no panacea: it is largely circumscribed by tort law and thus always struggles with diffuse harms,[417] long-tail risks, and harms without identifiable causation chains or clear assignment of liability.

The strength of the framework developed lies in its nuance. As a demonstration, I apply it to the topical frontier AI industry. I find there is great potential for a net regulatory effect, but which won't be realized without a carefully engineered intervention.[418] Guided by the framework, one such intervention is outlined: mandate foundation model developers carry deep coverage for specific catastrophic risks, and bar pure captives from fulfilling the mandate. By catalyzing the formation of specialized risk management institutions, either in the form of an industry mutual or joint underwriting association, such a mandate would centralize safety research efforts, establish industry coordination mechanisms, and bring mature risk assessment expertise to bear on novel dangers (not to mention increase financial resilience against major accidents).

[414] *See e.g.* the Financial Services Information Sharing and Analysis Center (FS-ISAC). For more see Financial Services Sector Risk Management Plan (2025), https://home.treasury.gov/system/files/216/Financial-Services-Sector-Risk-Management-Plan.pdf.

[415] Kneuper and Yandle, *supra* note 305.

[416] Véronique Bruggeman, Michael G. Faure & Tobias Heldt, *Insurance Against Catastrophe: Government Stimulation of Insurance Markets for Catastrophic Events*, SSRN Journal (2013), http://www.ssrn.com/abstract=2213772.

[417] The tort system simply isn't sensitive to many of such harms, it's often unclear who the injurers are, and often victims are only liminally aware of the harm inflicted on them. For example: who should be accused of what tort for the slow death and consolidation of news rooms across America due to the complex interaction between (among other things) changing business models, online advertising dynamics, and the idiosyncrasies of social media algorithms? The tort system is simply ill-suited for handling such harms.

Further research should thoroughly compare and synthesize policy strategies for managing frontier AI risks. Studying how governments and industry can help insurers in assessing highly uncertain and severe catastrophic risks is also of interest. Information sharing collaborations,[419] and particularly programs that aim at simply *identifying* novel threats, appear promising.[420]

The design of government backstops for catastrophic risk coverage will also require great care. Properly designed, such programs promise to internalize uninsurable externalities akin to a Pigouvian tax and raise minimum safety standards; poorly designed, they create severe moral hazard. Mechanism design innovations such as prediction markets might help the government risk-price premiums, but considerable work remains to make these viable.[421]

Finally, the model of insurers' regulatory effect could use further development. In particular, it could be extended to incorporate bilateral accident models and would benefit from greater empirical verification.

---

[418] As is the case in many other insurance markets. E.g. cf. Logue and Shniderman, *supra* note 330.

[419] This is in line with current efforts from the U.S. government *Trump AI Plan Calls for Cybersecurity Assessments, Threat Info-Sharing | Cybersecurity Dive*, https://www.cybersecuritydive.com/news/white-house-artificial-intelligence-action-plan-cybersecurity-trump/753856/ (last visited Sept. 21, 2025).

[420] Carpenter et al., *supra* note 280.

[421] Trout, *supra* note 97.

## APPENDIX A: CASE STUDIES OF INSURERS PRICING CATASTROPHIC RISKS

This appendix examines insurers' historical track record with pricing novel catastrophic risks across four domains: cyber, pandemics, nuclear meltdowns, and terrorism. The evidence reveals a critical pattern: insurers demonstrate competence in pricing *known unknowns* – risks they recognize but face uncertainty quantifying – while failing ruinously with *unknown unknowns* that fall outside their consideration set entirely.

For want of space and more granular data, I focus on *aggregate losses* (total claims paid across a line over a given period) and *loss ratios* (aggregate losses divided by premiums earned, expressed as percentages). Loss ratios above 90% indicate underpricing or unexpectedly high claims; ratios below 40% suggest overpricing that may harm competitiveness. Target loss ratios typically range between 40% and 60%, varying by business mix and market conditions.

### *Cyber*

The cyber insurance market is still considered young, showing continued growth and experimentation in policy language.[422]

Insiders disagree over whether we've seen the worst of catastrophic cyber and whether these have been effectively priced in[423] *Prima facie*, estimates of catastrophic cyber costs vary substantially across studies (Table 2). However, they may not vary by more than an order of magnitude after accounting for definitional differences: insured losses versus total economic losses; aggregate cyber losses over a year versus losses accumulated over a five-year period (including from slowed economic growth); the Probable Maximum Loss (PML)[424] based on historical data versus handcrafted scenarios.

---

[422] Xie, Lee, and Eling, *supra* note 355 at 716, 719.

[423] Kevin Williams, *CrowdStrike Losses May Be Biggest Test yet of Cybersecurity Insurance Risk Warning from Warren Buffett*, CNBC, July 2024, https://www.cnbc.com/2024/07/24/crowdstrike-biggest-test-yet-for-cyber-insurance-buffett-warned-about.html; Tom Johansmeyer, *Surprising Stats: The Worst Economic Losses from Cyber Catastrophes*, THE LOOP (Mar. 12, 2024), https://theloop.ecpr.eu/surprising-stats-the-worst-economic-losses-from-cyber-catastrophes/; *Cf.* AM BEST, *supra* note 360 at 2.

[424] A Probable Maximum Loss (PML) is an estimate of the largest loss an insurer could reasonably expect from a single catastrophic event over a given time horizon. E.g. the maximum loss that has a 1% chance of being exceeded over a 1 year time horizon, i.e. a loss insurers expect to occur once in 100 years. Another common benchmark: 1 in 250 year losses (99.6th percentile catastrophe).

| Year of Estimate | Type of Estimate | Geography/Sector | Estimate (billions) |
|---|---|---|---|
| 2014 | Insured losses<br>Undefined PML | Global | £20 [425] |
| 2015 | Economic losses over 5yrs<br>Select scenarios | U.S. | $1024 [426] |
| 2018 | Economic losses<br>1-in-100 year PML | Global<br>(Financial Sector) | $270-350 [427] |
| 2019 | Insured losses<br>Select scenarios | U.S. | $23.8 [428] |
| 2022 | Insured losses<br>Select scenarios | U.K. | £29 [429] |
| 2023 | Economic losses over 5yrs<br>Select scenarios | U.S. | $1100 [430] |
| 2023 | Insured losses<br>1-in-250 year PML | U.S. | $30 [431] |

*Table 2. Estimates of catastrophic cyber costs. Currencies are not standardized and figures are not inflation-adjusted.*

---

[425] UK CYBER SECURITY - THE ROLE OF INSURANCE IN MANAGING AND MITIGATING THE RISK (2015), https://assets.publishing.service.gov.uk/media/5a80f0c5ed915d74e62314f7/UK_Cyber_Security_Report_Final.pdf.

[426] BUSINESS BLACKOUT: THE INSURANCE IMPLICATIONS OF A CYBER ATTACK ON THE US POWER GRID (2015), https://www.jbs.cam.ac.uk/centres/risk/publications/technology-and-space/lloyds-business-blackout-scenario/.

[427] ANTOINE BOUVERET, CYBER RISK FOR THE FINANCIAL SECTOR: A FRAMEWORK FOR QUANTITATIVE ASSESSMENT (2018), https://www.imf.org/en/Blogs/Articles/2018/06/22/blog-estimating-cyber-risk-for-the-financial-sector.

[428] LOOKING BEYOND THE CLOUDS: A CYBER INSURANCE INDUSTRY CATASTROPHE LOSS STUDY (2019), https://www.guycarp.com/insights/2019/10/looking-beyond-the-clouds-a-cyber-insurance-industry-catastrophe-loss-study.html.

[429] FISCAL RISKS AND SUSTAINABILITY – JULY 2022 (2022), https://obr.uk/frs/fiscal-risks-and-sustainability-july-2022/.

[430] *Lloyd's Systemic Risk Scenario Reveals Global Economy Exposed to $3.5trn from Major Cyber Attack*, (2023), https://www.lloyds.com/insights/media-centre/press-releases/lloyds-systemic-risk-scenario-reveals-global-economy-exposed-to-3.5trn-from-major-cyber-attack.

[431] COALITION INC, ACTIVE CYBER RISK MODELING (2023), https://web.coalitioninc.com/download-active-cyber-risk-model-2023-03.html.

Turning away from predictions, the historical data suggests insurers are pricing catastrophic cyber risk fairly well. The 2017 *WannaCry* and *NotPetya* attacks (representing $60 million and $3 billion in insured losses, respectively)[432] did not destabilize loss ratios. The industry maintained a 32.4% loss ratio that year, down from prior years.[433] Accordingly, premiums remained stable through 2017-2019 (Figure 4). Similarly, early reports indicate the 2024 *CrowdStrike* supply chain failure (representing $400 million to $1.5 billion in insured losses)[434] has not triggered substantial premium increases.[435] This suggests that, overall, insurers successfully priced in major incidents like *NotPetya*, *WannaCry* and *Crowdstrike*.[436]

Analysis of Excess and Surplus (E&S) insurers – those writing the largest, most complex cyber policies – reinforces this conclusion. During 2015-2017, only the 95th percentile of insurers experienced loss ratios above 100%.[437] Even the worst hit insurers rarely exceeded loss ratios of ~130% and typically recovered within one or two years.[438] This means even the insurers with the most complex cyber exposures are successfully pricing the risk, all things considered: we aren't seeing loss ratios orders of magnitude off their targets.

---

[432] TOM JOHANSMEYER, COULD NOTPETYA'S TAIL BE GROWING? (2019), https://www.verisk.com/4a25ed/siteassets/media/pcs/pcs-cyber-catastrophe-notpetyas-tail.pdf; Luke Gallin, *Merck Reaches Settlement with Insurers over $1.4bn NotPetya Cyber Attack - Reinsurance News*, REINSURANCENE.WS (Jan. 10, 2024), https://www.reinsurancene.ws/merck-reaches-settlement-with-insurers-over-1-4bn-notpetya-cyber-attack/.

[433] NATIONAL ASSOCIATION OF INSURANCE COMMISSIONERS, REPORT ON THE CYBER INSURANCE MARKET 3 (2022), https://content.naic.org/sites/default/files/cmte-c-cyber-supplement-report-2022-for-data-year-2021.pdf.

[434] *Insured Losses from CrowdStrike Outage Could Reach $1.5 Bln, CyberCube Says*, REUTERS, July 25, 2024, https://www.reuters.com/business/finance/insured-losses-crowdstrike-outage-could-reach-15-bln-cybercube-says-2024-07-25/.

[435] *Cyber Risk Insurance Market Remains Buyer-Friendly - Aon Global 2025 Cyber Risk Report*, HTTPS://CYBER-RISK-REPORT.AONRISINGRESILIENT.COM/, https://cyber-risk-report.aonrisingresilient.com/cyber-risk-insurance-market-remains-buyer-friendly/ (last visited Sept. 14, 2025).

[436] NB: the 2020 SolarWinds attack, while of major significance for the government and national security community, was largely uninsured. See e.g. Samit Shah, *The Financial Impact of SolarWinds Breach*, BITSIGHT (2021), https://www.bitsight.com/blog/the-financial-impact-of-solarwinds-a-cyber-catastrophe-but-insurance-disaster-avoided.

[437] Xiaoying Xie, Charles Lee & Martin Eling, *Cyber Insurance Offering and Performance: An Analysis of the U.S. Cyber Insurance Market*, 45 GENEVA PAP RISK INSUR ISSUES PRACT 690, 710–713 (2020).

[438] NATIONAL ASSOCIATION OF INSURANCE COMMISSIONERS, REPORT ON THE CYBERSECURITY INSURANCE MARKET (2021), https://content.naic.org/sites/default/files/index-cmte-c-Cyber_Supplement_2020_Report.pdf; NATIONAL ASSOCIATION OF INSURANCE COMMISSIONERS, *supra* note 433; NATIONAL ASSOCIATION OF INSURANCE COMMISSIONERS, REPORT ON THE CYBER INSURANCE MARKET (2024), https://content.naic.org/sites/default/files/cmte-h-cyber-wg-2024-cyber-ins-report.pdf.

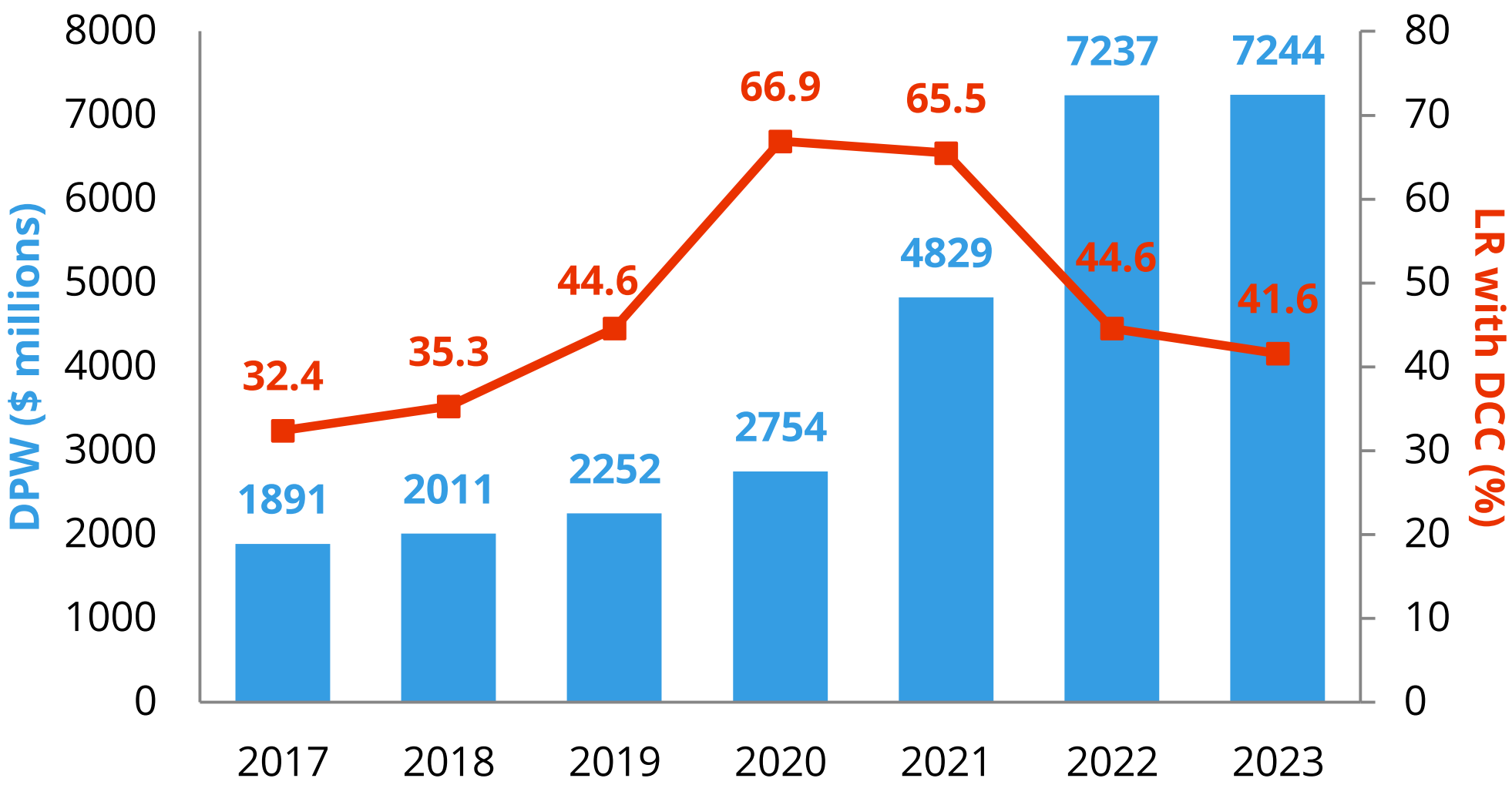

*Figure 4. Direct Premiums Written and Average Loss Ratios (with Defense and Cost Containment included)*[439]

Skeptics of cyber insurers' abilities will point out that loss ratios dramatically increased in 2020 and 2021, prompting insurers to quickly increase premiums to compensate for excessive losses.[440] However, the literature and commentary suggest this was a (widely unforeseen) side effect of the COVID-19 pandemic.[441] That loss ratios and premiums have since returned to pre-pandemic levels (Figure 4) lends further weight to this interpretation. I therefore treat these excess losses as a result of failing to foresee and price in *pandemic* risk, as discussed below.

Overall, cyber insurance has been and continues to be a profitable line of insurance. With the exception of the COVID-19 surge, loss ratios have generally been on target, despite major incidents. This indicates *aggregate* pricing models are working. That says little of the accuracy of *policyholder-specific* risk pricing, of course.[442] It should also be noted that this profitability reflects both accurate pricing and *careful policy terms*, including extensive exclusions and sublimits for catastrophic scenarios.[443]

---

[439] Data collated from NATIONAL ASSOCIATION OF INSURANCE COMMISSIONERS, *supra* note 433 fig. 1, 2; and AM BEST, *supra* note 360 at exhibit 1a.

[440] Schwarcz and Wolff, *supra* note 6 at 28.

[441] Harjinder Singh Lallie et al., *Cyber Security in the Age of COVID-19: A Timeline and Analysis of Cyber-Crime and Cyber-Attacks during the Pandemic*, 105 COMPUT SECUR 102248 (2021); Bernardi Pranggono & Abdullahi Arabo, *COVID-19 Pandemic Cybersecurity Issues*, 4 INTERNET TECHNOLOGY LETTERS e247 (2021); NATIONAL ASSOCIATION OF INSURANCE COMMISSIONERS, *supra* note 438.

[442] Schwarcz and Wolff, *supra* note 6 § II.B.2.

[443] NATIONAL ASSOCIATION OF INSURANCE COMMISSIONERS, *supra* note 438 at 8; Eling, *supra* note 332 at 212.

### *Pandemics*

The recent COVID-19 pandemic is a fascinating case study, revealing starkly different pricing outcomes depending on the line of insurance. Life insurers – who explicitly model pandemics – weathered COVID-19 fairly well.[444] By contrast, Property and Casualty (P&C) as well as Cyber insurers experienced substantial losses, having failed to recognize their exposure to pandemics.

Life insurance is directly exposed to pandemics. As such, life insurers and their reinsurers plan for excess mortality events using historical epidemic data and epidemiological models.[445] In fact, pandemic risk is explicitly factored into solvency capital requirements under regulatory frameworks like Solvency II.[446] While insurers acknowledge uncertainty, pandemics are squarely within their risk models. As such, COVID-19 did not surprise life insurers, and these lines weathered it fairly well.[447]

The same cannot be said for Cyber or P&C lines. As noted above, Cyber insurers saw an unexpected increase in claims. This was driven by a surge in opportunistic cyber attacks, exploiting knock-on effects from the pandemic and the government's response:[448] the pandemic generally increased fear and anxiety, which cyber attackers could exploit; the lock-downs meant citizenry were more dependent on public advisories which could be faked and exploited; the shift to remote work *en masse* created security vulnerabilities IT departments were unprepared for.

P&C insurers were also caught off guard, taking heavy losses on their Business Interruption (BI) lines.[449] This largely stemmed from failing to foresee extended lockdowns, business restrictions, and ensuing supply chain failures. However, it also had to do with unclear policy language, with many insurers believing they had excluded pandemic risk only to be ruled against in court.

In sum, insurers get a mixed report card with pandemic risk. The lines most obviously exposed (e.g. Life) handled it well, while lines that were indirectly exposed (e.g. Cyber and BI) entirely failed to price in the risk. In fairness, the elements of the COVID-19 pandemic that led to these unexpected losses (e.g. extended lockdowns and shift to remote work) were, in fact, unprecedented and unforeseen by nearly all social institutions (markets, businesses, governments, as well as insurers).

---

[444] I leave out health insurance since "health insurance systems vary across different countries" and it is therefore "difficult to give a comprehensive overview of how the health insurance sector was impacted." For more however, see Gunther Kraut, Paulina La Bonté & Andreas Richter, *Pandemic Risk Management and Insurance*, *in* HANDBOOK OF INSURANCE: VOLUME I 251, § 4.3 (Georges Dionne ed., 2025), https://doi.org/10.1007/978-3-031-69561-2_9.

[445] *See e.g.* Feng and Garrido, *supra* note 288; Nkeki and Iroh, *supra* note 288.

[446] Kraut, La Bonté, and Richter, *supra* note 444 at 263.

[447] *Id.* § 4.2.

[448] Pranggono and Arabo, *supra* note 441; Lallie et al., *supra* note 441.

[449] Kraut, La Bonté, and Richter, *supra* note 444 §§ 3, 4.1.

### *Nuclear*

Insurance against commercial nuclear catastrophes demonstrates successful long-term pricing of a low-probability, high-severity risk.

When the U.S. government first started offering licenses for commercial nuclear power plants in 1946, utility companies largely refused to enter the market absent affordable insurance coverage. This lack of coverage was part of the impetus for the Price-Anderson Act of 1957.[450] The Act clarified and channeled liability, mandated carrying insurance, and provided a government backstop for excess losses in the event of a major incident. In exchange, insurers pooled their capacity and formed a joint underwriting association, American Nuclear Insurers (ANI), to offer continuous specialized coverage. This includes coverage for third-party liability arising from meltdowns or major radiological releases.

This line of insurance has certainly been profitable, including similar pools in other markets.[451] ANI's state-sanctioned monopoly structure provides stability and allows participating insurers to spread potential catastrophic losses across future premium streams over extended periods, atypical for most insurance lines.[452] However, like Cyber, policy limits protect insurers from truly catastrophic losses.

That insurers have been steadily retaining more risk, ceding less to their reinsurers, suggests they are becoming more confident in their underwriting,[453] despite the continued dearth of loss data.[454] That said, after adjusting for inflation, ANI has not substantially increased its maximum coverage over 1957 levels.[455]

The evidence suggests that insurers make considerable effort to accurately risk-price each reactor site. Through inspections of insured sites, ANI determines an Engineering Rating Factor (ERF), which can add surcharges to the base premium.[456] Available data shows the ERF does correlate with reports by the Nuclear Regulatory Commission and a safety index from the Institute of Nuclear Power Operations (despite this last typically being private).[457]

All in all, on aggregate *and* at the policy-specific level, insurers appear to successfully price what nuclear risk they bear. Telling of insurers' long-run *ex ante* accuracy is a 1981 study. By analyzing premiums insurers were charging, it found insurers were estimating the frequency of a serious incident (core damage frequency) to be roughly 1-in-400 reactor years.[458] This was two orders of magnitude higher than the latest government-commissioned report at the time (the Rasmussen Report), which gave an estimate of 1-in-20,000 reactor years.[459] By analyzing historical accident data, a retrospective study from 1992 demonstrates that insurers' more conservative numbers had in fact been more accurate.[460]

---

[450] United States Nuclear Regulatory Commission, *supra* note 94 § 1.1.

[451] *Id.* at xx and § 2.3.3.

[452] *Id.* § 2.3.1.

[453] *Id.* § 2.3.3.

[454] *Id.* § 2.3.5.1.

[455] *Id.* § 2.3.3.

[456] Gudgel, *supra* note 256 at 158.

[457] *Id.* at 159–160.

### *Terror*

The unprecedented terrorist attacks of September 11, 2001, represent a catastrophic risk insurers entirely failed to foresee.

Before 9/11, terrorism insurance was typically included as part of standard commercial insurance policies at no additional cost. Curiously, earlier terrorist attacks on the order of hundreds of millions of dollars, both in the U.S. and elsewhere, did not prompt significant reaction from U.S. insurers.[461]

9/11 did. The $32.5 billion (2001 USD) in insured losses from 9/11's approximately $80 billion total damages represented "the costliest event ever in the history of insurance" at the time.[462] Facing depleted capital reserves, insurers and reinsurers rapidly excluded terrorism risk or dramatically increased premiums and reduced coverage even for non-terrorism insurance. In a telling example, Chicago's O'Hare Airport saw its premium-to-coverage ratio increase more than 275-fold.[463]

It's widely understood that insurers (as well as governments and markets) completely failed to foresee or price in catastrophic terrorist attacks: 9/11 came as a shock. Since then, where rates aren't set nationally by regulators (such as in France), a market for modeling terrorism risk has appeared,[464] and there is evidence of at least sectoral risk-pricing.[465] Anecdotally, underwriters express great uncertainty in the models, but also recognize that they are better than nothing.

---

[458] P. L. CHERNICK ET AL., DESIGN, COSTS, AND ACCEPTABILITY OF AN ELECTRIC UTILITY SELF-INSURANCE POOL FOR ASSURING THE ADEQUACY OF FUNDS FOR NUCLEAR POWER PLANT DECOMMISSIONING EXPENSE. TECHNICAL REPORT 80 (1981), https://inis.iaea.org/records/p4v10-nhz04.

[459] NC RASMUSSENN, REACTOR SAFETY STUDY: AN ASSESSMENT OF ACCIDENT RISKS IN US COMMERCIAL NUCLEAR POWER PLANTS, WASH-1400 (NUREG-75/014) 8 (1975), https://www.nrc.gov/docs/ML1533/ML15334A199.pdf.

[460] NUCLEAR REGULATORY COMMISSION, CHANGES IN PROBABILITY OF CORE DAMAGE ACCIDENTS INFERRED ON THE BASIS OF ACTUAL EVENTS (1992), https://www.nrc.gov/docs/ML2011/ML20114B527.pdf; *Cf.* TB Cochran, *Reassessing the Frequency of Partial Core Melt Accidents*, NRDC REPORT, NEWYORK, USA (2011), https://web.archive.org/web/20120508065832/http://www.energypolicyblog.com/2011/04/27/reassessing-the-frequency-of-partial-core-melt-accidents/.

[461] Erwann Michel-Kerjan & Paul A. Raschky, *The Effects of Government Intervention on the Market for Corporate Terrorism Insurance*, 27 EUROPEAN JOURNAL OF POLITICAL ECONOMY S122, § 3.1 (2011).

[462] Erwann Michel-Kerjan & Burkhard Pedell, *Terrorism Risk Coverage in the Post-9/11 Era: A Comparison of New Public–Private Partnerships in France, Germany and the U.S.*, 30 GENEVA PAP RISK INSUR ISSUES PRACT 144, 145 (2005).

[463] Michel-Kerjan and Pedell, *supra* note 332 at 6–7.

### *Conclusions*

This examination of catastrophic risk pricing suggests there is a critical distinction between how insurers handle *known unknowns* and *unknown unknowns*. Known unknowns represent risks that insurers recognize exist but whose probability and severity are uncertain. Examples here include catastrophic cyber accumulation, core damage in nuclear power, and excess mortality from pandemics. For these identified risks, insurers demonstrate competent pricing, maintaining sustainable and profitable loss ratios.

By contrast, unknown unknowns represent risks that fall entirely outside insurers' consideration set. Examples here include the cascade effects of COVID-19 on Cyber and BI insurance, and the scale of coordinated terrorist attacks prior to 9/11. Here insurers were blindsided. That said, it's not clear they failed any worse than other major social institutions, such as governments, markets, or businesses. This suggests these failures reflect fundamental epistemological limits rather than a particular institutional limit.

---

[464] Erwann Michel-Kerjan & Burkhard Pedell, *Terrorism Risk Coverage in the Post-9/11 Era: A Comparison of New Public–Private Partnerships in France, Germany and the U.S.*, 30 GENEVA PAP RISK INSUR ISSUES PRACT 144, n. 59 (2005).

[465] Erwann Michel-Kerjan & Burkhard Pedell, *How Does the Corporate World Cope with Mega-Terrorism? Puzzling Evidence from Terrorism Insurance Markets*, 18 JOURNAL OF APPLIED CORPORATE FINANCE 61, 16 (2006).